\documentclass{aa}

\usepackage{graphicx}
\usepackage{txfonts}

\usepackage{natbib}

\usepackage{hyperref}
\hypersetup{linkcolor=blue,citecolor=blue,colorlinks=true}

\usepackage{booktabs}

\newcommand{\eq}{Eq.~}

\newcommand{\Fig}{Fig.~}

\newcommand{\Sec}{Section~}

\newcommand{\App}{Appendix~}

\newcommand{\Tab}{Tab.~}

\begin{document}

    \title{Leptohadronic multimessenger modeling of 324 gamma-ray blazars\thanks{\Tab\ref{tab:all_parameters} is available in Appendix B of this preprint and in electronic form via \url{https://github.com/xrod/lephad-blazars} or at the CDS via \url{https://cdsarc.cds.unistra.fr/viz-bin/cat/J/A+A/681/A119}.}}

   \author{X. Rodrigues 
          \inst{1,2}
          \and
          V. S. Paliya
            \inst{3}
          \and
            S. Garrappa
            \inst{2}
          \and
            A. Omeliukh
            \inst{2}
          \and
            A. Franckowiak
            \inst{2}
          \and
            W. Winter
            \inst{4}
            }

    \institute{European Southern Observatory, Karl-Schwarzschild-Straße 2, 85748 Garching bei München, Germany
                \and
                Astronomical Institute, Fakultät für Physik und Astronomie, 
                Ruhr-Universität Bochum, 44780 Bochum, Germany
                \and
              Inter-University Centre for Astronomy and Astrophysics (IUCAA), SPPU Campus, 411007, Pune, India
              \and
              Deutsches Elektronen-Synchrotron DESY, 15738 Zeuthen, Germany
    }

   \date{Submitted 24 July 2023 | Accepted 20 October 2023}

 
  \abstract
  {The origin of the diffuse astrophysical neutrino flux observed by the IceCube experiment is still under debate. Multiple associations have been reported between high-energy neutrino events and individual blazars, such as the source TXS~0506+056, which are active galaxies with relativistic jets pointing toward Earth. From a theoretical perspective, the properties of these sources as neutrino emitters are not yet well understood.} 
   {By systematically modeling the effect of cosmic-ray protons on the multiwavelength data from the largest sample of bright gamma-ray blazars to date, we expect to learn about the multi-messenger nature of the active galaxy population as a whole, as well as the relationship between neutrino production and the multiwavelength spectrum of these sources.}
   {We predict the emitted multiwavelength and neutrino spectrum using a self-consistent numerical radiation model applied individually to each source in the sample. We then study the properties of the full population and identify empirical relations.
   We focus on public multiwavelength data from the radio to the gamma-ray bands from a sample of 324 blazars detected by the \textit{Fermi} Large Area Telescope (LAT), most of which are flat-spectrum radio quasars (FSRQs). This amounts to 34\% of all FSRQs in the latest \textit{Fermi} catalog. 
   }
   {We demonstrate that the optical and gigaelectronvolt gamma-ray broadband features are generally well described by electron emission, which helps for the location of the emission region relative to the central black hole to be constrained. For 33\% of the blazars in our sample, a description of the observed X-ray spectrum benefits from an additional component from proton interactions, in agreement with recent studies of individual IceCube candidate blazars. We show that, on average, blazars that are brighter in gigaelectronvolt gamma rays have a higher neutrino production efficiency but a lower best-fit baryonic loading. The predicted neutrino luminosity shows a positive correlation both with the observed flux of gigaelectronvolt  gamma rays and with the predicted flux of megaelectronvolt gamma rays. We also estimate the diffuse neutrino flux from gamma-ray blazars by extrapolating the result to the \textit{Fermi} population, and we show that it may be at the level of $\sim$20\% of the diffuse neutrino flux observed by IceCube, in agreement with current limits from stacking analyses. We discuss the implications of our results for future neutrino searches and suggest promising sources for potential detections.}
   {}


   \maketitle


\section{Introduction}
\label{sec:intro}

The origin of the cosmic rays and the astrophysical neutrinos is currently one of the central issues in high-energy astrophysics. The IceCube experiment, located at the South Pole, observes a diffuse flux of astrophysical neutrinos with energies between 100~TeV and a few petaelectronvolts~\citep{IceCube:2013low,IceCube:2014stg,IceCube:2015qii,IceCube:2015gsk,IceCube:2013cdw}, thought to be emitted when cosmic rays interact with ambient radiation (p$\gamma$) or matter (pp). Crucially, although these hadronic processes should be accompanied by the emission of high-energy gamma rays, so far the cosmic neutrino sky does not seem to significantly correlate with gamma-ray-emitting sources. This suggests that the photons emitted in cosmic ray interactions are being attenuated or otherwise losing their energy, either in the astrophysical source or en route to Earth~\citep{Murase:2015xka,Hooper:2016jls,Halzen:2018iak,Murase:2019pef,Halzen:2020clx}.

One of the most promising candidate source classes is that of blazars, which are active galactic nuclei (AGN) whose relativistic jet is pointing in a direction close to our line of sight~\citep{1995PASP..107..803U}. Because of this, the radiation that is emitted by the jet is relativistically boosted and so are the neutrinos potentially produced in the jet. 

At the population level, gamma-ray-bright blazars have been excluded as a major contributor to the IceCube diffuse flux and limits have been set on their collectively emitted flux of teraelectronvolt-petaelectronvolt neutrinos~\citep{IceCube:2016qvd,IceCube:2023htm}, although these limits depend on the average spectral shape of the emitted neutrino energy flux. \citet{Palladino:2018lov} have shown that for the blazar population to contribute significantly to the diffuse sub-petaelectronvolt IceCube flux, a numerous population of very low-luminosity sources would have to be very highly loaded in cosmic rays, while the sources that are brightest in gamma rays must necessarily be poor cosmic-ray accelerators, given the overall lack of associations mentioned above. Such a trend has also been suggested by multi-epoch blazar modeling~\citep{Petropoulou:2019zqp}. Furthermore, \citet{Petropoulou:2022sct} have recently proposed a model of magnetically loaded jets that may provide a natural explanation for this inverse relation between gamma-ray luminosity and cosmic-ray loading.

In spite of their subdominant contribution to the diffuse IceCube astrophysical flux, some observations seem to point increasingly toward some level of neutrino emission from blazars. From one side, this comes in the form of statistical correlations with certain blazar subclasses. For instance, \citet{Giommi:2020hbx} have demonstrated an association at the $3.2\sigma$ level between IceCube tracks and the positions of high- and intermediate-synchrotron peaked BL Lacs~\citep[HBLs and IBLs respectively, see also][]{Padovani:2014bha,Padovani:2016wwn}. More recently, \citet{Buson:2022fyf,Buson:2023irp} have highlighted a significant spatial association between a blazar catalog and the IceCube sky map with a p value $\sim10^{-7}$, while 
 subsequently \citet{Bellenghi:2023yza} obtained no significant correlation when utilizing an updated dataset.

Another important source of evidence are sporadic spatial associations between high-energy IceCube events and the positions of individual blazars. The blazar with the highest significance thus far is TXS~0506+056, a BL Lac object for which neutrino emission has been established at the $3.5\sigma$ level. This includes two observations: 1) a 290~TeV neutrino was detected by IceCube in 2017 in spatial coincidence with the source and in temporal coincidence with a six-month gamma-ray flare~\citep{TXS_MM}; and 2) archival IceCube data show an excess of neutrinos with energies of tens of teraelectronvolts around the source's position during 2014 and 2015, although this coincided with a low state in gamma rays~\citep{TXS_orphanflare}. Furthermore, after ten years of observations, the second hottest spot in the IceCube sky was coincident with the position of TXS 0506+056~\citep{Aartsen:2019fau}, which can be regarded as additional evidence in support of this blazar as a neutrino source candidate. The hottest spot from the same analysis is consistent with the position of the Seyfert galaxy NGC~1068~\citep{IceCube:2022der}, which also hosts an active black hole but is not a blazar.

Beside TXS~0506+056, IceCube has also detected high-energy events in spatial coincidence with other individual blazars of different classes, among which are PKS~1424-41~\citep{2016NatPh..12..807K,Gao:2016uld}, GB6~J1040+0617~\citep{2019ApJ...880..103G}, 3HSP~J095507.9+355101~\citep{Giommi:2020viy,2020ApJ...902...29P}, PKS~1502+106~\citep{Franckowiak:2020qrq}, and PKS~0735+178~\citep{Sahakyan:2022nbz}. These individual associations have been subsequently dissected by theoretical works by means of numerical leptohadronic modeling. For the 2017 flare of TXS~0506+056, some models suggest that the high-energy gamma rays co-emitted with the neutrinos interact through pair production with lower-energy photons in the jet, generating electromagnetic cascades whose energy is ultimately emitted by the source in the X-ray range~\citep[e.g.,][]{Cerruti:2018tmc,Gao:2018mnu,Keivani:2018rnh,Sahu:2020eep}. It is worth noting that while the very-high-energy gamma rays produced through hadronic processes cascade down to the X-ray regime, in the gigaelectronvolt gamma-ray range the source is optically thin in these models, which allows them to explain the bright state in gigaelectronvolt gamma rays observed during neutrino emission. The peak of this gigaelectronvolt emission as well as the peak of the optical emission are typically dominated by leptonic processes, while a potential hadronic signature is limited to the X-ray and, potentially, teraelectronvolt ranges.

As shown by~\citet{Padovani:2019xcv}, data show that TXS 0506+056 possesses a broad line region (BLR) surrounding the active black hole, which is a feature typical of flat-spectrum radio quasars (FSRQs). This leads to the reclassification of TXS~0506+056 as a masquerading BL Lac. This was also shown by~\citet{Sahakyan:2022nbz} to be the case  for the recent IceCube candidate PKS~0735+178, which motivated the authors to include a BLR when modeling the source. More recent modeling results by~\citet{VERITAS:2023eso} also support the existence of a BLR photon field interacting with the jet emission region in this source.

The presence of external photon fields form the BLR can lead both to enhanced neutrino production and  more extensive electromagnetic cascades, thus spreading the hadronic gamma rays in energy down to the megaelectronvolt or even X-ray regime. This external field model has been suggested as a viable scenario to explain the 2017 neutrino event from TXS 0506+056 and the simultaneous multiwavelength data~\citep{Keivani:2018rnh,Petropoulou:2019zqp,Zhang:2019htg} and has since been successfully applied to other candidate sources~\citep[e.g.,][]{Padovani:2019xcv,Rodrigues:2020fbu,Petropoulou:2020pqh,Oikonomou:2021akf,Gasparyan:2021oad,Sahakyan:2022nbz}. This model generally predicts a connection between neutrino emission and the fluxes in either X-rays or megaelectronvolt gamma rays, depending on the spectral shape of the external fields assumed and the bulk Lorentz factor of the jet. 

Under certain conditions, theoretical works have shown that electromagnetic cascades can be extremely efficient in attenuating gigaelectronvolt gamma rays, to the point where neutrino emission could theoretically be correlated with a flare in the megaelectronvolt band while a low flux could simultaneously be observed by the \textit{Fermi} Large Area Telescope (LAT) in gigaelectronvolt gamma rays. This scenario has been tested on the 2014-2015 emission from TXS 0506+056, by invoking high-density external photon fields originating in a BLR~\citep[][]{Reimer:2018vvw,Rodrigues:2018tku,Petropoulou:2019zqp} or in the black hole corona, with the neutrino emission zone lying in the close vicinity of the black hole \citep[e.g.,][]{Xue:2020kuw}. \citet{Petropoulou:2020pqh} have also tested a similar corona model for the extreme blazar 3HSP~J095507.9+355101, leading to very efficient neutrino production accompanied by bright emission of megaelectronvolt photons. On the one hand, this type of model could in principle help explain the general lack of associations between IceCube events and \textit{Fermi}-LAT sources, which inspired the idea of gamma-ray-suppressed neutrino flares~\citep{Kun:2023uld}. On the other hand, such high optical thicknesses generally require an environment with extremely high photon densities or Doppler boosts, which are possible only in very specific regions of parameter space. Under less extreme conditions, we should more likely expect some level of gigaelectronvolt emission during neutrino flares, as shown by the more ``conventional'' one-zone models cited previously. Furthermore, during hadronic flares there should be an overall increase in gamma-ray emission in the jet and it is unclear whether the additional attenuation is always sufficient to counterbalance this effect. 


In this work, we perform individual modeling of 324 gamma-ray blazars from the Candidate
Gamma-ray Blazar Survey (CGRaBS) catalog~\citep{Healey:2007gb}, 237 of which are FSRQs. This corresponds to 34\% of all 619 FSRQs in the latest \textit{Fermi}-LAT catalog by~\citet{Fermi-LAT:2019pir}. 

The sources in this sample have relatively high gamma-ray luminosities, typically above $10^{45}~\mathrm{erg}\,\mathrm{s}^{-1}$. On the one hand, the majority of the models cited above predict these sources to be efficient neutrino emitters \citep[e.g.,][]{Murase:2014foa}; on the other hand, as discussed above, bright blazars are the most constrained by IceCube stacking analyses~\citep{IceCube:2016qvd,IceCube:2023htm}, which can place strong limits on their overall emission. The objective of this work, besides characterizing the physical parameters of each source in the sample through model fitting, is to provide physics-driven predictions of their potential neutrino emission, both at the individual level and as a population.

We model each source in the sample by numerically simulating the radiative  interactions of cosmic-ray electrons and protons accelerated in the relativistic jet, in a one-zone framework. We then compare the predicted multiwavelength emission from each source to multiwavelength observations based on public data ranging from the radio band up to gamma rays, compiled by \citet{Paliya:2017xaq}. We fit the data at frequencies above 300 GHz and thus constrain the source parameters. 

In those cases where a proton contribution helps us explain the multiwavelength data, we present the emitted neutrino spectrum calculated self-consistently; in those cases where the multiwavelength data are well explained solely by electron emission, we set upper limits on the neutrino emission from the source.  In the case of FSRQs, we additionally consider external photon fields from thermal and broad line emission present in the BLR. We then connect the individual results of the modeling of each source to draw conclusions on the sample as a whole, with an emphasis on potential proton acceleration and neutrino emission. Compared to previous leptohadronic studies of blazar samples \citep[see e.g.,][]{Tavecchio:2009zb,Boettcher:2013wxa,Petropoulou:2015upa,Cerruti:2014iwa,Oikonomou:2019djc,Liodakis:2020dvd}, this work presents the largest self-consistently modeled sample of blazar AGN to date. 

Some of the IceCube candidates mentioned above are included in this sample and we discuss their potential as neutrino emitters in the context of this model, while also comparing our results with the literature. We place blazar TXS~0506+056, which is not included in our sample, in the context of the results obtained. Finally, we show how several blazars in this sample may be promising source candidates for future neutrino detectors such as IceCube-Gen2~\citep{IceCube-Gen2:2020qha}. 

The paper is organized as follows: in \Sec\ref{sec:methods} we describe the sample, multiwavelength data, physical model, and fitting method. In \Sec\ref{sec:results} we present the best-fit parameters, the predicted cosmic-ray acceleration by each source, and the respective flux of emitted neutrinos. In \Sec\ref{sec:neutrinos} we discuss the impact of our results in the context of future multi-messenger searches and suggest potential candidates in the sample for neutrino source searches that may help support or exclude the leptohadronic paradigm for high-luminosity blazars. In \Sec\ref{sec:limitations} we discuss some limitations of the model and we summarize our main conclusions in \Sec\ref{sec:conclusion}.

\section{Methods} \label{sec:methods}

\subsection{Source sample}
\label{sec:methods_data}

The CGRaBS catalog~\citep{Healey:2007gb} is a flux-limited ($F_{\rm 8.4~GHz}>65$ mJy) sample of 1625 radio-loud AGN thought as potential gamma-ray emitting quasars to be detected with the \textit{Fermi}-LAT. From this catalog, \citet{Paliya:2017xaq} selected 324 sources that were present in any of the \textit{Fermi}-LAT catalogs available at the time and which also have multifrequency observations. From these observations, they put together a multiwavelength SED for each source. This includes near-infrared data from ESO's GROND~\citep{Greiner:2008ms}, optical-ultraviolet data from the \textit{Swift} Ultra-Violet Optical Telescope~\citep[UVOT,][]{Roming:2005hv}, X-ray data from NuSTAR~\citep{NuSTAR:2013yza}, XMM-Newton~\citep{Matthews:2000jw}, Chandra~\citep{Weisskopf:2000tx}, and the \textit{Swift} X-ray Telescope~\citep[XRT,][]{Burrows:2005gfa} and Burst Alert Telescope~\citep[BAT,][]{Krimm:2013lwa}, as well as gamma-ray data from the \textit{Fermi}-LAT~\citep{Fermi-LAT:2009ihh}. They have also relied on archival data provided by the ASI Data Center (ASDC). Regarding \textit{Fermi}-LAT data, 310 sources are included in the 3LAC~\citep{Fermi-LAT:2015bhf}, which was the most up-to-date at the time of the analysis by \citet{Paliya:2017xaq}, 8 in the 2LAC~\citep{2011ApJ...743..171A}, and two in the 1LAC~\citep{Abdo:2010ge}. We refer the reader to \citet{Paliya:2017xaq} for more details on the multiwavelength data.

\begin{figure*}[htpb!]
\includegraphics[width=\textwidth,trim={0 2.5cm 0 0}, clip]{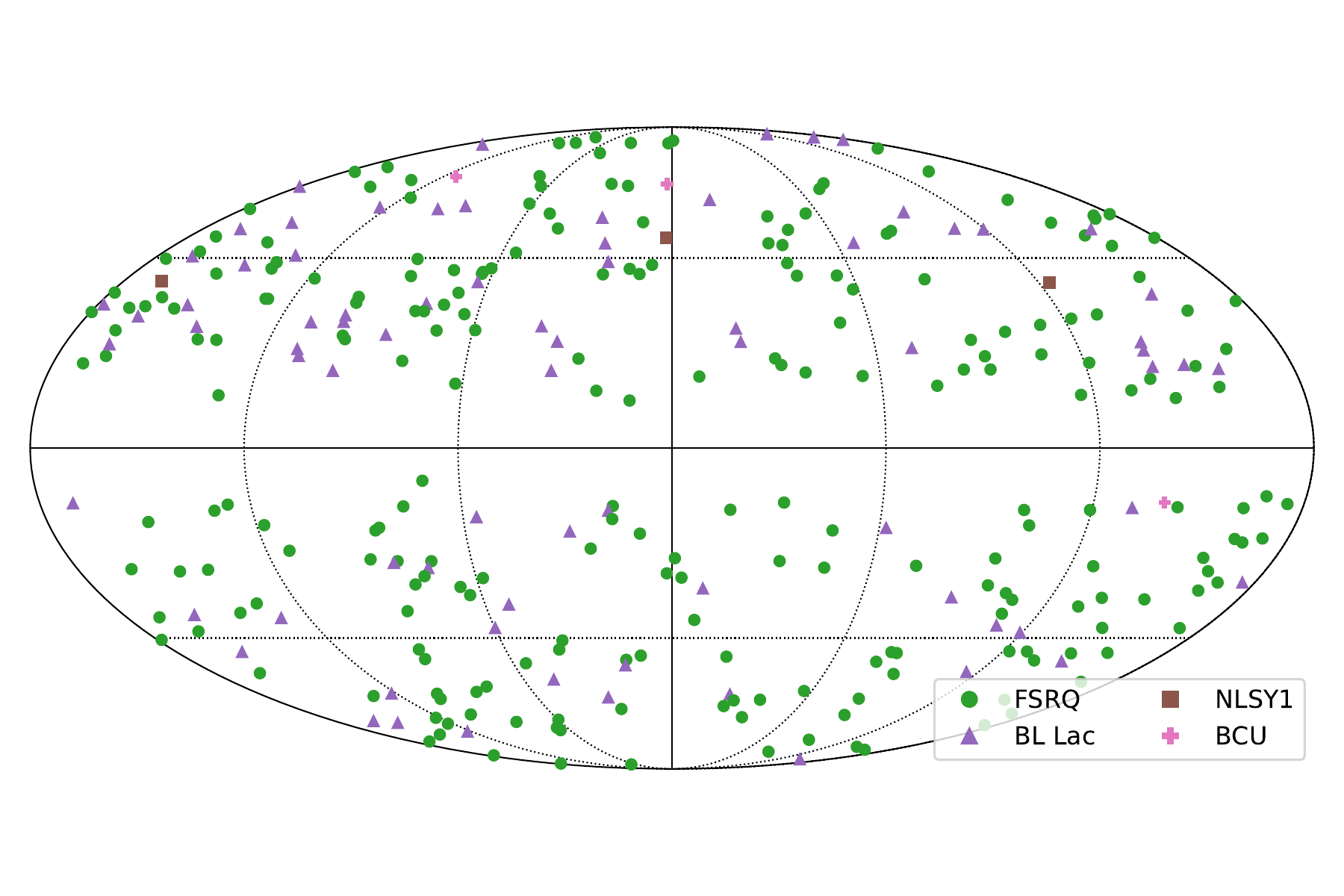}
\includegraphics[width=0.5\textwidth,trim={0 0 0 0}, clip]{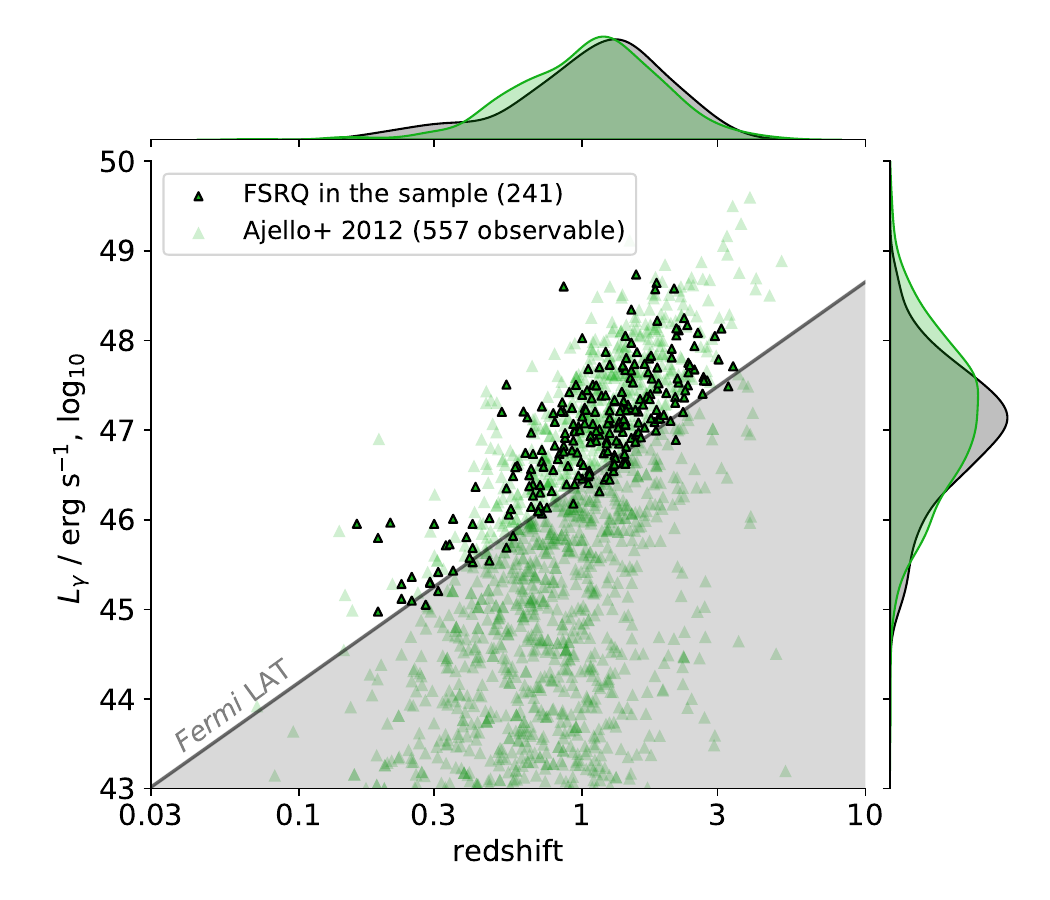}
\includegraphics[width=0.5\textwidth,trim={0 0 0 0}, clip]{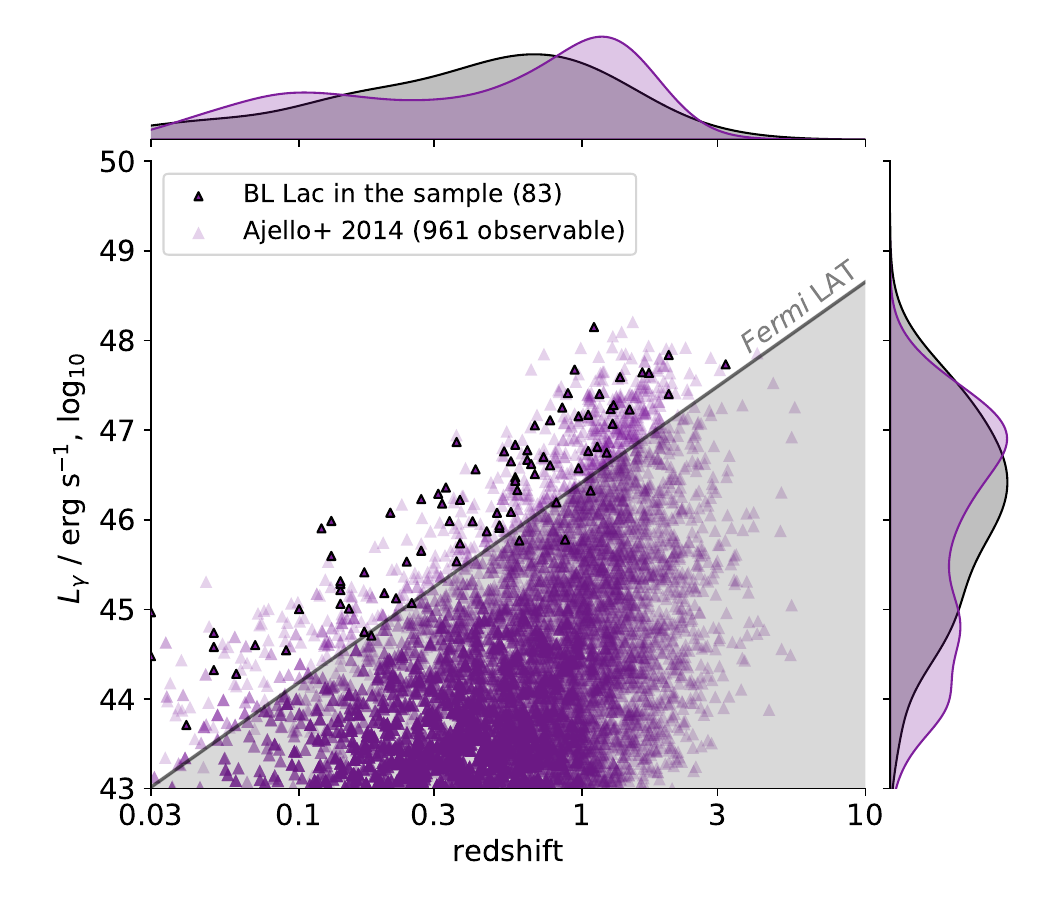}
\caption{Characterization of the sample. \textit{Upper panel:} Sky map with the positions of the sources studied in this work, which include 237 FSRQs, 88 BL Lacs, three narrow-line Seyferts, and three objects classified as BCUs in the \textit{Fermi} catalogs. \textit{Lower panels:} distribution of FSRQs \textit{(left)} and BL Lacs \textit{(right)} in the sample as a function of redshift and gamma-ray luminosity (dark points). For comparison, we show as lighter points to the distribution of the respective overall population, based on the luminosity functions by \citet{Ajello:2011zi,Ajello:2013lka}. The gray area marks the subthreshold region for the \textit{Fermi}-LAT. The black distributions alongside the margins refer to the sample; the green (left) and purple (right) distributions refer to the population of sources detected by the \textit{Fermi}-LAT (i.e., excluding the putative sources below the sensitivity threshold).} 
\label{fig:source_distribution}
\end{figure*}

The upper panel of \Fig\ref{fig:source_distribution} shows a sky map with the positions of the objects in the sample. This amounts to a total of 324 blazars, including 237 FSRQs, 38 low-frequency-peaked BL Lacs (LBLs), 22 intermediate-frequency-peaked BL Lacs (IBLs), 17 high-frequency-peaked BL Lacs (HBLs), 3 narrow-line Seyfert 1 galaxies (NLS1), and 3 sources classified as blazar candidates of uncertain type (BCUs) in the \textit{Fermi} 3LAC or 4LAC~\citep{Fermi-LAT:2015bhf,Fermi-LAT:2019pir}. Regarding the physical model, we apply to NLS1 galaxies the same geometric assumptions as used to describe FSRQs \citep[see e.g.,][]{2019ApJ...872..169P}, while for BCUs we make the conservative assumption that they are described by the simpler BL Lac model (cf.~\Sec\ref{sec:methods_model} where these differences in the modeling are described). 

In the lower panels of \Fig\ref{fig:source_distribution}, we show the distribution of sources as a function of redshift and gamma-ray luminosity. The sources in the sample are those shown as dark markers (FSRQs in the left and BL Lacs on the right). For comparison, we show as light markers a source distribution sampled from the luminosity function of FSRQs on the left~\citep{Ajello:2011zi} and BL Lacs on the right~\citep{Ajello:2013lka}. Above the \textit{Fermi}-LAT sensitivity, represented here as a line corresponding to a flux of $4\times10^{-12}\,\mathrm{erg}\,\mathrm{cm}^{-2}\mathrm{s}^{-1}$, this distribution of light markers follows the population of detected sources. Under that line, it is an extrapolation obtained by Monte Carlo sampling of the luminosity functions derived by \citet{Ajello:2011zi,Ajello:2013lka}. As we can see, the current sample covers a range of luminosities and redshifts comparable to that of the source population detected by the \textit{Fermi}-LAT. At the same time, as we can see in the histograms alongside the margins of the lower plots, the present sample (black histograms) does not follow exactly the population of \textit{Fermi}-LAT-detected sources (green and purple histograms). In particular, our FSRQ sample somewhat over-represents sources with $L_\gamma\sim10^{47}\,\mathrm{erg}\,\mathrm{s}^{-1}$ and under-represents the source population for $L_\gamma\gtrsim3\times10^{47}\,\mathrm{erg}\,\mathrm{s}^{-1}$. Regarding BL Lacs, we see that sources lying at $z\gtrsim1$ as well as $L_\gamma\lesssim\times10^{45}\,\mathrm{erg}\,\mathrm{s}^{-1}$ are under-represented in our sample. These considerations will play a role in the extrapolation of our results to the \textit{Fermi}-LAT population in \Sec\ref{section:neutrinos_diffuse}.

The redshift of each source is adopted from the respective \textit{Fermi}-LAT catalog, while the black hole mass and accretion disk luminosity values are those reported by \citet{Paliya:2017xaq}. In each case, these were deduced following one of the following three approaches:  1) by modeling the big blue bump, if detected at infrared to ultraviolet frequencies, with the standard optically thick, geometrically thin accretion disk model \citep{1973A&A....24..337S}; 2) from the optical emission line and continuum luminosities by adopting the virial technique \citep[cf.][]{2011ApJS..194...45S,2021ApJS..253...46P}, or 3) when both of the above-mentioned methods were not possible, the disk luminosity was estimated assuming $L_{\rm disk}\sim40\,L_{\gamma}^{0.93}$~\citep{2012MNRAS.421.1764S}. In such cases, a black hole mass was assumed $M_{\rm BH}=5\times10^8$~M$_{\odot}$.

In this work we perform individual leptohadronic modeling of each of the 324 blazars in this sample. The steps of this procedure are detailed in the following sections.

\subsection{One-zone radiation model}
\label{sec:methods_model}

We self-consistently model the emission of photons and secondary particles by a population of nonthermal electrons and protons interacting in the blob, using the time-dependent numerical framework AM$^3$~\citep{Gao:2016uld}.

We simulate the radiative zone in the jet as a spherical blob of radius\footnote{We use primed symbols to refer to the quantities as measured in the rest frame of the relativistic jet. Unprimed quantities refer to the rest frame of the black hole, unless explicitly mentioned that they are given in the observer's frame.} $R^\prime_\mathrm{b}$ permeated by a uniform magnetic field of strength $B^\prime$ and moving along the jet with a bulk Lorentz factor of $\Gamma_\mathrm{b}$, as shown in \Fig\ref{fig:model}. We assume the jet is observed at an angle $\theta_{\mathrm{obs}}=1/\Gamma_\mathrm{b}$ relative to its axis, resulting in a Doppler factor $\delta_{D}=\Gamma_\mathrm{b}$. We refer to the distance of the blob to the supermassive black hole as the dissipation radius, $R_\mathrm{diss}$.

Electrons (protons) are assumed to be accelerated to simple power-law distributions, $dN_{e(p)}/d\gamma^\prime_\mathrm{e}\propto \gamma^{\prime-\alpha_{\mathrm{e(p)}}}$, from a Lorentz factor  $\gamma_\mathrm{e(p)}^{\prime\mathrm{min}}$ up to $\gamma_\mathrm{e(p)}^{\prime\mathrm{max}}$. The total power injected in electrons (protons) into the radiation zone is given by an injection luminosity $L^\prime_\mathrm{e(p)}$. These and other parameters of the model are listed in \Tab\ref{tab:parameters}. We also assume that all particles, both charged and neutral,
escape the radiation zone advectively, that is, the escape timescale is the light-crossing time of the blob, $t^\prime_\mathrm{esc}=R^\prime_\mathrm{b}/c$.

\begin{table}[]
    \caption{List of model parameters.}
    \label{tab:parameters}
    \begin{tabular}{ll}
    \toprule
    Parameter & Search range\\
    \midrule
    $R^\prime_\mathrm{blob}$ / cm & [$10^{14}$,$10^{17}$]\\
    $B^\prime$ / gauss & [$10^{-1}$,$10^{1}$]\\
    $\Gamma_\mathrm{b}$ & [3.0,35.0]\\
    $R_\mathrm{diss}/R_\mathrm{BLR}$ & [1.0,4.0]\\
    $\gamma_\mathrm{e}^{\prime\mathrm{min}}$ & [$10^{1}$,$10^{3}$]\\
    $\gamma_\mathrm{e}^{\prime\mathrm{max}}$ & [$10^{3}$,$10^{7}$]\\
    $\alpha_\mathrm{e}$ & [1.0,3.0]\\
    $L^\prime_\mathrm{e}$ / erg s$^{-1}$ & [$10^{39}$,$10^{45}$]\\
    
    $\gamma_\mathrm{p}^{\prime\mathrm{max}}$ & [$10^{5}$,$10^{7}$]\\
    $L^\prime_\mathrm{p}/L^\prime_\mathrm{e}$ & [$10^{-2}$,$10^{5}$]\\
    \midrule
    $\gamma_\mathrm{p}^{\prime\mathrm{min}}$ & fixed to $10^2$\\
    $\alpha_\mathrm{p}$ & fixed to 1.0\\
    \midrule
    $L_\mathrm{disk}$& deduced by~\citet{Paliya:2017xaq}\\
    $R_\mathrm{BLR}$ & derived from $L_\mathrm{disk}$\\
    $R_\mathrm{torus}$ & derived from $L_\mathrm{disk}$\\
    \bottomrule
    \end{tabular}
    
    Note: The first eight parameters are searched by optimizing the leptonic model, while the proton properties are searched with the full leptohadronic model. The remaining parameters are phenomenologically constrained.
\end{table}

In every source simulation the time-dependent solver is evolved until a steady-state is reached, that is, an equilibrium between the injection and escape (or cooling) of electrons and protons. The emitted photon spectrum then becomes constant in time. At high enough energies, these photons interact efficiently with the EBL, leading to an attenuation of the gamma-ray fluxes at very high energies. We adopt the EBL model by \citet{Dominguez:2010bv} and calculate this energy-dependent flux attenuation as a final step after each the source simulation. For this we use the energy-dependent attenuation length values tabulated in Fermipy~\citep{Wood:2017yyb}.

\subsection{External radiation fields}
\label{sec:external}
Out of the 324 sources in our sample 241 are FSRQs or narrow-line Seyferts, which typically possess a bright accretion disk and a BLR surrounding the supermassive black hole. As an estimate of the disk luminosity $L_\mathrm{disk}$ and black hole mass $M_\mathrm{BH}$ of each source we rely on the values derived by \citet{Paliya:2017xaq}, as discussed in \Sec\ref{sec:methods_data}. Given the disk luminosity, we model the photon spectrum emitted by the disk as a multi-temperature Shakura-Sunyaev distribution, whose profile depends on the value of $M_\mathrm{BH}$ estimated for that same source~\citep{Shakura:1976xk}.

\begin{figure*}[htpb!]
\includegraphics[width=\textwidth,trim={0 0 0 0}, clip]{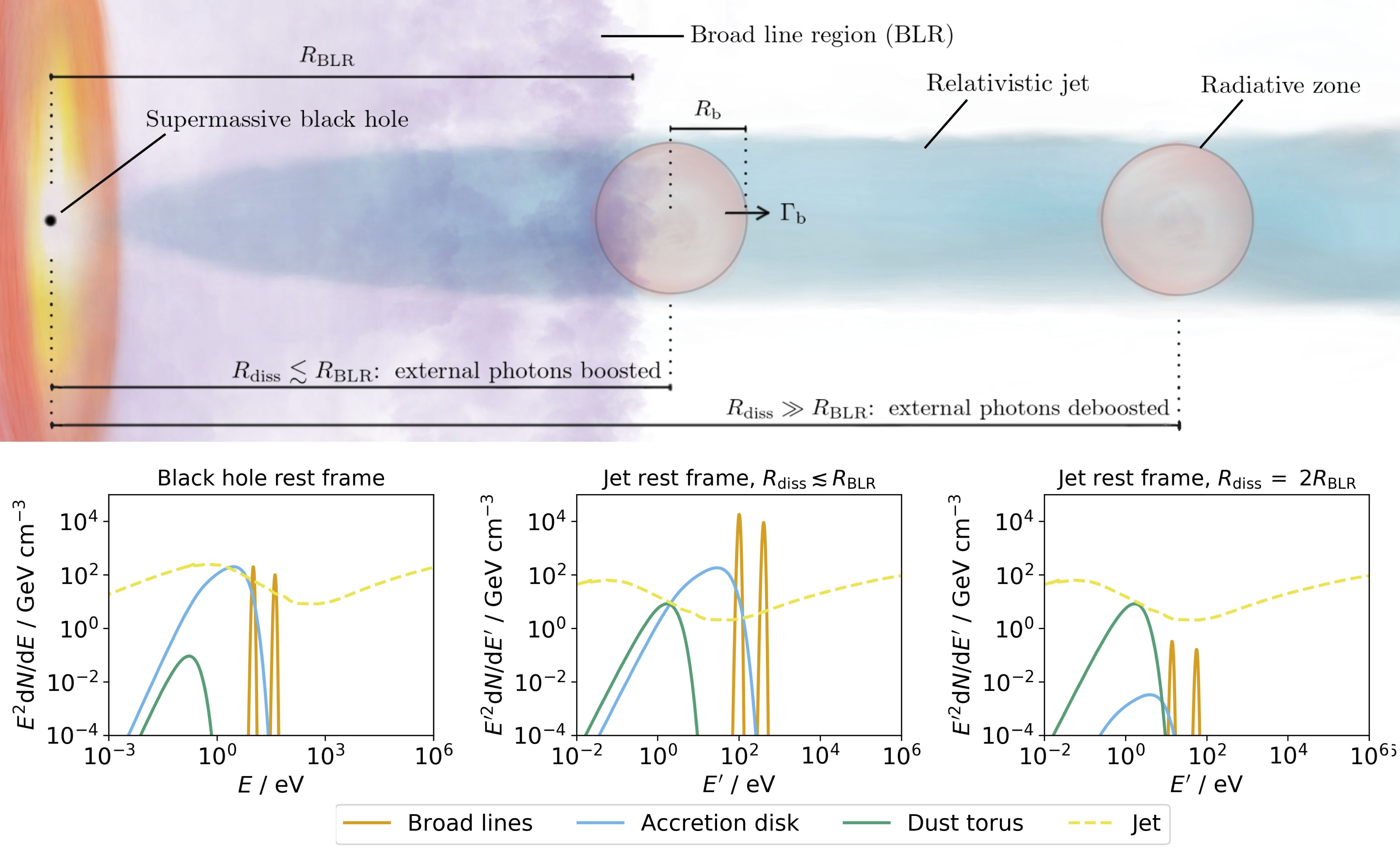}
\caption{Schematic representation of two possible geometries of the one-zone FSRQ model. The broad line region is represented by the purple clouds. In the case of the blob represented on the left, its proximity to the BLR implies a large relative Doppler factor, making the external fields appear highly boosted in the jet frame (compare left-hand plot, where the external fields are shown in the rest frame of the central engine, with the middle plot, where they are shown as energy density spectra in the jet frame). In the case represented on the right, due to the high dissipation radius $R_\mathrm{diss}$ the BLR photons impinge more from behind and the external fields from the BLR are less boosted, as shown in the lower right plot. The relativistic boosting of these external radiation fields follows the treatment by~\citet{Ghisellini:2009wa}. We note that on the left plot, the broad lines are isotropic in the black hole frame, while the disk emission is anisotropic except for the $\sim1\%$ that gets isotropized through Thomson scattering. This explains why in the middle plot the broad lines get more highly boosted than the disk emission.}
\label{fig:model}
\end{figure*}

Following \citet{Ghisellini:2009wa}, we model the BLR as a thin shell located at a radius $R_\mathrm{BLR}=10^{17}(L_\mathrm{disk}/10^{45}~\mathrm{erg}\,\mathrm{s}^{-1})^{1/2}~\mathrm{cm}$ around the supermassive black hole. We also consider a dusty torus of radius $R_\mathrm{torus}=2.5\times10^{18}(L_\mathrm{disk}/10^{45}~\mathrm{erg}\,\mathrm{s}^{-1})^{1/2}~\mathrm{cm}$.

We assume that 1\% of the luminosity of the thermal emission from the accretion disk is
Thomson-scattered in the BLR, leading to an isotropic component in the BLR volume that has the same spectral shape as the disk emission and a total luminosity of $0.01\,L_\mathrm{disk}$. A more significant
10\% of the disk luminosity is isotropized in the volume of the BLR through broad line emission~\citep{Greene:2005nj}.
Following previous studies ~\citep[e.g.,][]{Murase:2014foa,Rodrigues:2017fmu,Rodrigues:2018tku}, we consider the two lines that are typically strongest, namely the $\alpha$ lines of the Lyman series of hydrogen (with peak energy 10.2~eV and total luminosity $0.1\,L_\mathrm{disk}$) and helium (40.8~eV and $0.05\,L_\mathrm{disk}$). We model their spectra in the rest frame of the  black hole as Gaussian distributions with a half-width of 5\% of their respective peak energies. For the infrared emission from the torus we consider a fixed temperature of 500~K and a covering factor of 0.3. As a conservative assumption we ignore the contribution of potential corona emission, which in any instance would only contribute significantly to the blob system if it  lied extremely close to the black hole.

The luminosity of the BLR photons, $L_\mathrm{BLR}$, is converted into an energy density in the blob rest frame $u^\prime_\mathrm{BLR}$ through the following relation:
\begin{align}
u^\prime_\mathrm{BLR}(L_\mathrm{BLR},\Gamma_\mathrm{b},R_\mathrm{diss},R_\mathrm{BLR})=
\frac{{\delta_\mathrm{BLR}^{\mathrm{rel}}}^2(\Gamma_\mathrm{b},R_\mathrm{diss},R_\mathrm{BLR})L_\mathrm{BLR}}{4\pi \,R_\mathrm{ext}^2\,c},
\label{eq:BLR}
\end{align}
where $\delta^\mathrm{rel}_\mathrm{BLR}$ is the relative Doppler factor with which the radiation is boosted into the blob, which depends on the blob's relative position to the BLR. As illustrated in the diagram of \Fig\ref{fig:model}, in the cases where $R_\mathrm{diss}\leq R_\mathrm{BLR}$ we have $\delta^\mathrm{rel}_\mathrm{BLR}=\Gamma_\mathrm{b}$. For $R_\mathrm{diss} > R_\mathrm{BLR}$, the external photons impinge on the blob less frontally, leading to lower effective Doppler factors. For example, for a value of $\Gamma_\mathrm{b}=10$ and $R_\mathrm{diss} = 4R_\mathrm{BLR}$, we have ${\delta^\mathrm{rel}_\mathrm{BLR}}^2\approx0.1$ and the external fields from the BLR are unlikely to play a significant role in the blob emission. For the treatment of $\delta_\mathrm{BLR}^\mathrm{rel}$ we follow \citet{Ghisellini:2009wa}. Note that the energies of the BLR photons in the blob frame are also shifted by a factor $E^\prime=\delta^\mathrm{rel}_\mathrm{BLR}\,E$ and therefore also depend on both $\Gamma_\mathrm{b}$ and $R_\mathrm{diss}$.

For the infrared emission from the dust torus, \eq\ref{eq:BLR} also applies, with the substitutions $R_\mathrm{BLR}\rightarrow R_\mathrm{torus}$ and $\delta^\mathrm{rel}_\mathrm{BLR}\rightarrow\delta^\mathrm{rel}_\mathrm{torus}$. For all the cases tested here, the blob lies within the radius of the torus, so that $\delta^\mathrm{rel}_\mathrm{torus}=\Gamma_\mathrm{b}$.

\subsection{Parameter search}
\label{sec:methods_optimization}

\begin{figure*}[htpb!]
\centering
\includegraphics[width=0.9\textwidth,trim={0 1cm 0 0}, clip]{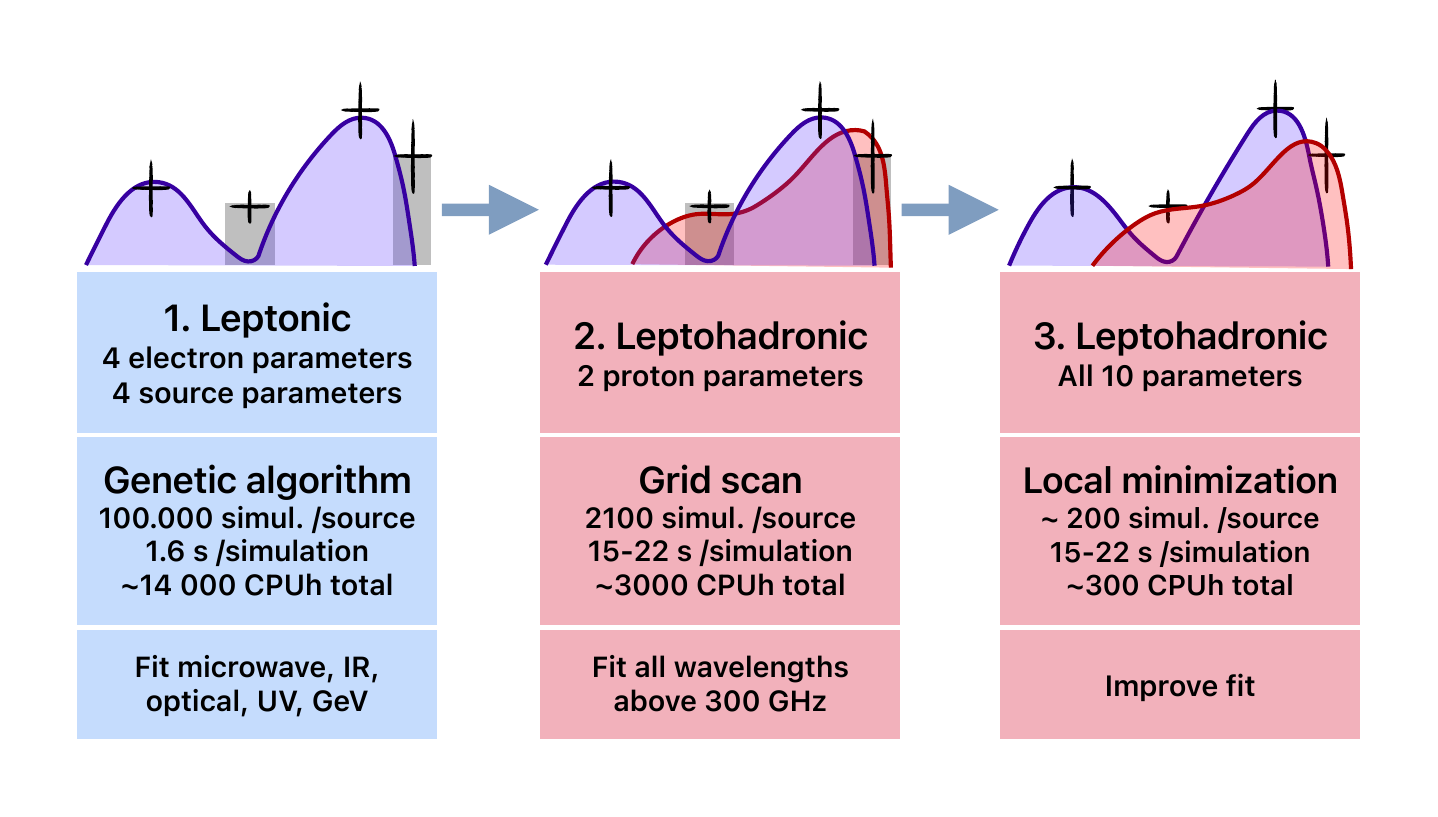}
\caption{Flowchart of the fitting procedure. The leptohadronic fit of the data from each source is performed as a three-step process. The initial step, where the optical and gamma ray broadband features are explained with a leptonic model, requires the most simulations because the vast parameter space is initially unconstrained. The number of CPU-hours quoted refers to the entire sample and is based on the code's performance of 1.6~s for a single leptonic simulation and $\sim$20~s for a leptohadronic simulation.}
\label{fig:fitting_method}
\end{figure*}

We search the parameter space of the leptohadronic model using a three-step method, represented schematically in \Fig\ref{fig:fitting_method}. First, for each source the  data in the infrared, optical, ultraviolet and gamma-ray bands are fit using a purely leptonic model. This is motivated by the fact that the optical and GeV gamma-ray broadband features of blazars can typically be well explained with leptonic models, even if protons are also assumed to be co-accelerated in the source, as shown by several leptohadronic modeling efforts of IceCube source candidates~\citep[e.g.,][]{Gao:2018mnu,Rodrigues:2018tku,Sahakyan:2022nbz}. At this stage we treat X-ray fluxes merely as upper limits for electron emission, although for some sources the X-ray data may actually be fit well by this emission. Radio fluxes (below 300~GHz) are considered as upper limits for the fit, because the relatively small radiation zone responsible for the emission of high-energy photons and neutrinos is necessarily optically thick to radio emission due to synchrotron self-absorption. These fluxes must therefore originate from larger emission regions, presumably located downstream in the relativistic jet.

By fitting the optical and GeV gamma-ray peaks with a leptonic model, we find the best-fit values for the four parameters describing the accelerated electron population and the four parameters describing the source properties. These are the first eight parameters listed in~\Tab\ref{tab:parameters}. We search this parameter space using a genetic algorithm \citep[e.g.,][]{goldberg89}, a metaheuristic whose stochastic nature is adequate for searching large and poorly constrained parameter spaces \citep[see][where a similar method was used]{Rodrigues:2018tku,Rodrigues:2020fbu}. Rather than requiring an initial guess, this method relies on a population of parameter sets that is originally randomly distributed within certain parameter boundaries. The values of these boundaries  are given in \Tab\ref{tab:parameters} for each parameter. In every iteration, the subsequent generations are optimized through the genetic heuristic.

The goodness of fit is judged by means of a chi-squared test between the predicted spectrum and the multiwavelength data in the frequency bands we wish to fit. For data points falling in the radio, X-rays, and gamma rays above 20~GeV, the model only adds to the chi-squared value if the predicted flux overshoots the observations. This allows us to negatively score leptonic solutions that overshoot the observed frequencies without actually fitting data in these bands:
\begin{align}
\begin{split}
    \chi^2_\mathrm{partial} =\frac{1}{N-N^\mathrm{lep}_\mathrm{par}+1}
    \Bigg[
&\sum\limits_i^{\mathrm{microwave~to~UV,}~\sim\mathrm{GeV}}\frac{(F_i-\bar F_i)^2}{\sigma_i^2}\,+ \\
    &\sum\limits_i^{\mathrm{radio,~X-ray,~>20\,GeV}}\frac{(F_i-\bar F_i)^2}{\sigma_i^2}\,\theta(F_i-\bar F_i) \Bigg],
\end{split}
\label{eq:chi2_lep}
\end{align}
where $N$ is the number of data points to fit, $N_\mathrm{par}^\mathrm{lep}=8$ is the number of free parameters of the leptonic model and $F_i$ and $\bar F_i$ are the predicted and observed photon flux values, respectively.

Having constrained the source geometry and the accelerated electrons, the only free variables left are those describing the accelerated protons. We fix the minimum proton Lorentz factor to $\gamma_\mathrm{p}^\mathrm{min}=10^2$ and the proton spectral index to $\alpha_\mathrm{p}=1.0$. The choice of fixing $\gamma_\mathrm{p}^\mathrm{min}$ is motivated by the fact that this parameter plays a minor role in the emission of secondary particles and radiation. In the case of $\alpha_\mathrm{p}$, this should play a role in the shape of the secondary photon emission; however, the peak proton energy is the leading factor in terms of neutrino emission.

We search the two remaining parameters, $\gamma_\mathrm{p}^{\prime\mathrm{max}}$ and $L_\mathrm{p}^\prime$, using a grid scan of $35\times70$ points, while fixing the remaining parameters to their best-fit values obtained in the previous step. The value ranges spanned by the grid scan are given in \Tab\ref{tab:parameters}. As we can see, we scan values of $10^5<\gamma_\mathrm{p}^{\prime\mathrm{max}}<10^7$. The motivation for this search range is as follows: for proton Lorentz factors below this range, hadronic interactions should be inefficient in this model; above this range, the emitted neutrinos have energies that are too high compared to the sensitivities of current experiments (cf. \Sec\ref{sec:limitations}).

In this second step, we wish to fit the X-ray and TeV gamma-ray data as well. The gamma rays emitted by proton interactions typically trigger electromagnetic cascades that lead to emission at lower frequencies, typically in the X-ray~\citep{Gao:2018mnu,Oikonomou:2021akf,Rodrigues:2020fbu,Sahakyan:2022nbz} or MeV gamma-ray band~\citep{Rodrigues:2018tku,Reimer:2018vvw,Petropoulou:2019zqp}. The best fit should therefore minimize the chi-squared value including X-rays and TeV gamma rays:
\begin{equation}   \chi^2_\mathrm{LepHad} =\frac{1}{N-N^\mathrm{LepHad}_\mathrm{par}+1}\sum\limits_i^{\mathrm{\nu>300~\mathrm{GHz}}}\frac{(F_i-\bar F_i)^2}{\sigma_i^2},
\label{eq:chi2}
\end{equation}
where $N^\mathrm{LepHad}_\mathrm{par}=10$ is the number of free parameters of the full leptohadronic model.

Finally, we perform a local minimization in order to improve the fit (right panel of \Fig\ref{fig:fitting_method}). Here, we consider all 10 best-fit leptohadronic parameter values and vary them continuously until the local minimum of $\chi^2_\mathrm{LepHad}$ is reached. The minimization is performed using iminuit~\citep{iminuit}. While the goal of the previous steps was to constrain the vast parameter space of each source, the local minimization step is instead more confined, yielding a more precise result. 

We consider the result with the minimum value of $\chi^2_\mathrm{LepHad}$ to be the best-fit result, since it best describes the multiwavelength dataset relative to other results. At the same time, its absolute value is not a direct indicator of the overall quality of the fit. This is because, as discussed in \Sec\ref{sec:methods_data}, the datasets are not simultaneous and often contain points spanning a wide range of flux values within the same frequency band, reflecting the source's intrinsic variability. For sources with high variability, the lowest possible value of $\chi^2_\mathrm{LepHad}$ may therefore lie considerably above unity.

Once we find the best-fit result for each source following the criteria described above, we then vary the proton luminosity parameter $L_\mathrm{p}^\prime$ in order to generate error estimates for the respective neutrino emission. We start with the best-fit value of $L_\mathrm{p}^\prime$ and vary it both upward and downward in logarithmic increments of $\log10(L_\mathrm{p}^\prime)=\pm0.1$, until we reach a value $\chi^2_\mathrm{LepHad}-\chi^{2\,\mathrm{best-fit}}_\mathrm{LepHad}=1$. This way we generate a $1\sigma$ uncertainty region for $L_\mathrm{p}^\prime$, which is the parameter to which the neutrino flux is most sensitive.

In the cases where the best fit is leptonic, that is, $L_\mathrm{p}^\prime=0$, we cannot define a best-fit maximum proton Lorentz factor $\gamma_\mathrm{p}^{\prime\,\mathrm{max}}$. In those cases, we test the injection of protons with $\gamma_\mathrm{p}^{\prime\,\mathrm{max}}=10^6$, which lies logarithmically in the middle of the range tested, and we determine the maximum value of $L_\mathrm{p}^\prime$ for which $\chi^2_\mathrm{LepHad}-\chi^{2\,\mathrm{best-fit}}_\mathrm{LepHad}\leq1$, by following the procedure described above. This represents the upper limit on the proton luminosity and the respective neutrino flux.

\section{Results}
\label{sec:results}

\subsection{Spectral fits}
\label{sec:results_seds}

Using the methods described in the previous section, we self-consistently calculate the predicted multiwavelength and neutrino spectra from the sample. The best-fit parameter values of each source are listed in full in \Tab\ref{tab:all_parameters} in descending order of the predicted neutrino flux. This includes the uncertainties on $L_\mathrm{p}^\prime$, obtained with the method described above, as well as the resulting uncertainties on the emitted neutrino flux $F_{\nu_\mu}$. We list first the sources for which the best-fit proton luminosity is incompatible with zero within 1$\sigma$, then those where it is nonzero but compatible with zero within 1$\sigma$, and finally those where only an upper limit can be established for the proton luminosity and the respective neutrino flux.

\begin{figure*}[htpb!]
\centering
\includegraphics[width=0.48\textwidth]{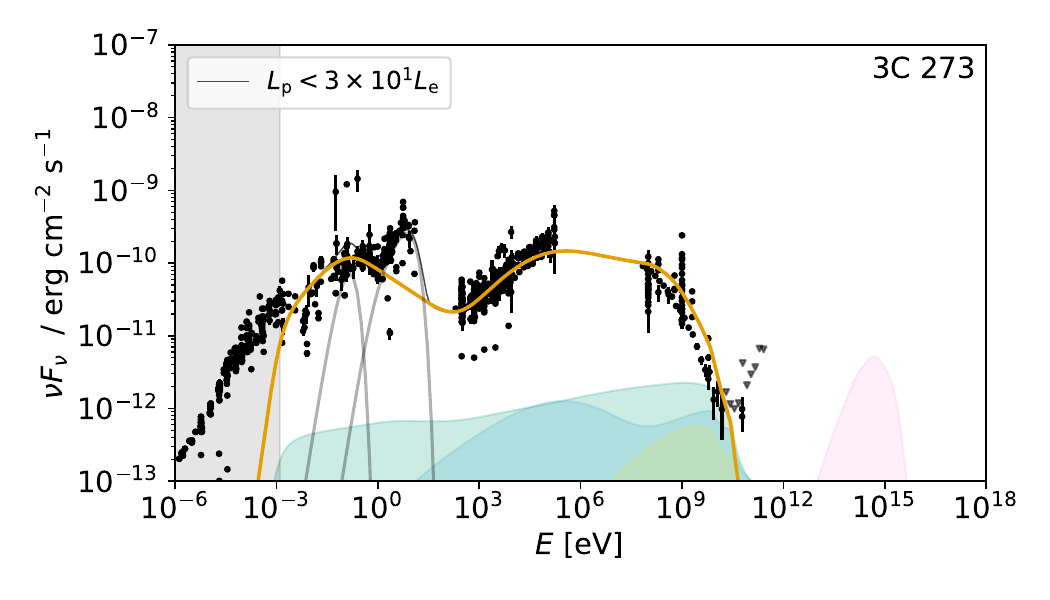}\includegraphics[width=0.48\textwidth]{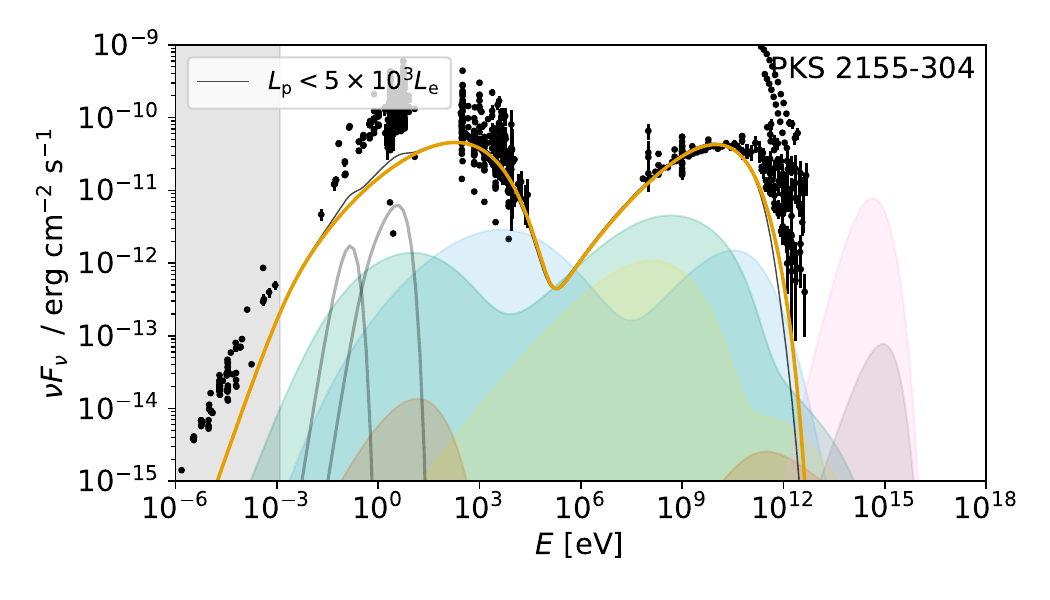}

\includegraphics[width=0.48\textwidth]{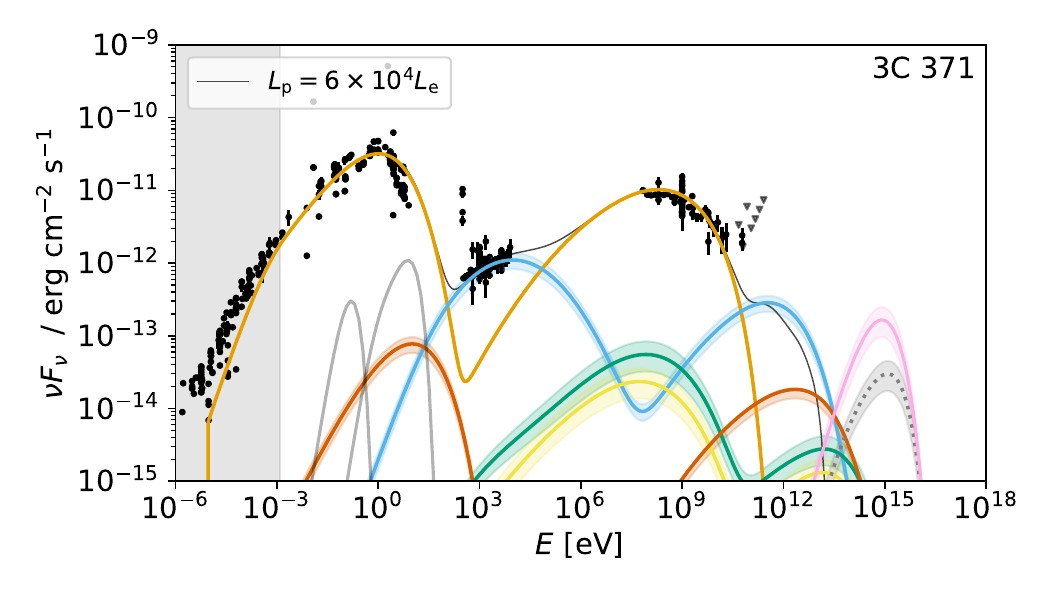}\includegraphics[width=0.48\textwidth]{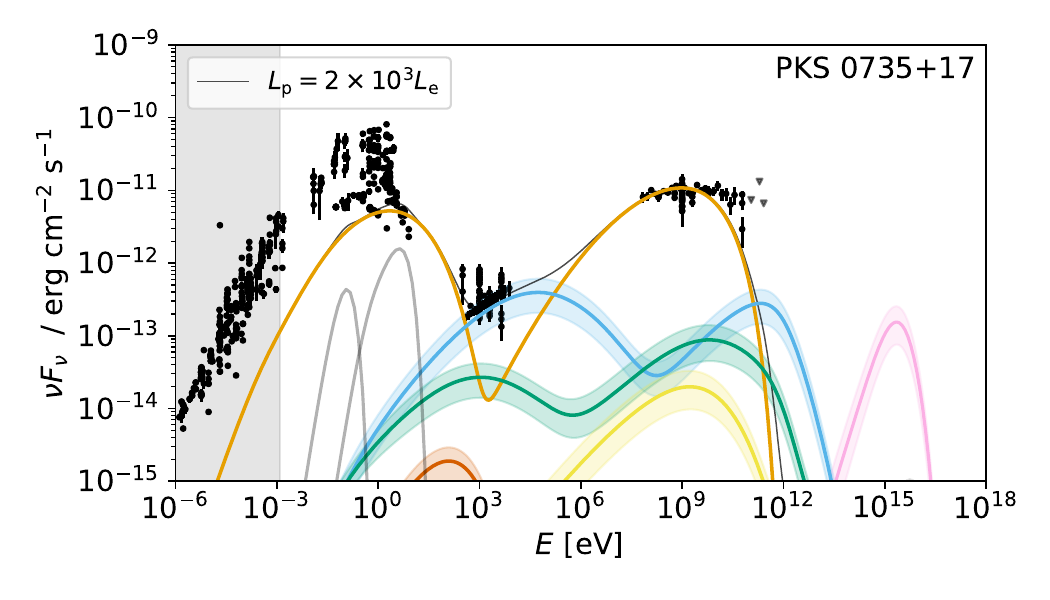}

\includegraphics[width=0.48\textwidth]{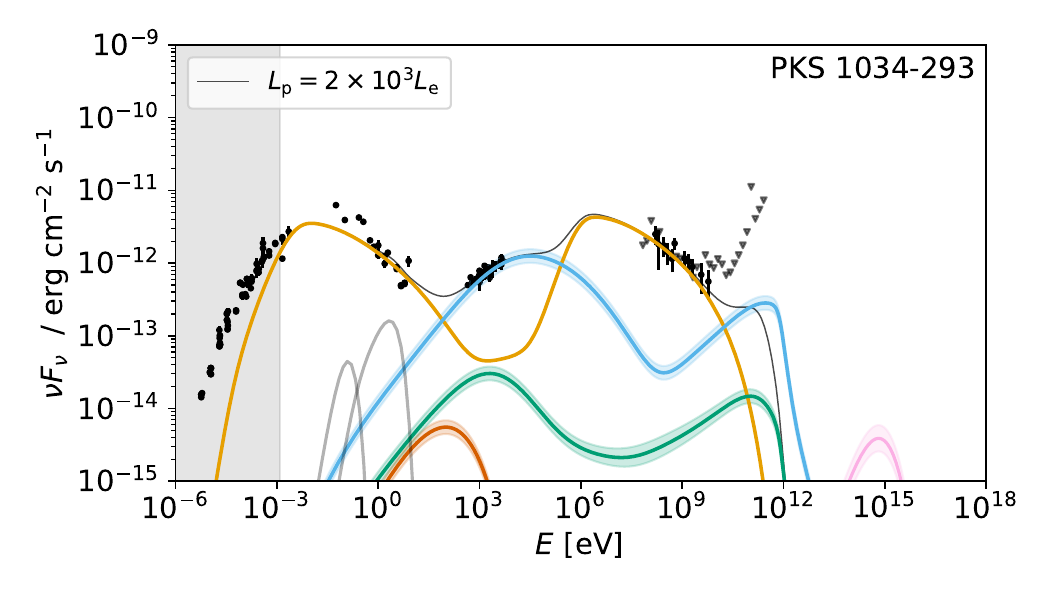}\includegraphics[width=0.48\textwidth]{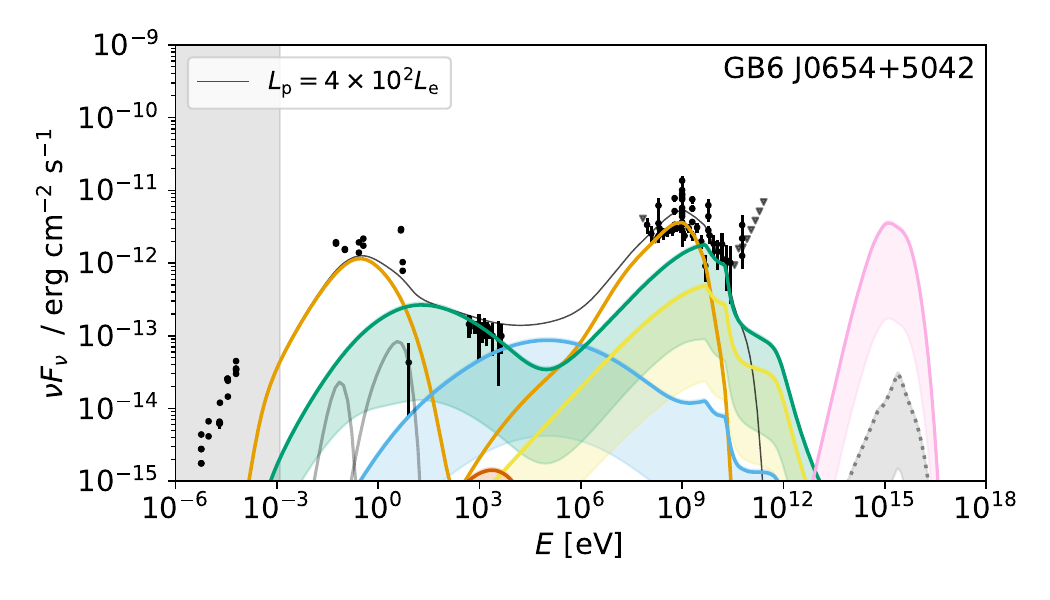}

\includegraphics[width=0.48\textwidth]{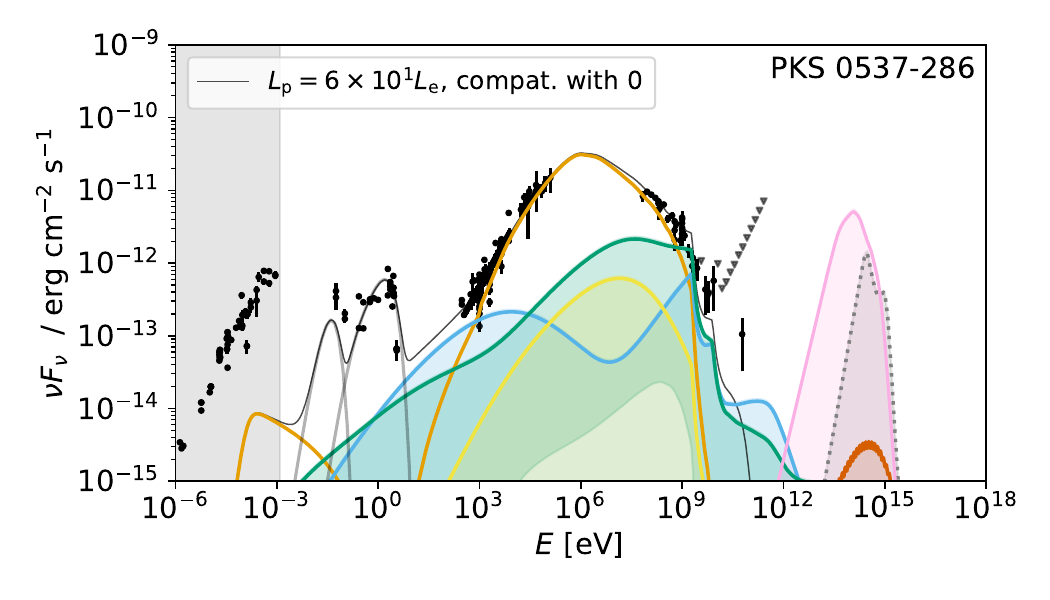}\includegraphics[width=0.48\textwidth]{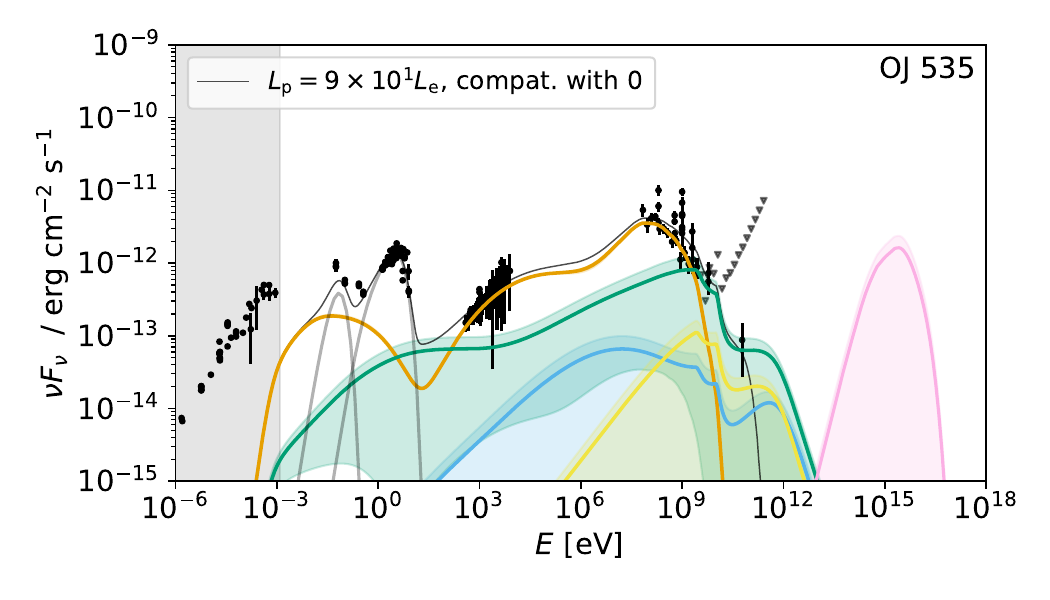}

\includegraphics[width=0.7\textwidth,trim={0 0 0 0}, clip]{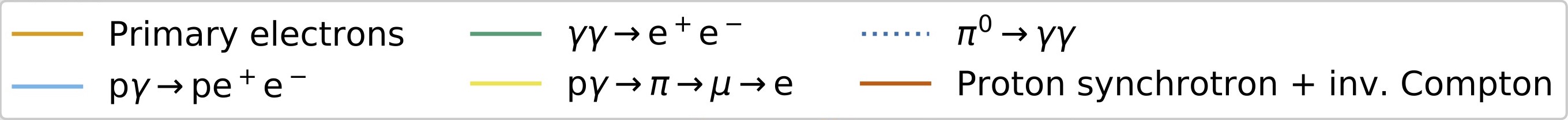}
\caption{Examples of best-fit results for 8 sources in the sample, given in the observer's frame. The total modeled photon emission (including attenuation on the EBL) is shown as a thin black curve, while the color curves represent the different components of that emission, as detailed in the caption. We note that the individual SED components are given without the effect of EBL attenuation, while for the total flux the EBL attenuation is taken into account. Additionally, we show as gray curves the thermal emission from the dust torus and the accretion disk. On the right-hand side of each plot we show either the best-fit neutrino flux (pink curve) or the upper limit for the predicted emission (pink shade). The black points represent the multiwavelength flux observations that are fitted in each case, while the gray inverted triangles represent upper limits. Finally, the gray shaded area represents the frequency range for which data are not fitted but rather considered as upper limits to the model, as explained in \Sec\ref{sec:methods_optimization}. As we can see, the origin of the X-ray and VHE gamma-ray fluxes may in some cases be dominated by leptonic emission (upper row), but can also be dominated by electromagnetic cascades triggered by proton interactions. The corresponding plot for all modeled sources, as well as the numerical data, can be found in the online repository \protect\hyperlink{https://github.com/xrod/lephad-blazars}{https://github.com/xrod/lephad-blazars}.}
\label{fig:components}
\end{figure*}

To support the discussion of the multiwavelength fits, we show in \Fig\ref{fig:components} the best-fit results for eight different sources. The total photon fluxes are shown as a thin black curve and are broken down into components of different physical origin, according to the legend.

In the two upper panels we can see on the left the results for the quasar 3C~273 and on the right for the HBL PKS~2155-304. In these cases, multiwavelength observations can be described purely by the emission from accelerated electrons, which is represented by the orange curve. Synchrotron emission from these primary electrons typically explains the nonthermal 
optical spectrum, together with thermal emission from the accretion disk and dust torus (gray curves). The \textit{Fermi}-LAT fluxes are typically explained by a dominant leptonic component. For these two sources, that leptonic component alone can actually explain entirely these observations, without the need for a proton component. In the case of FSRQs, this is often dominated by external Compton scattering of the BLR photons, while for BL Lacs it is always dominated by synchrotron-self Compton (SSC) in this model. For a total of 218 sources, or 66\% of the sample, leptonic emission can explain the fluxes across the multiwavelength spectrum. As indicated by the gray shaded area, observations below 300~GHz are considered as upper limits, as explained in \Sec\ref{sec:methods_optimization}; this is due to the fact that the compact region that we simulate is often optically thick to radio emission, which means these fluxes should originate in a larger zone that is not modeled here.

For these two sources, we can see that a subdominant component from proton interactions is allowed, but not required, to describe the data. In these cases, only an upper limit can be set for the proton luminosity, as determined by the $1\sigma$ criterion explained in \Sec\ref{sec:methods_optimization}. This maximum contribution to the emission is shown as blue, green, yellow, and orange bands (see caption on the bottom of the figure), while the respective maximum neutrino emission is shown as a pink band at $\sim$PeV energies.

In the case of the source PKS~2155-304 (upper right panel), we can see that at around 100~keV a potential contribution from secondaries coming from proton interactions (region shaded in blue) is tolerated within the $1\sigma$~level, but also here it is not necessary to describe the data. We therefore can only set a constraint on the maximum proton loading of $L_\mathrm{p}/L_\mathrm{e}<10^3$ and a corresponding maximum neutrino flux level, shown as a pink band. It is also worth noting the extreme variability of this source, which results in a large dispersion of the multiwavelength fluxes. This situation becomes more common the brighter the source, since it can be more easily detected by the multiwavelength experiments in different activity states. In these cases, the best-fit model reflects only one of the possible activity states of the source.

We can see that for PKS~2155-304 the synchrotron emission from primary electrons undershoots the observed fluxes in the infrared, optical, and ultraviolet, which does not happen for a source like 3C~273. This is due to the use of a single chi-squared criterion to optimize the results over the entire multiwavelength spectrum, as described in \Sec\ref{sec:methods_optimization}. Since PKS~2155-304 displays large variability in the optical, ultraviolet, and X-ray ranges, the fit becomes less sensitive to those wavelengths compared to gamma rays, where the spectrum has a well-defined shape, small error bars, and a narrower overall spread in flux. The same applies to the source PKS~0735+17, shown in the right plot on the second row of \Fig\ref{fig:components}, discussed in the following paragraph. In general, a more refined description of the low-energy bump would require a time-dependent analysis of the data using quasi-simultaneous datasets. Another possibility would be a more sophisticated algorithm that optimized simultaneously different individual features of the multiwavelength spectrum, such as the synchrotron peak frequency and Compton dominance. With the state-of-the-art methods, fitting a large number of sources requires a robust algorithm, and a more sophisticated fitting procedure lies therefore outside the scope of this work (see also \Sec\ref{sec:limitations} where other methodological limitations are discussed in depth.)

Turning now to the second row of \Fig\ref{fig:components} (BL Lacs 3C~371 and PKS~0735+178), we see that in these cases Bethe-Heitler pair production by protons (p$\gamma\rightarrow$pe$^{+}$e$^{-}$, shown in blue) describes the X-ray fluxes, filling the ``gap'' between the two broadband features. These X-ray fluxes originate in synchrotron emission by the secondary pairs, while inverse Compton (the high-energy bump in the same blue curve) does not play a role in explaining observations in these two cases. The source PKS~0735+178 has recently been studied by~\citet{Sahakyan:2022nbz}, although in that case the authors used a dataset selected in the time domain. We compare these results in \App\ref{app:literature}.

We see that in both these sources the best-fit neutrino spectrum is incompatible with zero because the proton contribution dominated the emission in the X-ray band. This neutrino best fit is shown as a pink curve, accompanied by the respective error band from varying $L_\mathrm{p}^\prime$ following the procedure discussed in \Sec\ref{sec:methods_optimization}. The accompanying gamma-ray emission (gray dotted curve visible in the case of 3C~371) is subsequently attenuated by the EBL and therefore does not appear in the final SED prediction, shown as a thin black curve.

In other cases, the proton contribution helps describe the data not only in the X-ray band but also in the gamma-ray band above 100~GeV. This is exemplified by the two BL Lacs on the third row of \Fig\ref{fig:components}. In the left-hand plot, pairs from the Bethe-Heitler process help explain both the X-ray flux and a hardening of the \textit{Fermi}-LAT spectrum. On the right-hand plot, pair production by high-energy gamma rays (green curve) also helps explain the X-ray emission as well as the high-energy part of the \textit{Fermi}-LAT spectrum.

Overall, for a total of 62 sources, or 20\% of the sample, the best-fit scenario involves proton emission to a level that is incompatible with zero within the $1\sigma$ region, either because it is necessary to explain X-ray data or high-energy gamma rays.

Finally, on the last row of \Fig\ref{fig:components}, we show two cases where the best-fit scenario includes a proton contribution but it is compatible with zero within the $1\sigma$ region. As we can see in these two cases, pair production by protons and by the gamma rays emitted from photo-pion production helps explain the spectral shape in X-rays and gamma rays above $\sim$10~GeV, but only marginally. The best-fit results of 44 sources fall in this category, which represents 14\% of the sample. In this case we can still estimate a best-fit neutrino flux, but this estimate is not as as significant as for the group of sources discussed above. We therefore list these sources in \Tab\ref{tab:all_parameters} separately from the above. We also do not take into account the best-fit proton luminosities when  discussing the neutrino emission sample properties.

\subsection{Estimates on cosmic-ray loading and neutrino emission}
\label{sec:results_neutrinos}

\begin{figure*}[htpb]
\includegraphics[width=\textwidth]{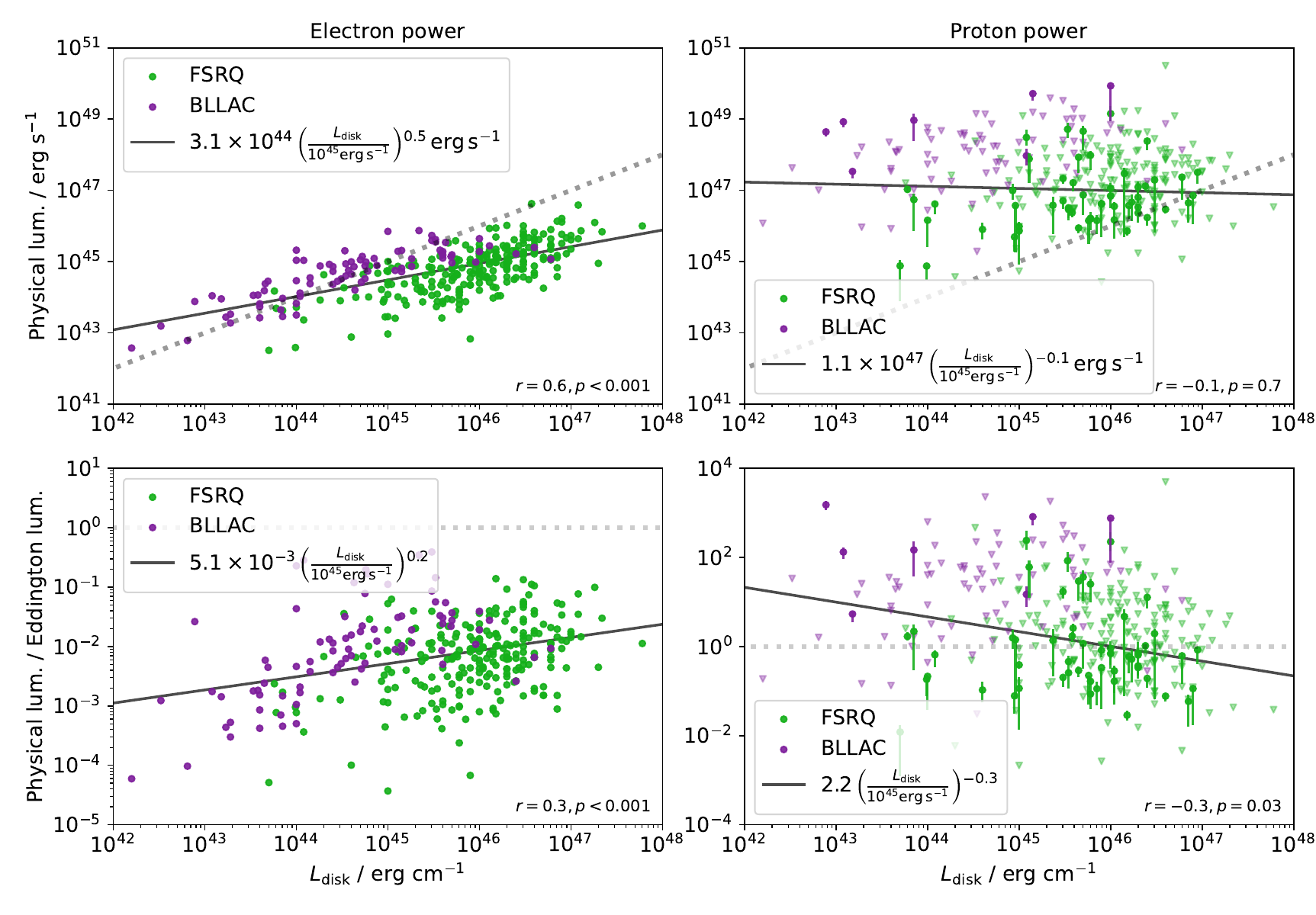}
\caption{Best-fit values of the physical luminosity injected in electrons \textit{(left)} and protons \textit{(right)}, as a function of each source's disk luminosity. The physical luminosities in the rest frame of the black hole are given by $L^\mathrm{phys}_{\mathrm{e,p,}B}=\Gamma_\mathrm{b}^2L^\prime_{\mathrm{e,p,}B}/2$. In the upper plots the physical luminosity is given in $\mathrm{erg}\,\mathrm{s}^{-1}$ and the dotted lines indicate $L^\mathrm{phys}=L_\mathrm{disk}$; in the lower panels it is given in units of the Eddington luminosity $L_\mathrm{Edd}$ and the dotted lines indicate $L^\mathrm{phys}=L_\mathrm{Edd}$. The best-fit power laws (excluding the upper limits) are shown as solid black lines. For protons, these relations are calculated using only the best-fit results that are incompatible with $L_\mathrm{p}^{\prime}=0$. These are shown as solid points, while the inverted triangles represent results compatible with $L_\mathrm{p}^{\prime}=0$ within 1$\sigma$. We also cite the result of the Pearson correlation coefficient $r$, applied to the logarithm of the corresponding variables, and its respective p value. As we can see, all variables present a statistically meaningful moderate power-law correlation with the disk luminosity, except the proton power in the upper right plot.}
\label{fig:power_vs_ldisk}
\end{figure*}

We now turn to the estimated properties of the proton population and subsequent predictions for neutrino emission. In \Fig\ref{fig:power_vs_ldisk} we show the best-fit values of the physical luminosity, $L^\mathrm{phys}_\mathrm{e,p}=L^\prime_\mathrm{e,p}\,\Gamma^2/2$, on the left for electrons and on the right for protons,  plotted against the disk luminosity of each source. For comparison, we show as a dotted line the relation $L^\mathrm{phys}=L_\mathrm{disk}$ and as a solid line the best-fit power law relation for the entire sample. In the lower panels, the physical luminosity is plotted in units of the source's Eddington luminosity $L_\mathrm{Edd}$, estimated for each source based on $M_\mathrm{BH}$. 

As found by previous leptonic and leptohadronic modeling studies \citep[e.g.,][]{Boettcher:2013wxa,Paliya:2017xaq}, we can see that the physical luminosity in electrons loaded onto the jet  follows closely the luminosity radiated by the disk (upper left plot). By fitting a power law we see that in this case it scales specifically with the square root of $L_\gamma$. In terms of the Eddington luminosity (lower left plot), this translates to a sub-Eddington electron injection rate of between $10^{-3}$ and $10^{-1}~L_\mathrm{Edd}$ for most sources.

In the right-hand panels of \Fig\ref{fig:power_vs_ldisk} we show the same values for protons. Here we show as solid points only those results where the proton luminosity is incompatible with zero within $1\sigma$ from the best fit. Best-fit results where the proton luminosity is either zero or compatible with zero within $1\sigma$ are shown as inverted triangles. We see that the maximum allowed proton powers are almost exclusively above $L_\mathrm{disk}$ (upper right plot). In terms of each source's Eddington luminosity, we can see in the lower right plot that there is an overall negative correlation of $L_\mathrm{p}^\mathrm{phys}/L_\mathrm{Edd}\sim L_\mathrm{disk}^{-0.3}$. Taking as an example a disk luminosity of $L_\mathrm{disk}=10^{45}\,\mathrm{erg}\,\mathrm{s}^{-1}$, the model predicts an average proton luminosity of  $L_\mathrm{p}^\mathrm{phys}=2.2L_\mathrm{Edd}$. This average behavior was extrapolated for the subset of sources for which a nonzero proton contribution is necessary in this model (solid points). Most of these  sources have $L_\gamma>10^{45}\,\mathrm{erg}\,\mathrm{s}^{-1}$ and have therefore, for the most part, proton powers $L_\mathrm{p}^\mathrm{phys}\lesssim L_\mathrm{Edd}$.  In a few cases, the best fit lies above the Eddington limit, reaching up to $L_\mathrm{p}^\mathrm{phys}\sim10^3L_\mathrm{Edd}$ in three sources. While jet loading at super-Eddington rates may be possible for short periods of flaring activity, it would be unrealistic to expect this during longer, steady quiescent states. A time-dependent study of the multiwavelength data could help one assert whether the datasets are selecting for flaring states in these cases.

\begin{figure*}[htpb]
\centering

\includegraphics[trim={0 12mm 0 5mm}, clip, width=0.48\textwidth]{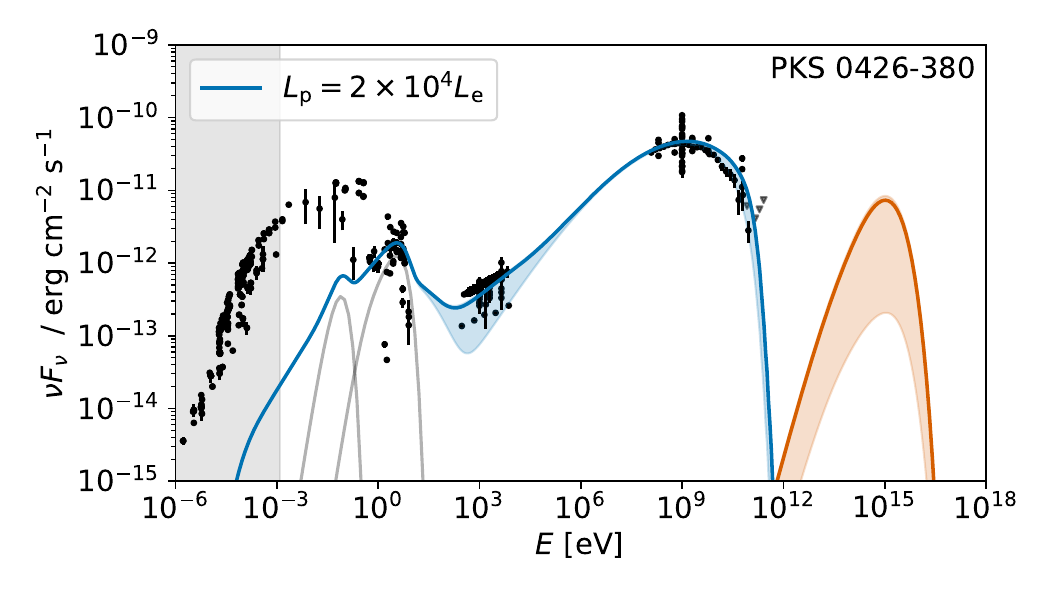}
\includegraphics[trim={0 12mm 0 5mm}, clip, width=0.48\textwidth]{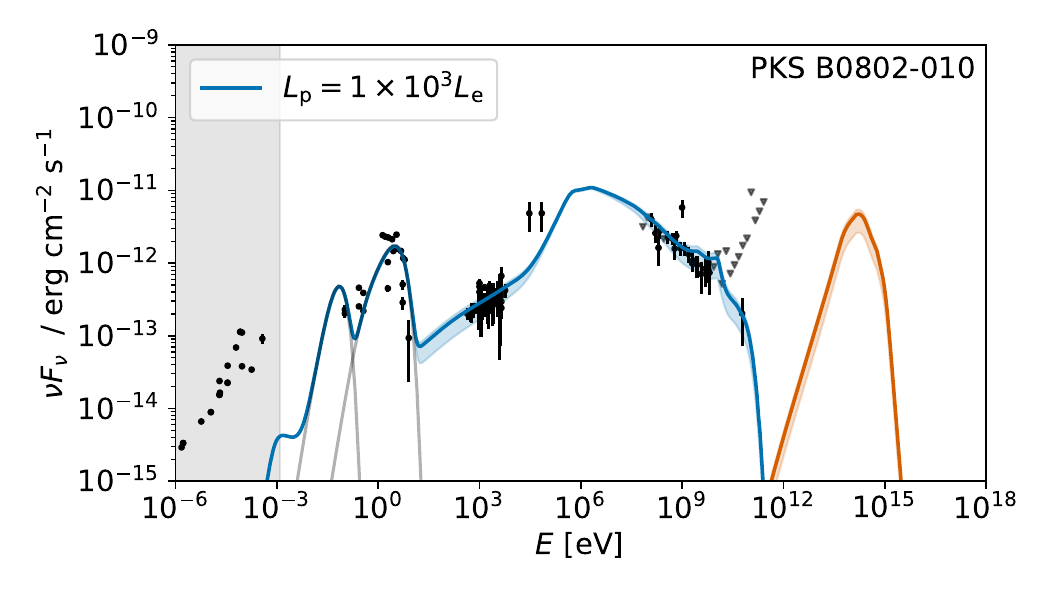}

\includegraphics[trim={0 12mm 0 5mm}, clip, width=0.48\textwidth]{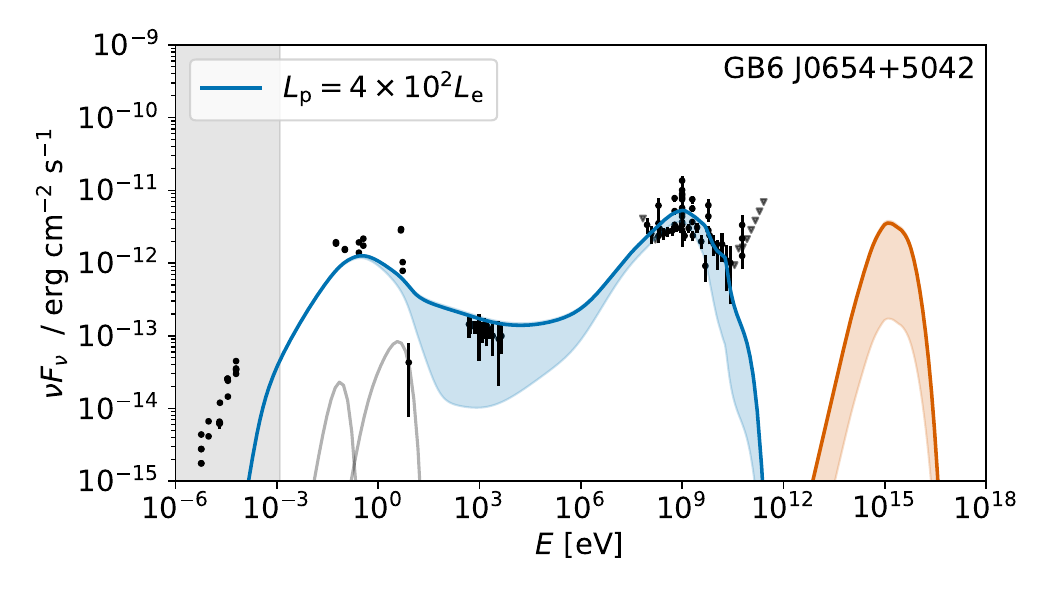}
\includegraphics[trim={0 12mm 0 5mm}, clip, width=0.48\textwidth]{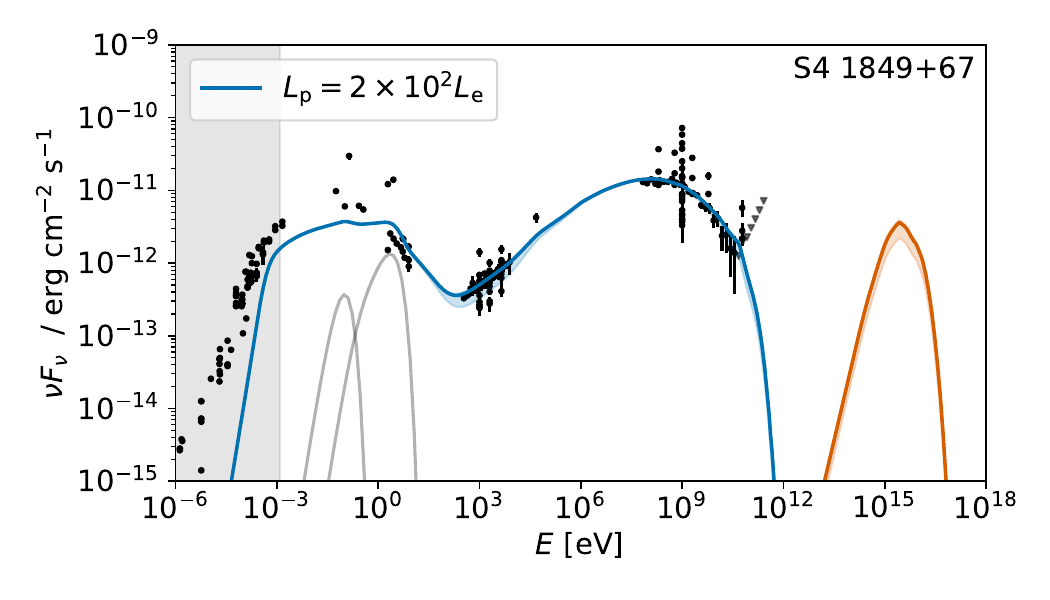}

\includegraphics[trim={0 12mm 0 5mm}, clip, width=0.48\textwidth]{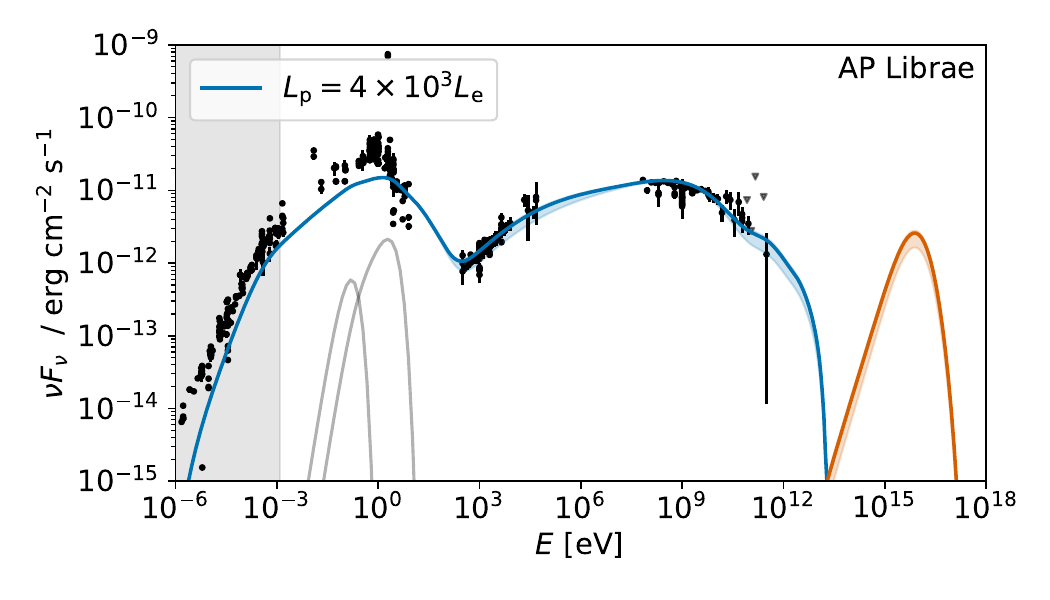}
\includegraphics[trim={0 12mm 0 5mm}, clip, width=0.48\textwidth]{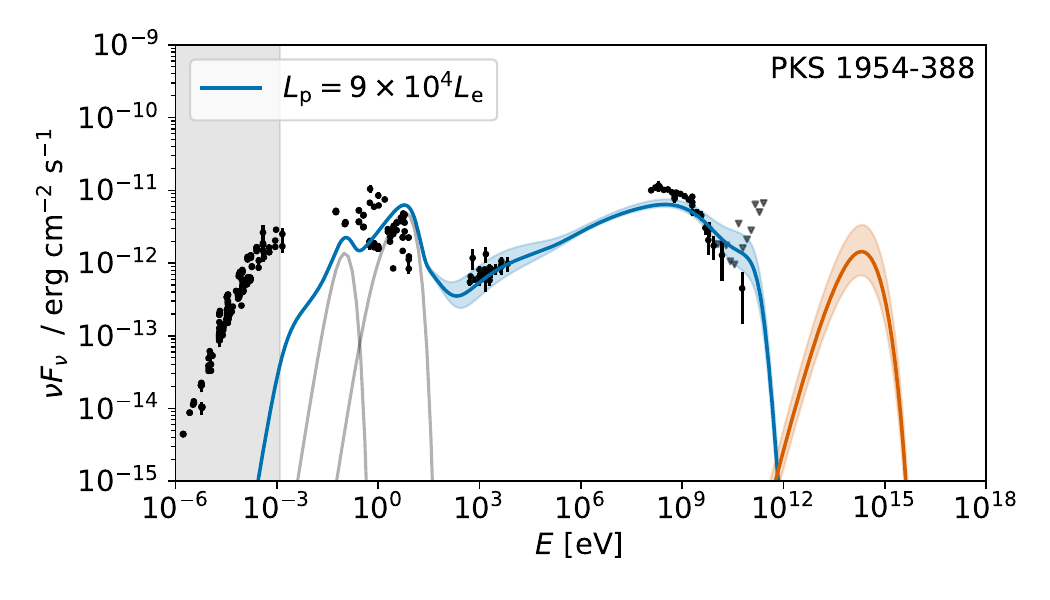}

\includegraphics[trim={0 12mm 0 5mm}, clip, width=0.48\textwidth]{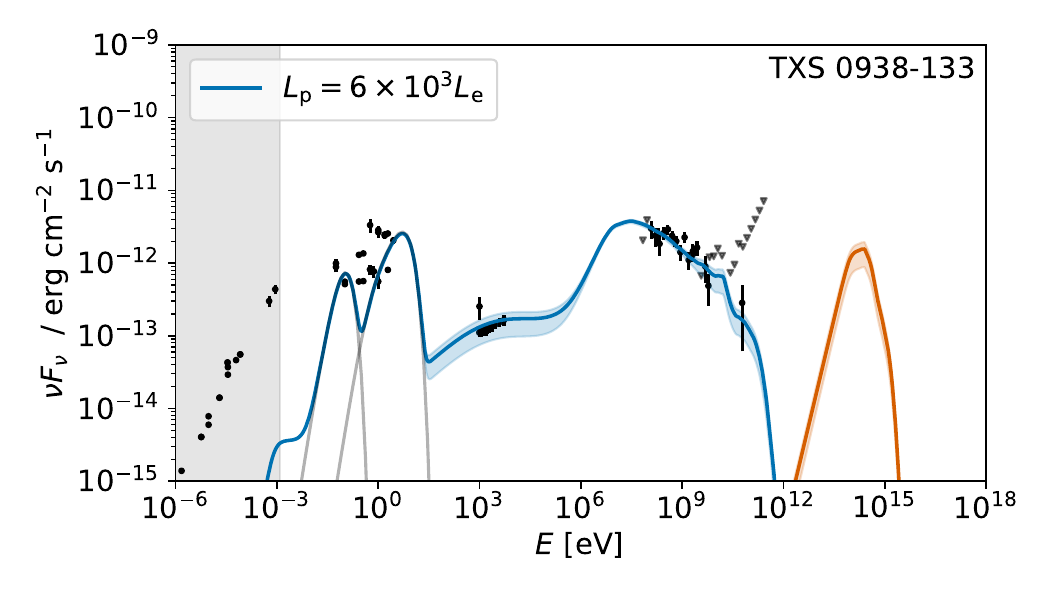}
\includegraphics[trim={0 12mm 0 5mm}, clip, width=0.48\textwidth]{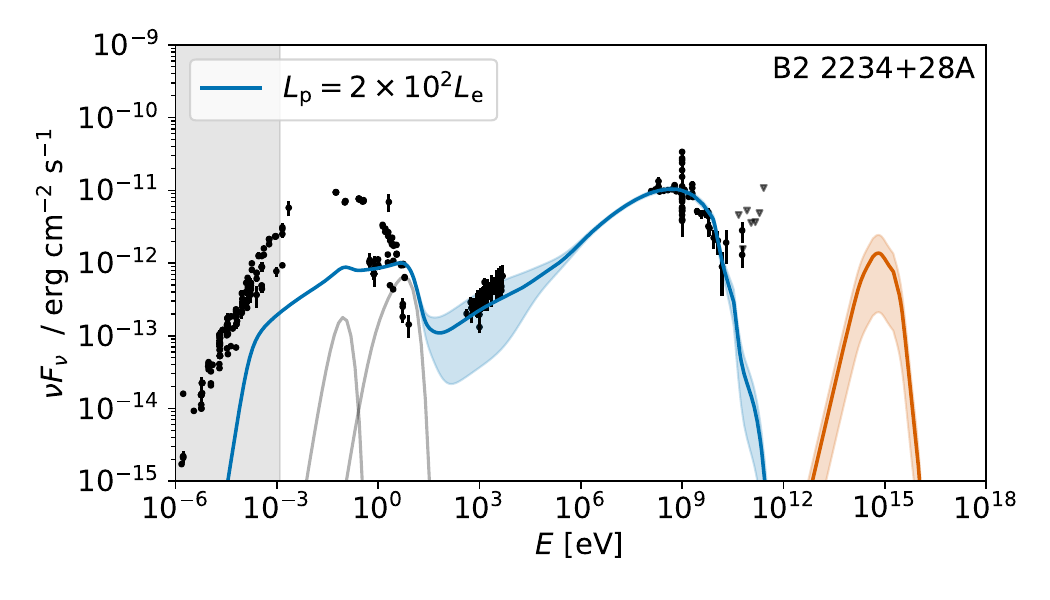}

\includegraphics[trim={0 5mm 0 5mm}, clip, width=0.48\textwidth]{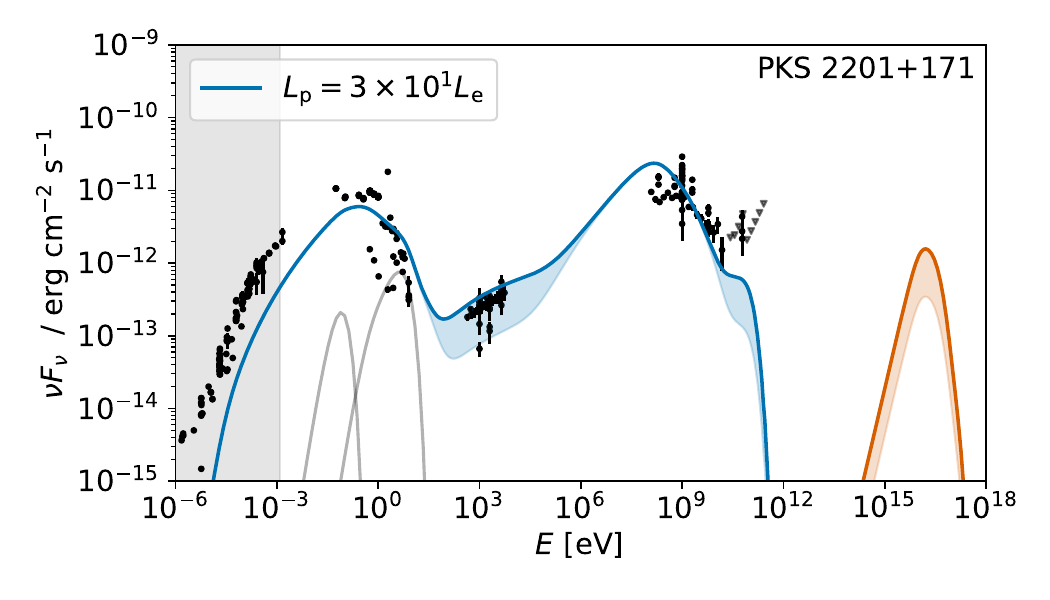}
\includegraphics[trim={0 5mm 0 5mm}, clip, width=0.48\textwidth]{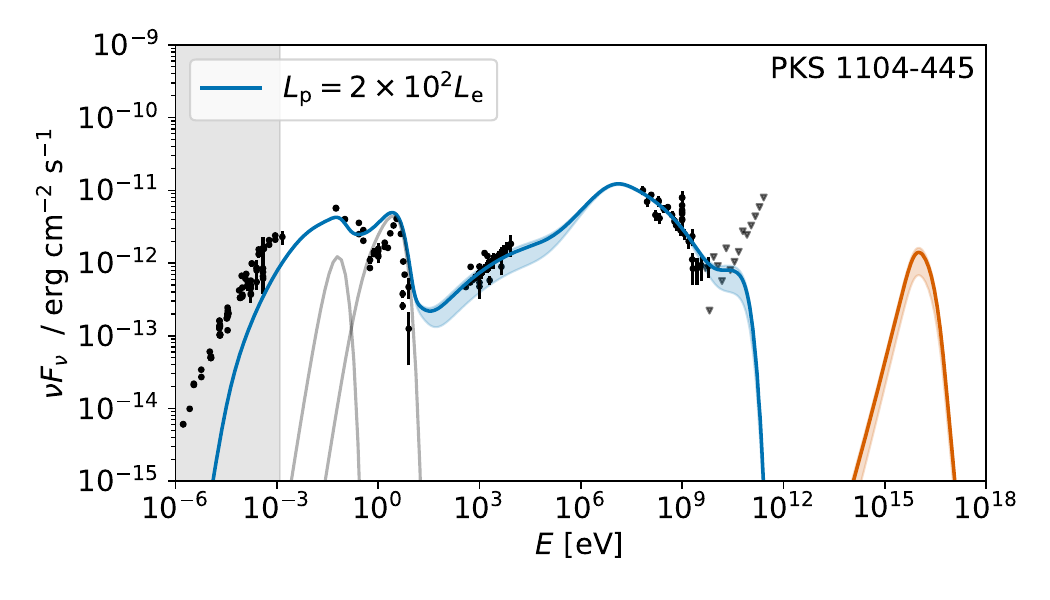}

\caption{Best-fit results of the 10 sources in the sample with the highest predicted neutrino fluxes. The plots are ordered by the value of the total predicted neutrino flux, as in~\Tab\ref{tab:all_parameters}, here from left to right and top to bottom. The blue curves show the multiwavelength emission predicted by the full leptohadronic model. The best-fit baryonic loading value is provided in the caption and can also be found in \Tab\ref{tab:all_parameters}. The orange curves on the right-hand side of each panel show the best-fit emitted all-flavor neutrino fluxes, as discussed in \Sec\ref{sec:results_neutrinos}. The corresponding plot for all modeled sources, as well as the numerical data, can be found in the online repository \protect\hyperlink{https://github.com/xrod/lephad-blazars}{https://github.com/xrod/lephad-blazars}.}
\label{fig:fits}
\end{figure*}

In \Fig\ref{fig:fits} we show the results for the ten sources with the highest predicted neutrino fluxes and for which the proton contribution is incompatible with zero. These correspond to the ten first sources in \Tab\ref{tab:all_parameters}. The first aspect worth noting is the variety of possible values of baryonic loading, defined here as $L_\mathrm{p}^\prime/L_\mathrm{e}^\prime$. In this subsample of ten sources, which all have comparable levels of neutrino emission, the baryonic loading ranges from 30 (PKS~2201+171) to $9\times10^4$ (PKS~1954-388). This is due to the drastically different neutrino production efficiencies, which depend on the best-fit source geometry parameters, as discussed later in this section.

The second aspect is the different size of the $1\sigma$ error bands. For example, in the cases of BL Lac GB6~J0654+5042 and FSRQ B2~2234+28A,  the error band covers a region that seems to strongly violate X-ray data. This is due to the large amount of gamma-ray data points in these cases, which implies that the chi-squared value is dominated by deviations in the gamma-ray band, making the fit less sensitive to deviations in the optical and X-ray bands. Still, what we see in all cases is that the variation in the predicted flux of neutrinos generally maps onto a variation of the predicted flux in X-rays and gamma rays above $\sim$10~GeV, due to the reasons discussed in the \Sec\ref{sec:results_seds}.

\begin{figure*}[htpb]
\includegraphics[width=\textwidth]{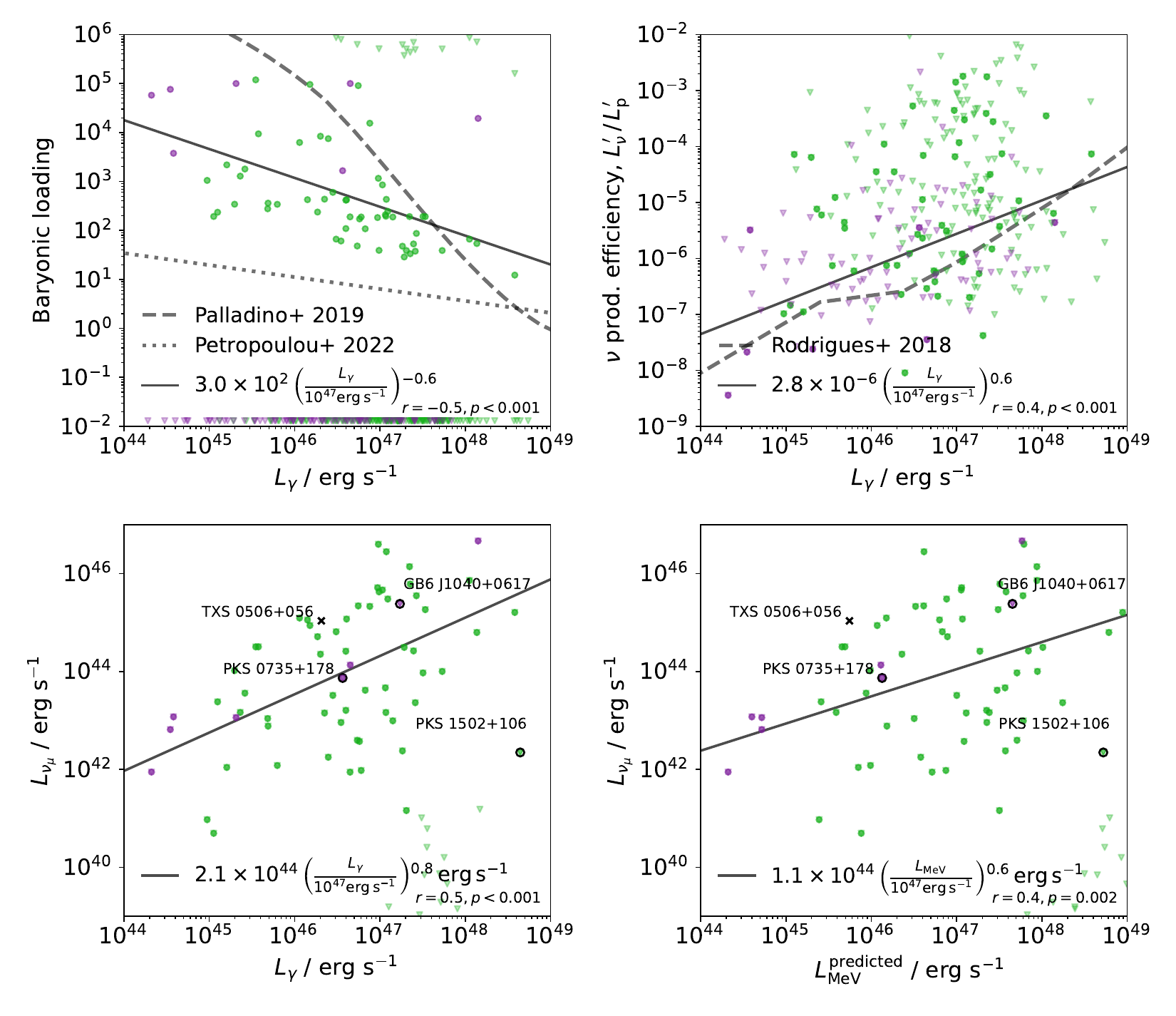}
\caption{Characterization of the best-fit baryonic loading, neutrino efficiency and predicted neutrino fluxes. \textit{Upper panels:} Distribution of the best-fit baryonic loading \textit{(left)} and neutrino production efficiency \textit{(right)} as a function of the source's gamma-ray luminosity between 100~MeV and 100~GeV. For comparison, in the left-hand-side plots we show the behavior predicted by the magnetically loaded jet model by \citet{Petropoulou:2022sct} and the phenomenological study by \citet{Palladino:2018lov}. \textit{Lower panels:} Predicted muon neutrino luminosity for each source as a function of the Fermi LAT luminosity \textit{(left)} and of the predicted MeV gamma-ray luminosity \textit{(right)}.  The color code follows the previous pictures in that FSRQs are shown in green and BL Lacs in purple. The cases where a purely leptonic model lies within 1$\sigma$ of the best-fit result are shown as inverted triangles. The points therefore represent sources for which the data favors leptohadronic model. The solid black lines show the best-fit power law relations. For reference, we show four recent IceCube candidate sources. Of these, only TXS~0506+056 is not contained in this sample, for which we adopt the values from previous studies. For the three other blazars marked, the values of neutrino and MeV gamma-ray luminosity are those predicted by this work.}
\label{fig:neutrino_with_gamma}
\end{figure*}

We now discuss the general trends of baryonic loading, neutrino production efficiency, and predicted neutrino luminosity. In the upper right panel of \Fig\ref{fig:neutrino_with_gamma} we show the estimated baryonic loading of each source as a function of its luminosity $L_\gamma$ in the range from 100~MeV to 100~GeV. The solid line shows the best-fit power law considering only the sources for which the proton luminosity is incompatible with zero (solid points). We observe that the baryonic loading of the sample scales with $~L_\gamma^{-0.6}$. For comparison, we show as a dashed curve the result of the phenomenological study by \citet{Palladino:2018lov}, who estimated the behavior of the baryonic loading necessary to explain the diffuse IceCube flux without violating constraints from lack of associations with high-luminosity sources. We see that the current model is compatible with that suggested by \citet{Palladino:2018lov} for sources with $L_\gamma\approx3\times10^{47}\,\mathrm{erg}\,\mathrm{s}^{-1}$, with a baryonic loading using the predictions of 100. For lower luminosity sources that work predicts higher baryonic loading values, which would be necessary to reach the IceCube diffuse flux level. Since this work focuses on sources with $L_\gamma\lesssim10^{45}\,\mathrm{erg}\,\mathrm{s}^{-1}$, we cannot extrapolate with confidence to that low-luminosity population. On the other hand, the work by \citet{Petropoulou:2022sct} predicts a lower baryonic loading that scales somewhat similarly with $L_\gamma$. The difference in absolute value comes from a fundamental difference in the physical model: in this work we consider magnetic field strengths mainly below the Gauss level and the presence of external fields as the main driver of neutrino production; in comparison, \citet{Petropoulou:2022sct} consider higher magnetic field strengths in the jet, assume magnetic reconnection as the main driver of cosmic ray acceleration, and do not model a specific source catalog.

In the upper right panel of \Fig\ref{fig:neutrino_with_gamma}, we show the modeled neutrino production efficiency of the population, which is shown to scale with $L_\gamma^{0.6}$. This trend is remarkably similar to that derived by \citet{Rodrigues:2017fmu}. On the one hand this is unsurprising, since the underlying geometric model and treatment of external fields was similar to the one applied here \citep[cf. also][]{Murase:2014foa}. On the other hand, \citet{Rodrigues:2017fmu} did not focus on any specific source sample instead using only generic principles of blazar emission based on average population properties. That work also did not include a computation of the photons emitted by the cosmic-ray interactions, which opened the possibility that for high enough values of baryonic loading the high-energy emission might overshoot the assumed gamma-ray luminosity. Here, by simulating a specific sample with real data, we can estimate the effect of hadronic interactions on the SED, as discussed in \Sec\ref{sec:results_seds}.

These opposite trends of the baryonic loading, which scales as $(L_\mathrm{p}^\prime/L_\mathrm{e}^\prime)\sim L_\gamma^{-0.6}$, and the neutrino efficiency, which scales as $(L_\mathrm{\nu}^\prime/L_\mathrm{p}^\prime)\sim L_\gamma^{0.6}$, emerge independently as a collective behavior of the sample from each individual source fit. From this we can already infer the behavior of $L_\nu$ with $L_\gamma$:

\begin{equation}
L_\nu^\prime\,\sim\,
\left(\frac{L_\mathrm{p}^\prime}{L_\mathrm{e}^\prime}\right) 
\left(\frac{L_\nu^\prime}{L_\mathrm{p}^\prime}\right) \left(\frac{L_\mathrm{e}^\prime}{L_\mathrm{disk}}\right)
\left(\frac{L_\mathrm{disk}}{L_\gamma}\right) L_\gamma \,=\, L_\gamma ^ {0.8},
\label{eq:luminosity_scaling}
\end{equation}
where the index 0.8 comes from introducing the relation $L_\mathrm{e}^\prime\sim L_\mathrm{disk}^{0.5}$, as shown in the upper left panel of \Fig\ref{fig:power_vs_ldisk}, and the additional relation $L_\mathrm{\gamma}\sim L_\mathrm{disk}^{0.7}$ found for this sample (as can be derived directly from the values of $L_\mathrm{disk}$ and $F_\gamma$ provided in \Tab\ref{tab:all_parameters}).

This relation between observed gamma-ray luminosity and predicted neutrino luminosity is demonstrated in the lower left plot of \Fig\ref{fig:neutrino_with_gamma}, where we show the predicted neutrino fluxes from the sample. For comparison, we show as a cross the case of TXS~0506+056 based on previous modeling results, since it is not contained in this sample (see \Sec\ref{sec:intro}). Other recent neutrino candidates that are in the sample are highlighted with black circles (see also \App\ref{app:literature} where we contrast the results with the literature on these sources). Finally, on the lower right plot of \Fig\ref{fig:neutrino_with_gamma} we show that $L_\nu\sim L_\mathrm{MeV}^{0.6}$, where $L_\mathrm{MeV}$ is the flux of gamma rays in the range between 0.1-100~MeV that is predicted by the model. Although we do not currently have data on that energy range, the model predicts that the neutrino luminosity scales with the emitted luminosity in MeV gamma rays, which strengthens the case for future MeV gamma-ray missions as neutrino-counterpart search machines ~\citep[e.g.,][]{DeAngelis:2021esn,Caputo:2022xpx}.

\section{Neutrino flux predictions and implications for multi-messenger searches}
\label{sec:neutrinos}

\subsection{Expected neutrino fluxes}
\label{section:neutrinos_fluxes}

\begin{figure*}[htpb]
\includegraphics[width=\textwidth]{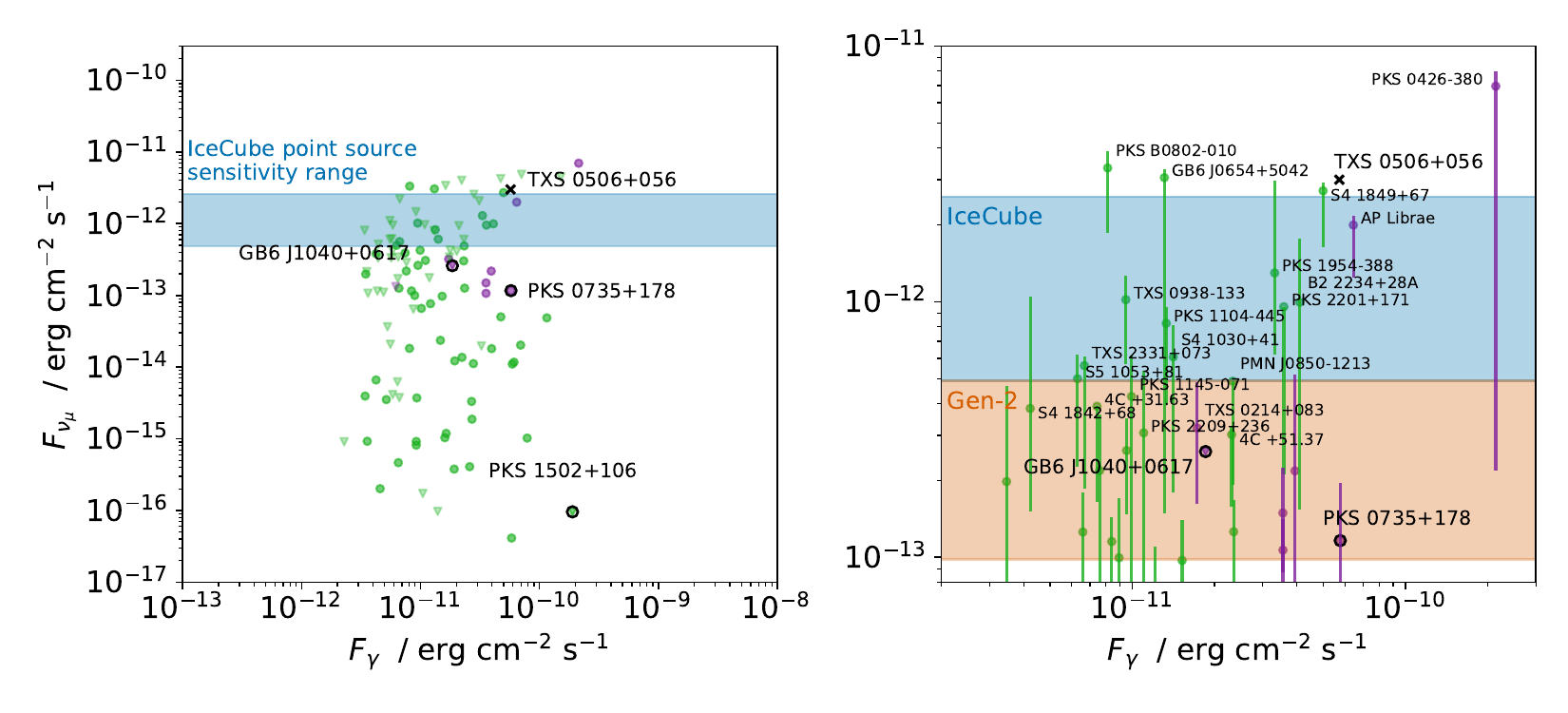}
\caption{Predicted muon neutrino flux from each source. Here we show only the cases where the neutrino flux is incompatible with zero within the $1\sigma$ region from the best fit. Previous neutrino candidates that are part of the sample are highlighted with a black circle and TXS 0506+056, which is not in the sample, is shown as a black cross for comparison. In both plots, we show in blue the IceCube point source sensitivity~\citep{IceCube:2016tpw}; on the right plot, we show additionally the sensitivity band for the future IceCube-Gen2~\citep{IceCube-Gen2:2020qha} and the names of the associated sources with the highest predicted fluxes (cf.~\Fig\ref{fig:fits}).}
\label{fig:neutrino_fluxes}
\end{figure*}

Using the predictions for the neutrino luminosity emitted by each source, we can deduce the corresponding flux of muon neutrinos at Earth and compare it to the sensitivities of present and future experiments. In \Fig\ref{fig:neutrino_fluxes} we show the predicted total muon neutrino flux observed at Earth as a function of each source's \textit{Fermi}-LAT flux. In both plots, we show in blue the IceCube sensitivity based on the seven-year point source analysis~\citep{IceCube:2016tpw}, which is plotted as a band due to its dependence on declination.

In the right-hand plot, we zoom into the highest fluxes and show additionally the names of the associated source for the objects with highest flux. We also show the sensitivity band of the future IceCube-Gen2 experiment, approximated as 3-5 times better compared to the current IceCube~\citep{IceCube-Gen2:2020qha}.
As we can see, the model suggests that seven sources in the sample may be within reach of the current IceCube, while more than twenty would be observable by the future IceCube-Gen2. Interestingly, both PKS~0735+178 and GB6~J040+0617, both of which have an associated high-energy IceCube alert \citep[see][]{Sahakyan:2022nbz,Fermi-LAT:2019hte}, appear here at a subthreshold level for detection by IceCube. This apparent discrepancy might be due, for example, to occasional flaring activity of these sources, with a corresponding temporary increase in neutrino fluxes that is not grasped by this model due to the lack of time dependence of the data (cf. discussion in \App\ref{app:literature}, in particular the upper right panel of \Fig\ref{fig:literature} where this point is clarified for PKS~0735+178).

\subsection{Predicted neutrino event rates}
\label{section:neutrinos_rates}

\begin{figure*}[htpb]
\includegraphics[width=\textwidth]{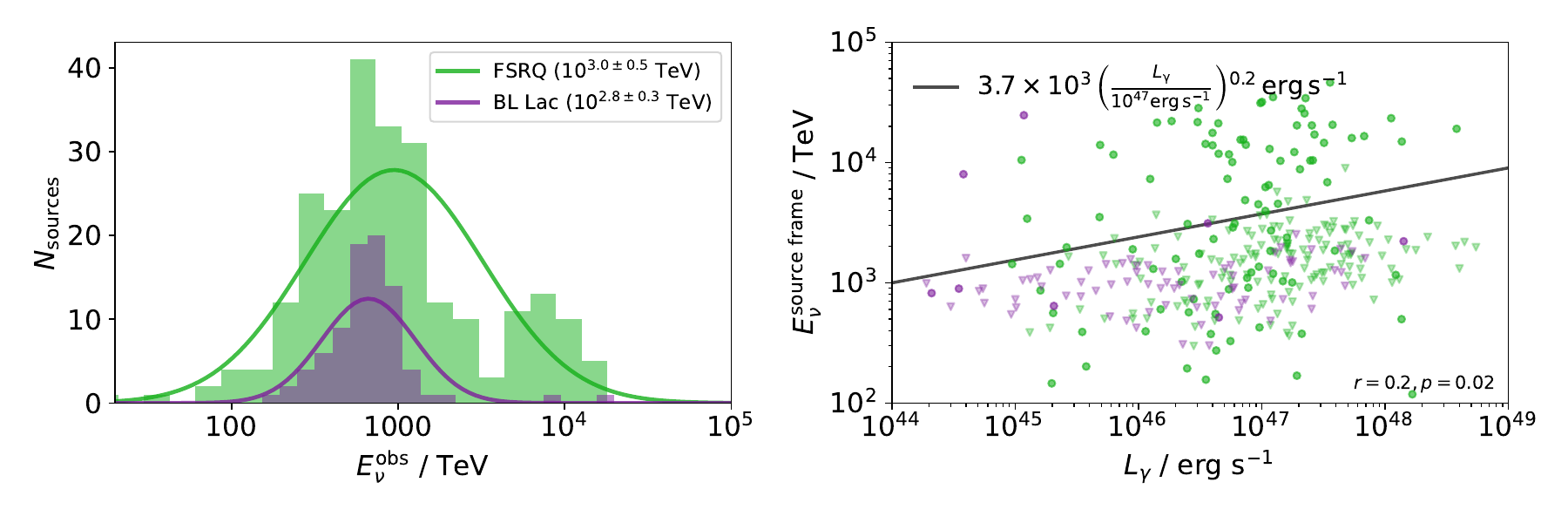}
\caption{Characterization of the neutrino spectra predicted by the model. \textit{Left:} Distribution of the predicted peak neutrino energy, given in the observer's frame. \textit{Right:} Distribution of the predicted neutrino energy in the observer's frame as a function of the source's gamma-ray luminosity $L_\gamma$. The best-fit power law is shown as a black line. This best-fit function includes only the results (shown as dots) and not the upper limits (shown as inverted triangles.)}
\label{fig:neutrino_epeak}
\end{figure*}

Additionally to the total neutrino flux, the number of expected neutrino detections depends on the energy of the emitted neutrinos and on the declination of each source. The former is summarized in \Fig\ref{fig:neutrino_epeak}. On the left-hand panel, we show the distribution of the predicted peak energy of the neutrino energy flux spectrum observed at Earth, $E_\nu^{\mathrm{obs}2}\mathrm{d}F_\nu\mathrm{d}E^\mathrm{obs}_\nu$. We can see that FSRQs have a broader distribution of peak energies that are centered around $E_\nu^\mathrm{obs}=1$~PeV but extend up to 10~PeV. BL Lacs, on the other hand, have a slightly lower dispersion and are centered around 630~TeV. Compared to the distribution of maximum proton energies in the sample (cf. lower right panel of \Fig\ref{fig:lephad_histrograms})), these distributions are much more peaked. This is due to the fact that the neutrino spectrum emerges from a convolution between the spectrum of cosmic-ray protons and the target ambient photons. In the case of FSRQs, those target photons can be the synchrotron photons in the jet or external photons from the BLR; this broader variety is the reason for the broader distribution of peak energies compared to BL Lacs, for which external photons are not available as targets in this model.  

In the right-hand panel of \Fig\ref{fig:neutrino_epeak}, we show the distribution as a function of gamma-ray luminosity, showing a positive correlation of $L_\nu^\mathrm{obs}\sim L_\gamma^{0.2}$, as expected from the arguments discussed in \Sec\ref{sec:results_neutrinos}. We note that in both the Gaussian fit on the left and the power-law fit on the right, we account for all sources that have a best-fit result involving protons, even those compatible with zero baryonic loading at the $1\sigma$ level. This is because given the physical properties of the source, which are determined by the best-fit estimate, the peak energy of the emitted neutrinos is determined regardless of the actual baryonic loading. Additionally to these, we show as inverted triangles the sources for which the best fit is leptonic.

Using the information on the neutrino peak energy and the declination of each source, we can now estimate the number of events in IceCube. For each source, we adopt the declination- and energy-dependent effective area from the seven-year point source analysis~\citep{IceCube:2016tpw} and convolve it with the predicted muon neutrino spectrum from the source, taking into account its specific declination.

\begin{figure*}[htpb]
\includegraphics[width=\textwidth]{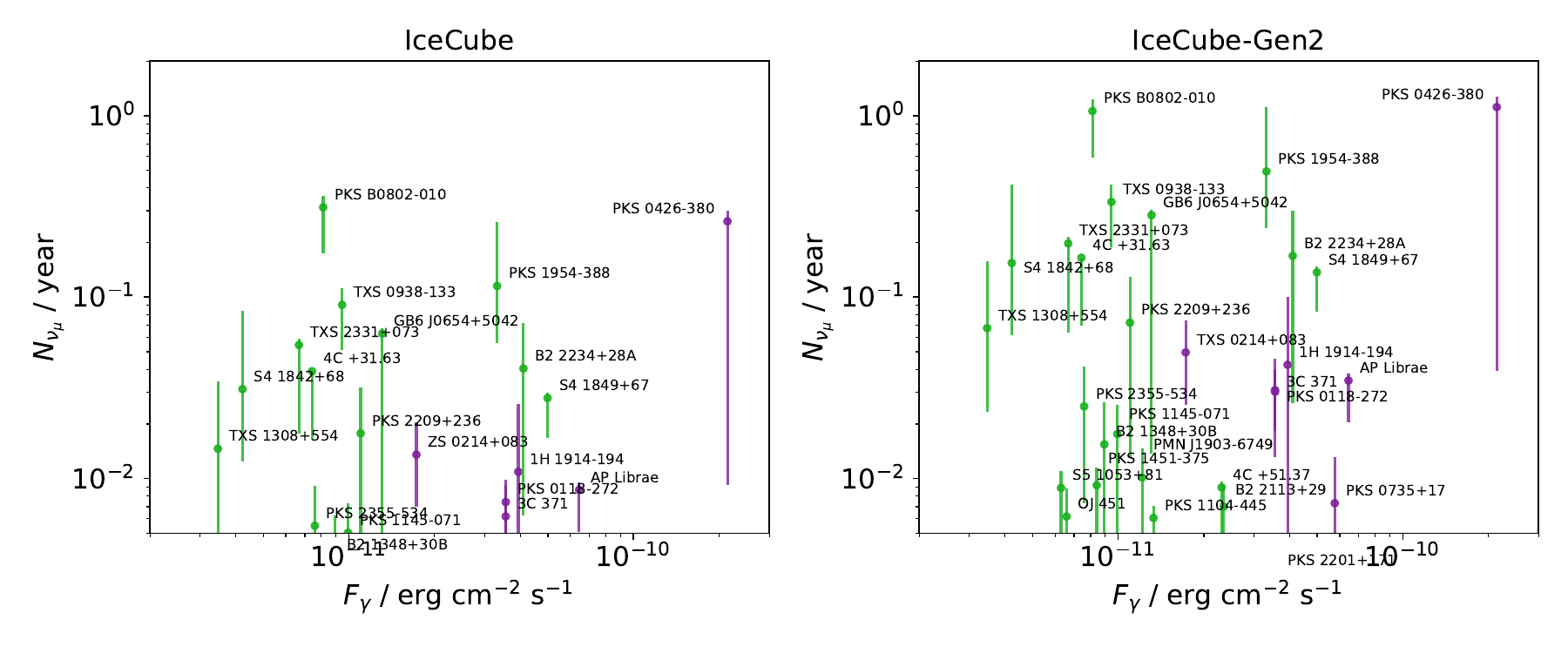}
\caption{Number of events per year in IceCube \textit{(left)} and in the future IceCube-Gen2 \textit{(right)} predicted by the model as function of each source's gamma-ray flux. Here we include only those sources for which the best-fit result is incompatible with zero baryonic loading within the 1$\sigma$ region. The name of each associated source is shown for reference (cf.~\Tab\ref{tab:all_parameters}). Green points denote FSRQs and purple points denote BL Lacs.}
\label{fig:neutrino_events}
\end{figure*}

The results of that analysis are shown in the left panel of  \Fig\ref{fig:neutrino_events}, Like in \Fig\ref{fig:neutrino_fluxes}, we only show those results where the baryonic loading is incompatible with zero within the 1$\sigma$ region from the best fit. As we can see, for a few sources the rate of events in IceCube approaches one per decade. These low rates reflect an estimate based on the average emission from the sources, while potential periods of hadronic flaring should lead to temporary neutrino outbursts that may explain sporadic associations. 

We then perform the same analysis for IceCube-Gen2. In that case, we consider the number of events in IceCube predicted for each source and scale it up by the ratio between the geometric effective areas of IceCube-Gen2 and the current IceCube~\citep{IceCube-Gen2:2020qha}. Given the different geometric setup of IceCube-Gen2, the relative improvement in effective area compared to IceCube is declination-dependent and ranges from a factor of 3 at the horizon up to a maximum factor of about 7~\citep[cf. dashed and dotted curves in shown in \Fig~24 by][]{IceCube-Gen2:2020qha}. This up-scaling therefore depends on the declination of each source.

These results are shown in the right panel of \Fig\ref{fig:neutrino_events}. We can see that several sources have a rate close or superior to about an event per decade, which suggests that IceCube-Gen2 will be able to resolve this subdominant population, even ignoring occasional flaring events.

Despite the positive correlation between gamma-ray and neutrino luminosity shown in \Fig\ref{fig:neutrino_with_gamma}, we see that there is a considerable number of relatively dim blazars with $L_\gamma\lesssim10^{-11}\,\mathrm{erg}\,\mathrm{cm}^{-2}\,\mathrm{s}^{-1}$ for which the model suggests a likely future detection.

\subsection{Diffuse neutrino flux estimates}
\label{section:neutrinos_diffuse}

\begin{figure}[htpb]
\includegraphics[trim={0 0.6cm 0 0}, clip,width=0.5\textwidth]{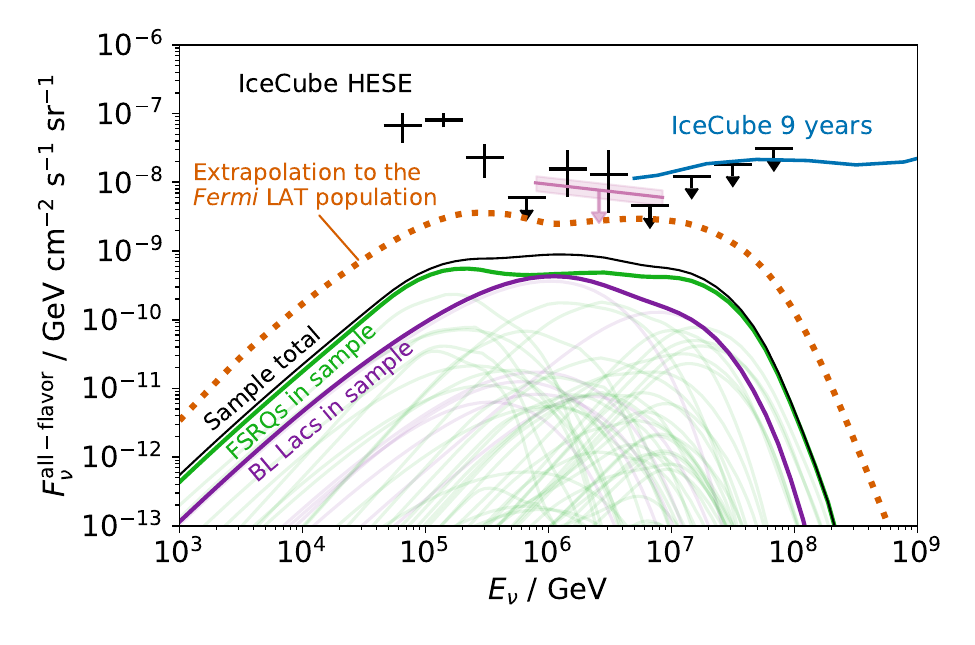}
\caption{All-flavor neutrino flux expected from the entire sample (solid black curve), separated into FSRQs (solid purple curve) and BL Lacs (solid green curve). Following the sampling method described in the main text, we then extrapolate the result to the entire \textit{Fermi}-LAT blazar sample and obtain the diffuse spectrum shown as a dotted curve. For visual reference, the individual contributions are represented as thin curves (see also \Tab\ref{tab:all_parameters}). The black data points show IceCube data from the analysis by~\citet{IceCube:2015gsk} and the blue curve shows the 9 year IceCube sensitivity~\citep{IceCube:2018fhm}. In pink we show the stacking limits derived for blazars by~\citet{IceCube:2016qvd}.}
\label{fig:neutrino_diffuse}
\end{figure}

Finally, we combine the neutrino flux predictions from the individual sources in our sample to estimate their contribution to the IceCube diffuse flux. This total contribution is shown as solid curves in \Fig\ref{fig:neutrino_diffuse}, in green for FSRQs and purple for BL Lacs. The sum of these two components is shown as a black solid curve. As we can see, in this model, our sample contributes to the diffuse flux at a level of about 5\%, reaching a peak flux of about $9\times10^{-10}\,\mathrm{GeV}\,\mathrm{cm}^{-2}\,\mathrm{s}^{-1}\,\mathrm{sr}^{-1}$ at energies of about 1~PeV. This diffuse flux extends, however, from 100~TeV to about 10 PeV. This follows in part from the range of proton energies tested in this model (cf.~\Tab\ref{tab:parameters}), since in these sources the neutrino energy follows closely that of the primary cosmic rays. We cannot exclude the possibility that cosmic rays are accelerated to ultra-high energies at least in some blazars, since this scenario is not tested in this model~\cite[see for example the discussion in][and the connection to the ultra-high-energy cosmic ray issue]{Rodrigues:2020pli}, leading potentially to a diffuse neutrino flux at energies higher than dozens of PeV.

We then estimate the potential contribution of the entire population of \textit{Fermi}-LAT blazars by extrapolating these results. Since our sample does not exactly follow the luminosity distribution of the \textit{Fermi}-LAT population (cf. \Fig\ref{fig:source_distribution}), rather than simply multiplying the flux with a source count factor, we estimate this flux by performing Monte Carlo sampling of 1) values of  $(L_\gamma,z)$ from the luminosity function by \citet{Ajello:2011zi,Ajello:2013lka}, as shown in \Fig\ref{fig:source_distribution}, and considering only those generated sources whose flux lies above the \textit{Fermi}-LAT sensitivity; 2) values of $L_{\nu_\mu}$ from the distribution of $L_{\nu_\mu}(L_\gamma)$ predicted by the model (lower left panel of \Fig\ref{fig:neutrino_with_gamma}); 3) $E_\nu^\mathrm{peak}$ from the distribution shown in the histogram of \Fig\ref{fig:neutrino_epeak}; 4) a neutrino spectral shape from the pool of the model results, depending on $L_{\nu_\mu}$. All these sampling steps are done independently for FSRQs and BL Lacs. We then sum over the spectra obtained and multiply the total spectrum by a factor given by the fraction of sources in the sample for which the result is incompatible with a leptonic solution, which is 23\% (8\%) for FSRQs (BL Lacs). This incorporates into the result the fact that most sources in the generated population should also be compatible with a purely leptonic description and should therefore not contribute to the diffuse neutrino flux.

The result of this extrapolation is shown as a dotted curve in \Fig\ref{fig:neutrino_diffuse}.
Compared to the sample itself, 
we can see that it is broader in energy, which arises from the fact that the distribution with $L_\gamma$ of \textit{Fermi}-LAT sources population (green and purple curves along the vertical axes of the bottom panels of \Fig\ref{fig:source_distribution}) are slightly broader  compared to our sample (black curves).

As we can see, the model predicts a diffuse flux from the \textit{Fermi}-LAT population at the level of 20\% of the observed IceCube flux. This is in agreement with current limits set by the IceCube collaboration for gamma-ray blazars through stacking analyses ~\citep{IceCube:2016qvd}. We can see this by comparing the dotted curve with the IceCube stacking limit, shown as a pink band with a downward arrow.

This supports a scenario where the bulk of the diffuse IceCube flux currently detected originates in a population of more abundant sources, potentially undetected in gamma rays. The sporadic associations we observe between blazars and high-energy IceCube events may then arise from hadronic flares with temporarily enhanced neutrino emission.

At the same time, these results
support the possibility that continued observations, as well as more sensitive future experiments, should help detect an increasingly significant diffuse contribution from gamma-ray blazars. In particular, it is worth noting that the diffuse neutrino flux predicted by the model has a considerable component up to hundreds of PeV dominated mainly by FSRQs, a regime that can be more easily probed by future experiments like IceCube-Gen2. 

\section{Limitations}
\label{sec:limitations}

Finally, we discuss a few caveats of our method and their implications on the interpretation of the results. The two first limitations arise from the fact that the multiwavelength datasets are not time-selected. Firstly, this leads to the fact that some of the datasets contain points with a vast dispersion in flux within the same wavelength band due to variability. This is especially the case for high-luminosity sources (cf. e.g., upper right plot of \Fig\ref{fig:components}). On the one hand, regardless of the variability in the data, the $\chi^2$ minimization method guarantees that the best fit describes an ``average'' activity state, which in accordance with the purpose of the study. On the other hand, is is difficult to quantify how this affects the uncertainties on the neutrino predictions in \Tab\ref{tab:all_parameters}; firstly, because the flux variability is different in different wavelength bands; secondly, because the potential contribution from the hadronic emission does not scale in a linear fashion across wavelengths. For example, if a source displays gamma-ray fluxes ten times higher than those captured by the best-fit SED (as we can see in \Fig\ref{fig:components} for blazar PKS~2155-304), this does not necessarily mean we would expect a neutrino flux ten times higher during such a flare \citep[cf. e.g.,][]{Oikonomou:2019djc,Rodrigues:2020fbu,Gasparyan:2021oad}. 
 
The second important implication of this lack of temporal selection of the data is that is is possible that in some cases the data in different bands represents different activity states of the source. For example, \textit{Fermi}-LAT data are obtained from regular surveying over several years and is therefore more likely to represent the different states of activity of a source; in contrast, an instrument like the \textit{Swift}-BAT is triggered more likely by flares. This issue could be improved, for example, by taking a dataset from pointing observations, which is typically the case for X-rays, and then selecting quasi-simultaneous observations form surveying experiments in other wavelengths. However, this would considerably limit the amount of multiwavelength data available to the point where a fit would be challenging for most sources.

The third limitation in the method is the lack of a dedicated analysis of the flux variability in the different wavelength bands. This is of particular importance for the infrared, optical, and ultraviolet, since in those bands the emission can be dominated either by thermal emission from the disk and the torus, or by nonthermal emission from the jet. In most cases, our fitting method is able to discriminate between these two scenarios based purely on the spectral shape. For example, in the fit to source OJ~535 (lower right plot of Fig.~\ref{fig:components}) we can see that the thermal components can fit well the data in this band, while in other cases the spectral shape is better fitted with a synchrotron spectrum. However, in cases like the quasar 3C~273 (upper left panel of Fig.~\ref{fig:components}), the spectral shape is ambiguous. In our model, the microwave fluxes are fit with a synchrotron component, which should lead to some variability in that band; however, the fluxes actually display little variability, within a factor of up to a few~\citep[e.g.,][]{Fernandes:2020jrs}, which seems to favor instead thermal emission from the core. Resolving these ambiguous cases would require a more detailed analysis of the light curves, which would be prohibitive for such a large catalog with the methods employed here. It is therefore important to keep this ambiguity in mind when interpreting the results, since in some cases a different fit could be possible. 

Another point worth mentioning is that we treat all BL Lac sources with the same model, assuming that the source does not contain a BLR. Given that two distinct IceCube candidates have been determined to be masquerading BL Lacs~\citep{Padovani:2019xcv,Sahakyan:2022nbz}, it is likely that this source subclass is more representative in terms of neutrino emission. It is a fact that most of the BL Lacs in our sample are LBLs, for which it is less likely that the jet emission could hide their broad line emission and therefore more likely that they are true BL Lac objects; however, the sample also includes 22 IBLs and 17 HBLs. For those sources, a masquerading BL Lac model including a BLR could eventually also be tested, likely leading to higher neutrino fluxes due to the additional target photons available for hadronic interactions. In that sense, the current model can be seen as a conservative approach where neutrino production in IBLs and HBLs is not enhanced by any external photon fields. It is particularly important to keep this fact in mind when interpreting the extrapolation shown in \Fig\ref{fig:neutrino_diffuse}.

Another important aspect to mention is the vastness of the parameter space of the one-zone model. First, this means that the range of parameters tested (cf.~\Tab\ref{tab:parameters}) is limited. For example, as discussed in \Sec\ref{section:neutrinos_diffuse} in the context of the diffuse flux predictions, the ultra-high-energy cosmic ray scenario is not tested in this work, since the maximum proton energy that was tested was 10~PeV. Another important factor is the magnetic field strength, which we limit to the 10 gauss level. As we know, in theoretical scenarios where the high-energy emission originates in proton synchrotron emission, the magnetic fields required are typically higher~\citep{Liodakis:2020dvd}, as are scenarios where the cosmic rays are accelerated through magnetic reconnection in magnetically dominated jets~\citep{Petropoulou:2022sct}. In that sense, the predictions made by this work are strictly limited to the paradigm provided by the conventional one-zone leptohadronic framework.

The large size of the parameter space of the model also leads to degenerate solutions with comparable levels of goodness-of-fit, making it challenging to explore the parameter space in a complete way. The use of a genetic algorithm followed by a local minimization provides a robust and highly parallelizable method for finding a minimum in a highly non-linear space of many parameters; however, this method comes with two important caveats:

\textit{A)} It does not guarantee the uniqueness of the result. In fact, as we discuss in \App\ref{app:literature}, other works that have previously modeled sources in this sample have described multiwavelength data with different parameters, leading to different predictions on neutrino emission. At the same time, this is often also due to different multiwavelength datasets, obtained through different selections in the time domain, or, in some cases, to differences in numerical simulation method itself. We therefore do not claim completeness in the way we explored the parameter space of the model, nor the uniqueness of our results within the leptohadronic framework.

\textit{B)} As discussed in \Sec\ref{sec:methods_optimization}, we start by scanning for solutions where the X-ray fluxes are not necessarily described by purely leptonic emission but the gamma-ray peak is (cf.~\Fig\ref{fig:fitting_method}). This derives directly from the hypothesis that hadronic cascades may in some cases dominate the X-ray emission, which is informed by previous models in the literature, as discussed in \Sec\ref{sec:intro}. In that sense, this criterion may be seen as a prior underlying the parameter search. That means that the solutions obtained for which hadronic cascades dominate the X-ray emission (see e.g., second and third rows in \Fig\ref{fig:components}) should be interpreted as being within the leptohadronic hypothesis and purely leptonic solutions may also be possible. \citet{Paliya:2017xaq} have in fact provided purely leptonic solutions for the entire sample; however, their model does not treat radiative energy losses in a time-dependent and self-consistent way as is done here.

Finally, in order to limit the search to a more manageable parameter space we have also fixed two parameters of the model, as described in \Sec\ref{sec:methods_model}: the jet viewing angle and the spectral index of the proton population.

Fixing the viewing angle to $\theta_\mathrm{obs}=1/\Gamma_\mathrm{b}$ is motivated by the effect of relativistic beaming, which compresses the majority of the jet's emission into a solid angle of $1/\Gamma_\mathrm{b}^2$. In the BL Lac model, the Doppler and Lorentz factors are in fact degenerate, since they impact only the energy and flux transformation from the rest frame of the jet into the observer's frame. However, it is important to note two caveats: firstly, in FSRQs the external fields transform into the jet frame with the Lorentz factor $\Gamma_b$, while the transformation of the jet emission into the observer's frame depends on both $\Gamma_\mathrm{b}$ and $\theta_\mathrm{obs}$. These two parameters are therefore not degenerate when external fields are present, which means that by fixing the observing angle we are effectively excluding a portion of the parameter space of the jet geometry. Secondly, the effect of the relativistic beaming on the boosting of the external fields into the jet frame and of the neutrino fluxes into the observer's frame should dramatically increase for jets observed on-axis~\citep{Boettcher:2023yvm}. Given that our approach does not include this possibility, it can be described as conservative regarding both the effects of external fields on neutrino production and electromagnetic cascades, as well as the predicted levels of neutrino emission.

Regarding the spectral index of the accelerated proton spectrum, this was fixed to $\alpha_\mathrm{p}=1$. This somewhat aggressive choice means that a relatively large fraction of the accelerated protons will be injected into the system at high energies, leading to the development of electromagnetic cascades and efficient neutrino emission. A softer injection spectrum, such as $\alpha_\mathrm{p}=2$ as expected from first-order Fermi acceleration, would require a higher overall proton injection power in order to produce the same photon flux from proton interactions. This would then result in higher estimates of the best-fit baryonic loading in those sources for which proton-triggered cascades dominate the X-ray flux. At the same time, an energy-dependent escape mechanism such as Bohm-like diffusion, which was not explored in this work, can lead to a harder steady-state proton spectrum compared to the injected one \citep[see e.g., Fig. 8 of][]{Rodrigues:2017fmu}. With such an escape mechanism, a soft accelerated proton spectrum would lead to a harder steady-state spectrum, assuming the numerical system is evolved long enough to achieve that steady state.


\section{Conclusion} \label{sec:conclusion}

We have performed a source-by-source analysis of a sample of 324 blazars from the CGRaBS catalog, all detected in gigaelectronvolt gamma rays. Of these, 237 are FSRQs. From the physical perspective, we have adopted an external field model, which has been successfully deployed in several studies in recent years to describe individual associations between IceCube events and individual blazars. The modeling was performed through numerical, self-consistent, cosmic-ray simulations where we calculated the time-dependent interactions of cosmic-ray electrons and protons in the relativistic jet. We estimated the best-fit model parameters by fitting the predicted multiwavelength emission to a set of public data from each source. We then predicted the emitted neutrino spectrum,  estimated self-consistently from the same simulation. 

We have described the multiwavelength observations of the 324 blazars from the infrared up to gamma rays and derived the best-fit parameters. For 106 sources, or 33\% of the sample, radiative emission from proton interactions either dominates the observed fluxes in X-rays or at least improves the fit compared to a purely leptonic scenario. In some cases, the description of high-energy gamma rays above tens of gigaelectronvolt  also benefits from a hadronic contribution. For the remaining 218 sources, we have constrained the leptonic model parameters and established upper limits on the baryonic loading of the jet and the emitted neutrino spectrum. Overall, we can say that the leptohadronic paradigm, which was previously applied to individual IceCube candidate blazars, has shown to provide a solution for the entire sample. At the same time, we cannot guarantee the uniqueness of these solutions due to the vast parameter space of the model.

For the sources that have a nonzero proton contribution, we have estimated the best-fit baryonic loading, neutrino production efficiency, emitted neutrino flux, and peak energy of the emitted neutrinos and we have studied possible correlations. We have found that the required proton powers are mostly sub-Eddington, with a few exceptions that require super-Eddington accretion. We have also demonstrated that the best-fit baryonic loading has a decreasing trend as a function of the blazar's gamma-ray luminosity. This conclusion is consistent with phenomenological expectations, given the stacking limits on the brightest sources derived by IceCube. Importantly, this conclusion was obtained here from a purely physics-driven approach through source-by-source modeling without these phenomenological considerations.

On average, the modeled neutrino luminosity has shown to depend positively on the source's gamma-ray luminosity in the \textit{Fermi}-LAT range, with a power-law relation $L_\nu\sim L_\gamma^{0.8}$, as well as on the predicted luminosity in megaelectronvolt  gamma rays. Having tested maximum energy values of the primary protons in the range from 100~TeV up to 10~PeV, we deduced neutrino spectra that peak mostly around 1~PeV for FSRQs, but with a considerable number of sources peaking as high as 10~PeV. For BL Lacs, the neutrino spectra peak mostly around 630~TeV. For about 30 sources in the sample, we predict a neutrino flux above $10^{-13}\,\mathrm{erg}\,\mathrm{cm}^{-2}\,\mathrm{s}^{-1}$, which lie above the sensitivity of the future IceCube-Gen2 experiment. Our results predict that within the first decade of operation, IceCube-Gen2 should have detected the ten brightest sources in this sample, even considering only an average quiescent state emission. If we add the effect of sporadic hadronic flares, the actual number of associations should be considerably higher.

We also conclude that this sample may contribute to the diffuse IceCube flux at the $\sim$5\% level. Furthermore, by extrapolating to the population of \textit{Fermi}-LAT sources, we derived a contribution of 20\% of the diffuse IceCube flux. This contribution is in agreement with upper limits derived in stacking analyses~\citep{IceCube:2016qvd}. While IceCube currently seems to detect sporadic neutrino outbursts from hadronic blazar flares, our result suggests that next-generation neutrino experiments such as IceCube-Gen2 can potentially probe the steady-state emission from this elusive population of cosmic-ray accelerators.

\begin{acknowledgements}

We thank P. Padovani, P. Giommi, and S. Kiehlmann for insightful comments on the manuscript. We acknowledge funding from the German Science Foundation DFG, via the Collaborative Research Center SFB1491 ``Cosmic Interacting Matters - From Source To Signal''.

\end{acknowledgements}


\begin{thebibliography}{}
    \expandafter\ifx\csname natexlab\endcsname\relax\def\natexlab#1{#1}\fi
    
    \bibitem[{Aartsen {et~al.}(2018)Aartsen, Ackermann, Adams, Aguilar, Ahlers,
      Ahrens, Al~Samarai, Altmann, Andeen, Anderson, Ansseau, Anton, Arg{\"u}elles,
      Arsioli, Auffenberg, Axani, Bagherpour, Bai, Barron, Barwick, Baum, Bay,
      Beatty, Becker, Becker~Tjus, BenZvi, Berley, Bernardini, Besson, Binder,
      Bindig, Blaufuss, Blot, Bohm, Boerner, Bos, Boeser, Botner, Bourbeau,
      Bourbeau, Bradascio, Braun, Brenzke, Bretz, Bron, Brostean-Kaiser, Burgman,
      Busse, Carver, Cheung, Chirkin, Christov, Clark, Classen, Coenders, Collin,
      Conrad, Coppin, Correa, Cowen, Cross, Dave, Day, de~Andr{\'e}, De~Clercq,
      Delaunay, Dembinski, DeRidder, Desiati, de~Vries, DeWasseige, DeWith,
      DeYoung, D{\'\i}az-V{\'e}lez, Di~Lorenzo, Dujmovic, Dumm, Dunkman, Dvorak,
      Eberhardt, Ehrhardt, Eichmann, Eller, Evenson, Fahey, Fazely, Felde,
      Filimonov, Finley, Flis, Franckowiak, Friedman, Fritz, Gaisser, Gallagher,
      Gerhardt, Ghorbani, Giommi, Glauch, Gluesenkamp, Goldschmidt, Gonzalez,
      Grant, Griffith, Haack, Hallgren, Halzen, Hanson, Hebecker, Heereman,
      Helbing, Hellauer, Hickford, Hignight, Hill, Hoffman, Hoffmann, Hoinka,
      Hokanson-Fasig, Hoshina, Huang, Huber, Hultqvist, Huennefeld, Hussain, In,
      Iovine, Ishihara, Jacobi, Japaridze, Jeong, Jero, Jones, Kalaczynski, Kang,
      Kappes, Kappesser, Karg, Karle, Katz, Kauer, Keivani, Kelley, Kheirandish,
      Kim, Kim, Kintscher, Kiryluk, Kittler, Klein, Koirala, Kolanoski, Koepke,
      Kopper, Kopper, Koschinsky, Koskinen, Kowalski, Krammer, Krings, Kroll,
      Krueckl, Kunwar, Neilson, Kuwabara, Kyriacou, Labare, Lanfranchi, Larson,
      Lauber, Leonard, Lesiak-Bzdak, Leuermann, Liu, Lozano~Mariscal, Lu,
      Luenemann, Luszczak, Madsen, Maggi, Mahn, Mancina, Maruyama, Mase, Maunu,
      Meagher, Medici, Meier, Menne, Merino, Meures, Miarecki, Micallef, Momente,
      Montaruli, Moore, Morse, Moulai, Nahnhauer, Nakarmi, Naumann, Neer,
      Niederhausen, Nowicki, Nygren, Pollmann, Olivas, {\'O}~Murchadha,
      O{\textquoteright}Sullivan, Padovani, Palczewski, Pandya, Pankova, Peiffer,
      Pepper, Perez de~los Heros, Pieloth, Pinat, Plum, Price, Przybylski, Raab,
      Raedel, Rameez, Rawlins, Rea, Reimann, Relethford, Relich, Resconi, Rhode,
      Richman, Robertson, Rongen, Rott, Ruhe, Ryckbosch, Rysewyk, Safa, Saelzer,
      Sahakyan, Sanchez~Herrera, Sandrock, Sandroos, Santander, Sarkar, Sarkar,
      Satalecka, Schlunder, Schmidt, Schneider, Schoenen, Schoeneberg, Schumacher,
      Sclanfani, Seckel, Seunarine, Soedingrekso, Soldin, Song, Spiczak, Spiering,
      Stachurska, Stamatikos, Stanev, Stasik, Stettner, Steuer, Stezelberger,
      Stokstad, Stoessl, Strotjohann, Stuttard, Sullivan, Sutherland, Taboada,
      Tatar, Tenholt, Ter-Antonyan, Terliuk, Tilav, Toale, Tobin, Toennis, Toscano,
      Tosi, Tselengidou, Tung, Turcati, Turley, Ty, Unger, Usner, Van~Driessche,
      Van~Eijk, van Eijndhoven, Vandenbroucke, Vanheule, van Santen, Vogel,
      Vraeghe, Walck, Wallace, Wallraff, Wandler, Wandkowsky, Waza, Weaver, Weiss,
      Wendt, Werthebach, Westerhoff, Whelan, Whitehorn, Wiebe, Wiebusch, Wille,
      Williams, Wills, Wolf, Wood, Wood, Woschnagg, Xu, Xu, Xu, Yanez, Yodh,
      Yoshida, \& Yuan}]{TXS_orphanflare}
    Aartsen, M., Ackermann, M., Adams, J., {et~al.} 2018, Science, 361, 147
    
    \bibitem[{{Aartsen} {et~al.}(2018){Aartsen}, {Ackermann}, {Adams},
      {et~al.}}]{TXS_MM}
    {Aartsen}, M., {Ackermann}, M., {Adams}, J., {et~al.} 2018, Science, 361
    
    \bibitem[{Aartsen {et~al.}(2020)Aartsen, Ackermann, Adams,
      {et~al.}}]{Aartsen:2019fau}
    Aartsen, M., Ackermann, M., Adams, J., {et~al.} 2020, Phys. Rev. Lett., 124,
      051103
    
    \bibitem[{Aartsen {et~al.}(2013{\natexlab{a}})}]{IceCube:2013low}
    Aartsen, M.~G. {et~al.} 2013{\natexlab{a}}, Science, 342, 1242856
    
    \bibitem[{Aartsen {et~al.}(2013{\natexlab{b}})}]{IceCube:2013cdw}
    Aartsen, M.~G. {et~al.} 2013{\natexlab{b}}, Phys. Rev. Lett., 111, 021103
    
    \bibitem[{Aartsen {et~al.}(2014)}]{IceCube:2014stg}
    Aartsen, M.~G. {et~al.} 2014, Phys. Rev. Lett., 113, 101101
    
    \bibitem[{Aartsen {et~al.}(2015{\natexlab{a}})}]{IceCube:2015gsk}
    Aartsen, M.~G. {et~al.} 2015{\natexlab{a}}, Astrophys. J., 809, 98
    
    \bibitem[{Aartsen {et~al.}(2015{\natexlab{b}})}]{IceCube:2015qii}
    Aartsen, M.~G. {et~al.} 2015{\natexlab{b}}, Phys. Rev. Lett., 115, 081102
    
    \bibitem[{Aartsen {et~al.}(2017{\natexlab{a}})}]{IceCube:2016tpw}
    Aartsen, M.~G. {et~al.} 2017{\natexlab{a}}, Astrophys. J., 835, 151
    
    \bibitem[{Aartsen {et~al.}(2017{\natexlab{b}})}]{IceCube:2016qvd}
    Aartsen, M.~G. {et~al.} 2017{\natexlab{b}}, Astrophys. J., 835, 45
    
    \bibitem[{Aartsen {et~al.}(2018)}]{IceCube:2018fhm}
    Aartsen, M.~G. {et~al.} 2018, Phys. Rev. D, 98, 062003
    
    \bibitem[{Aartsen {et~al.}(2021)}]{IceCube-Gen2:2020qha}
    Aartsen, M.~G. {et~al.} 2021, J. Phys. G, 48, 060501
    
    \bibitem[{Abbasi {et~al.}(2022)}]{IceCube:2022der}
    Abbasi, R. {et~al.} 2022, Science, 378, 538
    
    \bibitem[{Abbasi {et~al.}(2023)}]{IceCube:2023htm}
    Abbasi, R. {et~al.} 2023, Astrophys. J., 954, 75
    
    \bibitem[{Abdo(2010)}]{Abdo:2010ge}
    Abdo, A.~A. 2010, Astrophys. J., 715, 429
    
    \bibitem[{Acero {et~al.}(2015)}]{Fermi-LAT:2015bhf}
    Acero, F. {et~al.} 2015, Astrophys. J. Suppl., 218, 23
    
    \bibitem[{Acharyya {et~al.}(2023)}]{VERITAS:2023eso}
    Acharyya, A. {et~al.} 2023, Astrophys. J., 954, 70
    
    \bibitem[{{Ackermann} {et~al.}(2011){Ackermann}, {Ajello}, {Allafort},
      {Antolini}, {Atwood}, {Axelsson}, {Baldini}, {Ballet}, {Barbiellini},
      {Bastieri}, {Bechtol}, {Bellazzini}, {Berenji}, {Blandford}, {Bloom},
      {Bonamente}, {Borgland}, {Bottacini}, {Bouvier}, {Bregeon}, {Brigida},
      {Bruel}, {Buehler}, {Burnett}, {Buson}, {Caliandro}, {Cameron}, {Caraveo},
      {Casandjian}, {Cavazzuti}, {Cecchi}, {Charles}, {Cheung}, {Chiang},
      {Ciprini}, {Claus}, {Cohen-Tanugi}, {Conrad}, {Costamante}, {Cutini}, {de
      Angelis}, {de Palma}, {Dermer}, {Digel}, {Silva}, {Drell}, {Dubois},
      {Escande}, {Favuzzi}, {Fegan}, {Ferrara}, {Finke}, {Focke}, {Fortin},
      {Frailis}, {Fukazawa}, {Funk}, {Fusco}, {Gargano}, {Gasparrini}, {Gehrels},
      {Germani}, {Giebels}, {Giglietto}, {Giommi}, {Giordano}, {Giroletti},
      {Glanzman}, {Godfrey}, {Grenier}, {Grove}, {Guiriec}, {Gustafsson},
      {Hadasch}, {Hayashida}, {Hays}, {Healey}, {Horan}, {Hou}, {Hughes},
      {Iafrate}, {J{\'o}hannesson}, {Johnson}, {Johnson}, {Kamae}, {Katagiri},
      {Kataoka}, {Kn{\"o}dlseder}, {Kuss}, {Lande}, {Larsson}, {Latronico},
      {Longo}, {Loparco}, {Lott}, {Lovellette}, {Lubrano}, {Madejski}, {Mazziotta},
      {McConville}, {McEnery}, {Michelson}, {Mitthumsiri}, {Mizuno}, {Moiseev},
      {Monte}, {Monzani}, {Moretti}, {Morselli}, {Moskalenko}, {Murgia},
      {Nakamori}, {Naumann-Godo}, {Nolan}, {Norris}, {Nuss}, {Ohno}, {Ohsugi},
      {Okumura}, {Omodei}, {Orienti}, {Orlando}, {Ormes}, {Ozaki}, {Paneque},
      {Parent}, {Pesce-Rollins}, {Pierbattista}, {Piranomonte}, {Piron}, {Pivato},
      {Porter}, {Rain{\`o}}, {Rando}, {Razzano}, {Razzaque}, {Reimer}, {Reimer},
      {Ritz}, {Rochester}, {Romani}, {Roth}, {Sanchez}, {Sbarra}, {Scargle},
      {Schalk}, {Sgr{\`o}}, {Shaw}, {Siskind}, {Spandre}, {Spinelli}, {Strong},
      {Suson}, {Tajima}, {Takahashi}, {Takahashi}, {Tanaka}, {Thayer}, {Thayer},
      {Thompson}, {Tibaldo}, {Tinivella}, {Torres}, {Tosti}, {Troja}, {Uchiyama},
      {Vandenbroucke}, {Vasileiou}, {Vianello}, {Vitale}, {Waite}, {Wallace},
      {Wang}, {Winer}, {Wood}, {Wood}, \& {Zimmer}}]{2011ApJ...743..171A}
    {Ackermann}, M., {Ajello}, M., {Allafort}, A., {et~al.} 2011, \apj, 743, 171
    
    \bibitem[{Ajello {et~al.}(2012)}]{Ajello:2011zi}
    Ajello, M. {et~al.} 2012, Astrophys. J., 751, 108
    
    \bibitem[{Ajello {et~al.}(2014)}]{Ajello:2013lka}
    Ajello, M. {et~al.} 2014, Astrophys. J., 780, 73
    
    \bibitem[{Ajello {et~al.}(2020)}]{Fermi-LAT:2019pir}
    Ajello, M. {et~al.} 2020, Astrophys. J., 892, 105
    
    \bibitem[{Atwood {et~al.}(2009)}]{Fermi-LAT:2009ihh}
    Atwood, W.~B. {et~al.} 2009, Astrophys. J., 697, 1071
    
    \bibitem[{Banik {et~al.}(2020)Banik, Bhadra, Pandey, \&
      Majumdar}]{Banik:2019twt}
    Banik, P., Bhadra, A., Pandey, M., \& Majumdar, D. 2020, Phys. Rev. D, 101,
      063024
    
    \bibitem[{Bellenghi {et~al.}(2023)Bellenghi, Padovani, Resconi, \&
      Giommi}]{Bellenghi:2023yza}
    Bellenghi, C., Padovani, P., Resconi, E., \& Giommi, P. 2023, Astrophys. J.
      Lett., 955, L32
    
    \bibitem[{Boettcher(2023)}]{Boettcher:2023yvm}
    Boettcher, M. 2023 [\eprint[arXiv]{2308.01083}]
    
    \bibitem[{Boettcher {et~al.}(2013)Boettcher, Reimer, Sweeney, \&
      Prakash}]{Boettcher:2013wxa}
    Boettcher, M., Reimer, A., Sweeney, K., \& Prakash, A. 2013, Astrophys. J.,
      768, 54
    
    \bibitem[{Burrows {et~al.}(2005)}]{Burrows:2005gfa}
    Burrows, D.~N. {et~al.} 2005, Space Sci. Rev., 120, 165
    
    \bibitem[{Buson {et~al.}(2023)Buson, Tramacere, Oswald, Barbano,
      de~Clairfontaine, Pfeiffer, Azzollini, Baghmanyan, \& Ajello}]{Buson:2023irp}
    Buson, S., Tramacere, A., Oswald, L., {et~al.} 2023
      [\eprint[arXiv]{2305.11263}]
    
    \bibitem[{Buson {et~al.}(2022)}]{Buson:2022fyf}
    Buson, S. {et~al.} 2022, Astrophys. J. Lett., 933, L43, [Erratum:
      Astrophys.J.Lett. 934, L38 (2022), Erratum: Astrophys.J. 934, L38 (2022)]
    
    \bibitem[{Caputo {et~al.}(2022)}]{Caputo:2022xpx}
    Caputo, R. {et~al.} 2022, J. Astron. Telesc. Instrum. Syst., 8, 044003
    
    \bibitem[{Cerruti {et~al.}(2019)Cerruti, Zech, Boisson, Emery, Inoue, \&
      Lenain}]{Cerruti:2018tmc}
    Cerruti, M., Zech, A., Boisson, C., {et~al.} 2019, Mon. Not. Roy. Astron. Soc.,
      483, L12
    
    \bibitem[{Cerruti {et~al.}(2015)Cerruti, Zech, Boisson, \&
      Inoue}]{Cerruti:2014iwa}
    Cerruti, M., Zech, A., Boisson, C., \& Inoue, S. 2015, Mon. Not. Roy. Astron.
      Soc., 448, 910
    
    \bibitem[{De~Angelis {et~al.}(2021)}]{DeAngelis:2021esn}
    De~Angelis, A. {et~al.} 2021, Exper. Astron., 51, 1225
    
    \bibitem[{Dembinski \& et~al.(2020)}]{iminuit}
    Dembinski, H. \& et~al., P.~O. 2020
    
    \bibitem[{Dominguez {et~al.}(2011)}]{Dominguez:2010bv}
    Dominguez, A. {et~al.} 2011, Mon. Not. Roy. Astron. Soc., 410, 2556
    
    \bibitem[{Fernandes {et~al.}(2020)Fernandes, Pati\~no \'Alvarez, Chavushyan,
      Schlegel, \& Vald\'es}]{Fernandes:2020jrs}
    Fernandes, S., Pati\~no \'Alvarez, V.~M., Chavushyan, V., Schlegel, E.~M., \&
      Vald\'es, J.~R. 2020, Mon. Not. Roy. Astron. Soc., 497, 2066
    
    \bibitem[{Franckowiak {et~al.}(2020)Franckowiak, Garrappa, Paliya,
      {et~al.}}]{Franckowiak:2020qrq}
    Franckowiak, A., Garrappa, S., Paliya, V., {et~al.} 2020, Astrophys. J., 893,
      162
    
    \bibitem[{Gao {et~al.}(2019)Gao, Fedynitch, Winter, \& Pohl}]{Gao:2018mnu}
    Gao, S., Fedynitch, A., Winter, W., \& Pohl, M. 2019, Nature Astron., 3, 88
    
    \bibitem[{Gao {et~al.}(2017)Gao, Pohl, \& Winter}]{Gao:2016uld}
    Gao, S., Pohl, M., \& Winter, W. 2017, Astrophys. J., 843, 109
    
    \bibitem[{{Garrappa} {et~al.}(2019){Garrappa}, {Buson}, {Franckowiak},
      {Fermi-LAT Collaboration}, {Shappee}, {Beacom}, {Dong}, {Holoien},
      {Kochanek}, {Prieto}, {Stanek}, {Thompson}, {ASAS-SN Collaboration},
      {Aartsen}, {Ackermann}, {Adams}, {Aguilar}, {Ahlers}, {Ahrens}, {Alispach},
      {Andeen}, {Anderson}, {Ansseau}, {Anton}, {Arg{\"u}elles}, {Auffenberg},
      {Axani}, {Backes}, {Bagherpour}, {Bai}, {Barbano}, {Barwick}, {Baum}, {Bay},
      {Beatty}, {Becker}, {Becker Tjus}, {BenZvi}, {Berley}, {Bernardini},
      {Besson}, {Binder}, {Bindig}, {Blaufuss}, {Blot}, {Bohm}, {B{\"o}rner},
      {B{\"o}ser}, {Botner}, {Bourbeau}, {Bourbeau}, {Bradascio}, {Braun}, {Bretz},
      {Bron}, {Brostean-Kaiser}, {Burgman}, {Busse}, {Carver}, {Chen}, {Cheung},
      {Chirkin}, {Clark}, {Classen}, {Collin}, {Conrad}, {Coppin}, {Correa},
      {Cowen}, {Cross}, {Dave}, {de Andr{\'e}}, {De Clercq}, {DeLaunay},
      {Dembinski}, {Deoskar}, {De Ridder}, {Desiati}, {de Vries}, {de Wasseige},
      {de With}, {DeYoung}, {Diaz}, {D{\'\i}az-V{\'e}lez}, {Dujmovic}, {Dunkman},
      {Dvorak}, {Eberhardt}, {Ehrhardt}, {Eller}, {Evenson}, {Fahey}, {Fazely},
      {Felde}, {Filimonov}, {Finley}, {Franckowiak}, {Friedman}, {Fritz},
      {Gaisser}, {Gallagher}, {Ganster}, {Garrappa}, {Gerhardt}, {Ghorbani},
      {Glauch}, {Gl{\"u}senkamp}, {Goldschmidt}, {Gonzalez}, {Grant}, {Griffith},
      {G{\"u}nder}, {G{\"u}nd{\"u}z}, {Haack}, {Hallgren}, {Halve}, {Halzen},
      {Hanson}, {Hebecker}, {Heereman}, {Helbing}, {Hellauer}, {Henningsen},
      {Hickford}, {Hignight}, {Hill}, {Hoffman}, {Hoffmann}, {Hoinka},
      {Hokanson-Fasig}, {Hoshina}, {Huang}, {Huber}, {Hultqvist}, {H{\"u}nnefeld},
      {Hussain}, {In}, {Iovine}, {Ishihara}, {Jacobi}, {Japaridze}, {Jeong},
      {Jero}, {Jones}, {Kang}, {Kappes}, {Kappesser}, {Karg}, {Karl}, {Karle},
      {Katz}, {Kauer}, {Keivani}, {Kelley}, {Kheirand ish}, {Kim}, {Kintscher},
      {Kiryluk}, {Kittler}, {Klein}, {Koirala}, {Kolanoski}, {K{\"o}pke}, {Kopper},
      {Kopper}, {Koskinen}, {Kowalski}, {Krings}, {Kr{\"u}ckl}, {Kulacz}, {Kunwar},
      {Kurahashi}, {Kyriacou}, {Labare}, {Lanfranchi}, {Larson}, {Lauber}, {Lazar},
      {Leonard}, {Leuermann}, {Liu}, {Lohfink}, {Lozano Mariscal}, {Lu},
      {Lucarelli}, {L{\"u}nemann}, {Luszczak}, {Madsen}, {Maggi}, {Mahn}, {Makino},
      {Mallot}, {Mancina}, {Mari{\textcommabelow s}}, {Maruyama}, {Mase}, {Maunu},
      {Meagher}, {Medici}, {Medina}, {Meier}, {Meighen-Berger}, {Menne}, {Merino},
      {Meures}, {Miarecki}, {Micallef}, {Moment{\'e}}, {Montaruli}, {Moore},
      {Moulai}, {Nagai}, {Nahnhauer}, {Nakarmi}, {Naumann}, {Neer}, {Niederhausen},
      {Nowicki}, {Nygren}, {Obertacke Pollmann}, {Olivas}, {O'Murchadha},
      {O'Sullivan}, {Palczewski}, {Pandya}, {Pankova}, {Park}, {Peiffer},
      {P{\'e}rez de los Heros}, {Pieloth}, {Pinat}, {Pizzuto}, {Plum}, {Price},
      {Przybylski}, {Raab}, {Raissi}, {Rameez}, {Rauch}, {Rawlins}, {Rea},
      {Reimann}, {Relethford}, {Renzi}, {Resconi}, {Rhode}, {Richman}, {Robertson},
      {Rongen}, {Rott}, {Ruhe}, {Ryckbosch}, {Rysewyk}, {Safa}, {Sanchez Herrera},
      {Sandrock}, {Sandroos}, {Santander}, {Sarkar}, {Sarkar}, {Satalecka},
      {Schaufel}, {Schlunder}, {Schmidt}, {Schneider}, {Schneider}, {Schumacher},
      {Sclafani}, {Seckel}, {Seunarine}, {Silva}, {Snihur}, {Soedingrekso},
      {Soldin}, {Song}, {Spiczak}, {Spiering}, {Stachurska}, {Stamatikos},
      {Stanev}, {Stasik}, {Stein}, {Stettner}, {Steuer}, {Stezelberger},
      {Stokstad}, {St{\"o}{\ss}l}, {Strotjohann}, {Stuttard}, {Sullivan},
      {Sutherland}, {Taboada}, {Tenholt}, {Ter-Antonyan}, {Terliuk}, {Tilav},
      {Tomankova}, {T{\"o}nnis}, {Toscano}, {Tosi}, {Tselengidou}, {Tung},
      {Turcati}, {Turcotte}, {Turley}, {Ty}, {Unger}, {Unland Elorrieta}, {Usner},
      {Vand enbroucke}, {Van Driessche}, {van Eijk}, {van Eijndhoven}, {Vanheule},
      {van Santen}, {Vraeghe}, {Walck}, {Wallace}, {Wallraff}, {Wandkowsky},
      {Watson}, {Weaver}, {Weiss}, {Weldert}, {Wendt}, {Werthebach}, {Westerhoff},
      {Whelan}, {Whitehorn}, {Wiebe}, {Wiebusch}, {Wille}, {Williams}, {Wills},
      {Wolf}, {Wood}, {Wood}, {Woschnagg}, {Wrede}, {Xu}, {Xu}, {Xu}, {Yanez},
      {Yodh}, {Yoshida}, {Yuan}, \& {IceCube Collaboration}}]{2019ApJ...880..103G}
    {Garrappa}, S., {Buson}, S., {Franckowiak}, A., {et~al.} 2019, \apj, 880, 103
    
    \bibitem[{Garrappa {et~al.}(2019)}]{Fermi-LAT:2019hte}
    Garrappa, S. {et~al.} 2019, Astrophys. J., 880, 880:103
    
    \bibitem[{Gasparyan {et~al.}(2021)Gasparyan, B\'egu\'e, \&
      Sahakyan}]{Gasparyan:2021oad}
    Gasparyan, S., B\'egu\'e, D., \& Sahakyan, N. 2021, Mon. Not. Roy. Astron.
      Soc., 509, 2102
    
    \bibitem[{Ghisellini \& Tavecchio(2009)}]{Ghisellini:2009wa}
    Ghisellini, G. \& Tavecchio, F. 2009, Mon. Not. Roy. Astron. Soc., 397, 985
    
    \bibitem[{Giommi {et~al.}(2020{\natexlab{a}})Giommi, Glauch, Padovani, Resconi,
      Turcati, \& Chang}]{Giommi:2020hbx}
    Giommi, P., Glauch, T., Padovani, P., {et~al.} 2020{\natexlab{a}}, Mon. Not.
      Roy. Astron. Soc., 497, 865
    
    \bibitem[{Giommi {et~al.}(2020{\natexlab{b}})Giommi, Padovani, Oikonomou,
      Glauch, Paiano, \& Resconi}]{Giommi:2020viy}
    Giommi, P., Padovani, P., Oikonomou, F., {et~al.} 2020{\natexlab{b}}, Astron.
      Astrophys., 640, L4
    
    \bibitem[{Goldberg(1989)}]{goldberg89}
    Goldberg, D.~E. 1989, Genetic Algorithms in Search, Optimization, and Machine
      Learning (New York: Addison-Wesley)
    
    \bibitem[{Greene \& Ho(2005)}]{Greene:2005nj}
    Greene, J.~E. \& Ho, L.~C. 2005, Astrophys. J., 630, 122
    
    \bibitem[{Greiner {et~al.}(2008)}]{Greiner:2008ms}
    Greiner, J. {et~al.} 2008, Publ. Astron. Soc. Pac., 120, 405
    
    \bibitem[{Halzen \& Kheirandish(2020)}]{Halzen:2020clx}
    Halzen, F. \& Kheirandish, A. 2020, Nature Phys., 16, 498
    
    \bibitem[{Halzen {et~al.}(2019)Halzen, Kheirandish, Weisgarber, \&
      Wakely}]{Halzen:2018iak}
    Halzen, F., Kheirandish, A., Weisgarber, T., \& Wakely, S.~P. 2019, Astrophys.
      J. Lett., 874, L9
    
    \bibitem[{Harrison {et~al.}(2013)}]{NuSTAR:2013yza}
    Harrison, F.~A. {et~al.} 2013, Astrophys. J., 770, 103
    
    \bibitem[{Healey {et~al.}(2008)Healey, Romani, Cotter, Michelson, Schlafly,
      Readhead, Giommi, Chaty, Grenier, \& Weintraub}]{Healey:2007gb}
    Healey, S.~E., Romani, R.~W., Cotter, G., {et~al.} 2008, Astrophys. J. Suppl.,
      175, 97
    
    \bibitem[{Hooper(2016)}]{Hooper:2016jls}
    Hooper, D. 2016, JCAP, 09, 002
    
    \bibitem[{{Kadler} {et~al.}(2016){Kadler}, {Krau{\ss}}, {Mannheim}, {Ojha},
      {M{\"u}ller}, {Schulz}, {Anton}, {Baumgartner}, {Beuchert}, {Buson},
      {Carpenter}, {Eberl}, {Edwards}, {Eisenacher Glawion}, {Els{\"a}sser},
      {Gehrels}, {Gr{\"a}fe}, {Gulyaev}, {Hase}, {Horiuchi}, {James}, {Kappes},
      {Kappes}, {Katz}, {Kreikenbohm}, {Kreter}, {Kreykenbohm}, {Langejahn},
      {Leiter}, {Litzinger}, {Longo}, {Lovell}, {McEnery}, {Natusch}, {Phillips},
      {Pl{\"o}tz}, {Quick}, {Ros}, {Stecker}, {Steinbring}, {Stevens}, {Thompson},
      {Tr{\"u}stedt}, {Tzioumis}, {Weston}, {Wilms}, \&
      {Zensus}}]{2016NatPh..12..807K}
    {Kadler}, M., {Krau{\ss}}, F., {Mannheim}, K., {et~al.} 2016, Nature Physics,
      12, 807
    
    \bibitem[{Keivani {et~al.}(2018)Keivani, Murase, Petropoulou,
      {et~al.}}]{Keivani:2018rnh}
    Keivani, A., Murase, K., Petropoulou, M., {et~al.} 2018, Astrophys. J., 864, 84
    
    \bibitem[{Krimm {et~al.}(2013)}]{Krimm:2013lwa}
    Krimm, H.~A. {et~al.} 2013, Astrophys. J. Suppl., 209, 14
    
    \bibitem[{Kun {et~al.}(2023)Kun, Bartos, Becker~Tjus, Biermann, Franckowiak,
      Halzen, \& Mezo}]{Kun:2023uld}
    Kun, E., Bartos, I., Becker~Tjus, J., {et~al.} 2023, Astron. Astrophys., 679,
      A46
    
    \bibitem[{Liodakis \& Petropoulou(2020)}]{Liodakis:2020dvd}
    Liodakis, I. \& Petropoulou, M. 2020, Astrophys. J. Lett., 893, L20
    
    \bibitem[{Matthews {et~al.}(2001)Matthews, van Driel, \&
      Monnier-Ragaigne}]{Matthews:2000jw}
    Matthews, L.~D., van Driel, W., \& Monnier-Ragaigne, D. 2001, Astron.
      Astrophys., 365, 1
    
    \bibitem[{Murase(2020)}]{Murase:2019pef}
    Murase, K. 2020, PoS, ICRC2019, 965
    
    \bibitem[{Murase {et~al.}(2016)Murase, Guetta, \& Ahlers}]{Murase:2015xka}
    Murase, K., Guetta, D., \& Ahlers, M. 2016, Phys. Rev. Lett., 116, 071101
    
    \bibitem[{Murase {et~al.}(2014)Murase, Inoue, \& Dermer}]{Murase:2014foa}
    Murase, K., Inoue, Y., \& Dermer, C.~D. 2014, Phys. Rev. D, 90, 023007
    
    \bibitem[{Oikonomou {et~al.}(2019)Oikonomou, Murase, Padovani, Resconi, \&
      M\'esz\'aros}]{Oikonomou:2019djc}
    Oikonomou, F., Murase, K., Padovani, P., Resconi, E., \& M\'esz\'aros, P. 2019,
      Mon. Not. Roy. Astron. Soc., 489, 4347
    
    \bibitem[{Oikonomou {et~al.}(2021)Oikonomou, Petropoulou, Murase, Tohuvavohu,
      Vasilopoulos, Buson, \& Santander}]{Oikonomou:2021akf}
    Oikonomou, F., Petropoulou, M., Murase, K., {et~al.} 2021, JCAP, 10, 082
    
    \bibitem[{Padovani {et~al.}(2019)Padovani, Oikonomou, Petropoulou, Giommi, \&
      Resconi}]{Padovani:2019xcv}
    Padovani, P., Oikonomou, F., Petropoulou, M., Giommi, P., \& Resconi, E. 2019,
      Mon. Not. Roy. Astron. Soc., 484, L104
    
    \bibitem[{Padovani \& Resconi(2014)}]{Padovani:2014bha}
    Padovani, P. \& Resconi, E. 2014, Mon. Not. Roy. Astron. Soc., 443, 474
    
    \bibitem[{Padovani {et~al.}(2016)Padovani, Resconi, Giommi, Arsioli, \&
      Chang}]{Padovani:2016wwn}
    Padovani, P., Resconi, E., Giommi, P., Arsioli, B., \& Chang, Y.~L. 2016, Mon.
      Not. Roy. Astron. Soc., 457, 3582
    
    \bibitem[{{Paliya} {et~al.}(2020){Paliya}, {B{\"o}ttcher}, {Olmo-Garc{\'\i}a},
      {Dom{\'\i}nguez}, {Gil de Paz}, {Franckowiak}, {Garrappa}, \&
      {Stein}}]{2020ApJ...902...29P}
    {Paliya}, V.~S., {B{\"o}ttcher}, M., {Olmo-Garc{\'\i}a}, A., {et~al.} 2020,
      \apj, 902, 29
    
    \bibitem[{{Paliya} {et~al.}(2021){Paliya}, {Dom{\'\i}nguez}, {Ajello},
      {Olmo-Garc{\'\i}a}, \& {Hartmann}}]{2021ApJS..253...46P}
    {Paliya}, V.~S., {Dom{\'\i}nguez}, A., {Ajello}, M., {Olmo-Garc{\'\i}a}, A., \&
      {Hartmann}, D. 2021, \apjs, 253, 46
    
    \bibitem[{Paliya {et~al.}(2017)Paliya, Marcotulli, Ajello, Joshi, Sahayanathan,
      Rao, \& Hartmann}]{Paliya:2017xaq}
    Paliya, V.~S., Marcotulli, L., Ajello, M., {et~al.} 2017, Astrophys. J., 851,
      33
    
    \bibitem[{{Paliya} {et~al.}(2019){Paliya}, {Parker}, {Jiang}, {Fabian},
      {Brenneman}, {Ajello}, \& {Hartmann}}]{2019ApJ...872..169P}
    {Paliya}, V.~S., {Parker}, M.~L., {Jiang}, J., {et~al.} 2019, \apj, 872, 169
    
    \bibitem[{Palladino {et~al.}(2019)Palladino, Rodrigues, Gao, \&
      Winter}]{Palladino:2018lov}
    Palladino, A., Rodrigues, X., Gao, S., \& Winter, W. 2019, Astrophys. J., 871,
      41
    
    \bibitem[{Petropoulou(2014)}]{Petropoulou:2014pfa}
    Petropoulou, M. 2014, Astron. Astrophys., 571, A83
    
    \bibitem[{Petropoulou {et~al.}(2015)Petropoulou, Dimitrakoudis, Padovani,
      Mastichiadis, \& Resconi}]{Petropoulou:2015upa}
    Petropoulou, M., Dimitrakoudis, S., Padovani, P., Mastichiadis, A., \& Resconi,
      E. 2015, Mon. Not. Roy. Astron. Soc., 448, 2412
    
    \bibitem[{Petropoulou {et~al.}(2020{\natexlab{a}})Petropoulou, Murase,
      Santander, {et~al.}}]{Petropoulou:2019zqp}
    Petropoulou, M., Murase, K., Santander, M., {et~al.} 2020{\natexlab{a}},
      Astrophys. J., 891, 115
    
    \bibitem[{Petropoulou {et~al.}(2020{\natexlab{b}})Petropoulou, Oikonomou,
      Mastichiadis, Murase, Padovani, Vasilopoulos, \&
      Giommi}]{Petropoulou:2020pqh}
    Petropoulou, M., Oikonomou, F., Mastichiadis, A., {et~al.} 2020{\natexlab{b}},
      Astrophys. J., 899, 113
    
    \bibitem[{Petropoulou {et~al.}(2022)Petropoulou, Psarras, \&
      Giannios}]{Petropoulou:2022sct}
    Petropoulou, M., Psarras, F., \& Giannios, D. 2022, Mon. Not. Roy. Astron.
      Soc., 518, 2719
    
    \bibitem[{Petropoulou {et~al.}(2017)Petropoulou, Vasilopoulos, \&
      Giannios}]{Petropoulou:2016tro}
    Petropoulou, M., Vasilopoulos, G., \& Giannios, D. 2017, Mon. Not. Roy. Astron.
      Soc., 464, 2213
    
    \bibitem[{Reimer {et~al.}(2019)Reimer, Boettcher, \& Buson}]{Reimer:2018vvw}
    Reimer, A., Boettcher, M., \& Buson, S. 2019, Astrophys. J., 881, 46, [Erratum:
      Astrophys.J. 899, 168 (2020)]
    
    \bibitem[{Rodrigues {et~al.}(2018)Rodrigues, Fedynitch, Gao, Boncioli, \&
      Winter}]{Rodrigues:2017fmu}
    Rodrigues, X., Fedynitch, A., Gao, S., Boncioli, D., \& Winter, W. 2018,
      Astrophys. J., 854, 54
    
    \bibitem[{Rodrigues {et~al.}(2019)Rodrigues, Gao, Fedynitch, Palladino, \&
      Winter}]{Rodrigues:2018tku}
    Rodrigues, X., Gao, S., Fedynitch, A., Palladino, A., \& Winter, W. 2019,
      Astrophys. J., 874, L29
    
    \bibitem[{Rodrigues {et~al.}(2021{\natexlab{a}})Rodrigues, Garrappa, Gao,
      Paliya, Franckowiak, \& Winter}]{Rodrigues:2020fbu}
    Rodrigues, X., Garrappa, S., Gao, S., {et~al.} 2021{\natexlab{a}}, Astrophys.
      J., 912, 54
    
    \bibitem[{Rodrigues {et~al.}(2021{\natexlab{b}})Rodrigues, Heinze, Palladino,
      van Vliet, \& Winter}]{Rodrigues:2020pli}
    Rodrigues, X., Heinze, J., Palladino, A., van Vliet, A., \& Winter, W.
      2021{\natexlab{b}}, Phys. Rev. Lett., 126, 191101
    
    \bibitem[{Roming {et~al.}(2005)}]{Roming:2005hv}
    Roming, P. W.~A. {et~al.} 2005, Space Sci. Rev., 120, 95
    
    \bibitem[{Sahakyan {et~al.}(2022)Sahakyan, Giommi, Padovani, Petropoulou,
      B\'egu\'e, Boccardi, \& Gasparyan}]{Sahakyan:2022nbz}
    Sahakyan, N., Giommi, P., Padovani, P., {et~al.} 2022, Mon. Not. Roy. Astron.
      Soc., 519, 1396
    
    \bibitem[{Sahakyan {et~al.}(2023)Sahakyan, Harutyunyan, \&
      Israyelyan}]{Sahakyan:2023cas}
    Sahakyan, N., Harutyunyan, G., \& Israyelyan, D. 2023, Mon. Not. Roy. Astron.
      Soc., 521, 1013
    
    \bibitem[{Sahu {et~al.}(2020)Sahu, L\'opez~Fort\'\i{}n, \&
      Nagataki}]{Sahu:2020eep}
    Sahu, S., L\'opez~Fort\'\i{}n, C.~E., \& Nagataki, S. 2020, Astrophys. J., 898,
      103
    
    \bibitem[{{Sbarrato} {et~al.}(2012){Sbarrato}, {Ghisellini}, {Maraschi}, \&
      {Colpi}}]{2012MNRAS.421.1764S}
    {Sbarrato}, T., {Ghisellini}, G., {Maraschi}, L., \& {Colpi}, M. 2012, \mnras,
      421, 1764
    
    \bibitem[{{Shakura} \& {Sunyaev}(1973)}]{1973A&A....24..337S}
    {Shakura}, N.~I. \& {Sunyaev}, R.~A. 1973, \aap, 24, 337
    
    \bibitem[{Shakura \& Sunyaev(1976)}]{Shakura:1976xk}
    Shakura, N.~I. \& Sunyaev, R.~A. 1976, Mon. Not. Roy. Astron. Soc., 175, 613
    
    \bibitem[{{Shen} {et~al.}(2011){Shen}, {Richards}, {Strauss}, {Hall},
      {Schneider}, {Snedden}, {Bizyaev}, {Brewington}, {Malanushenko},
      {Malanushenko}, {Oravetz}, {Pan}, \& {Simmons}}]{2011ApJS..194...45S}
    {Shen}, Y., {Richards}, G.~T., {Strauss}, M.~A., {et~al.} 2011, \apjs, 194, 45
    
    \bibitem[{Tavecchio {et~al.}(2010)Tavecchio, Ghisellini, Ghirlanda, Foschini,
      \& Maraschi}]{Tavecchio:2009zb}
    Tavecchio, F., Ghisellini, G., Ghirlanda, G., Foschini, L., \& Maraschi, L.
      2010, Mon. Not. Roy. Astron. Soc., 401, 1570
    
    \bibitem[{{Urry} \& {Padovani}(1995)}]{1995PASP..107..803U}
    {Urry}, C.~M. \& {Padovani}, P. 1995, \pasp, 107, 803
    
    \bibitem[{Weisskopf {et~al.}(2000)Weisskopf, Tananbaum, van Speybroeck, \&
      O'Dell}]{Weisskopf:2000tx}
    Weisskopf, M.~C., Tananbaum, H.~D., van Speybroeck, L.~P., \& O'Dell, S.~L.
      2000, Proc. SPIE Int. Soc. Opt. Eng., 4012, 2
    
    \bibitem[{Wood {et~al.}(2018)Wood, Caputo, Charles, Di~Mauro, Magill, \&
      Perkins}]{Wood:2017yyb}
    Wood, M., Caputo, R., Charles, E., {et~al.} 2018, PoS, ICRC2017, 824
    
    \bibitem[{Xue {et~al.}(2021)Xue, Liu, Wang, Ding, \& Wang}]{Xue:2020kuw}
    Xue, R., Liu, R.-Y., Wang, Z.-R., Ding, N., \& Wang, X.-Y. 2021, Astrophys. J.,
      906, 51
    
    \bibitem[{Zhang {et~al.}(2020)Zhang, Petropoulou, Murase, \&
      Oikonomou}]{Zhang:2019htg}
    Zhang, B.~T., Petropoulou, M., Murase, K., \& Oikonomou, F. 2020, Astrophys.
      J., 889, 118
    
    \end{thebibliography}


\begin{appendix}

\section{Comparison with recent studies of individual sources}
\label{app:literature}

\begin{figure*}[htpb!]

\includegraphics[width=0.47\textwidth,trim={0 12mm 0 5mm}, clip]{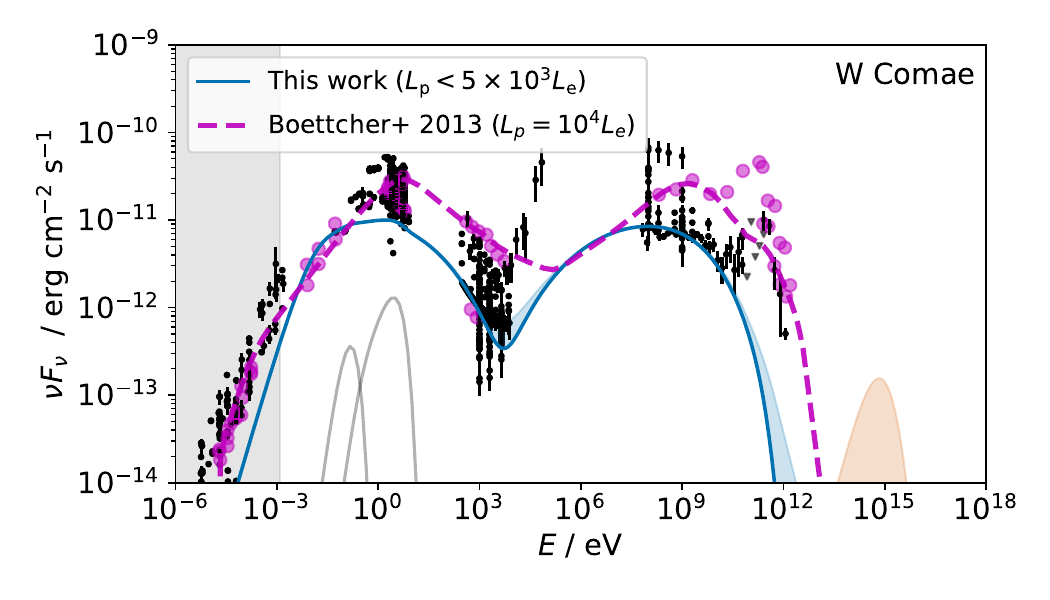}
\includegraphics[width=0.47\textwidth,trim={0 12mm 0 5mm}, clip]{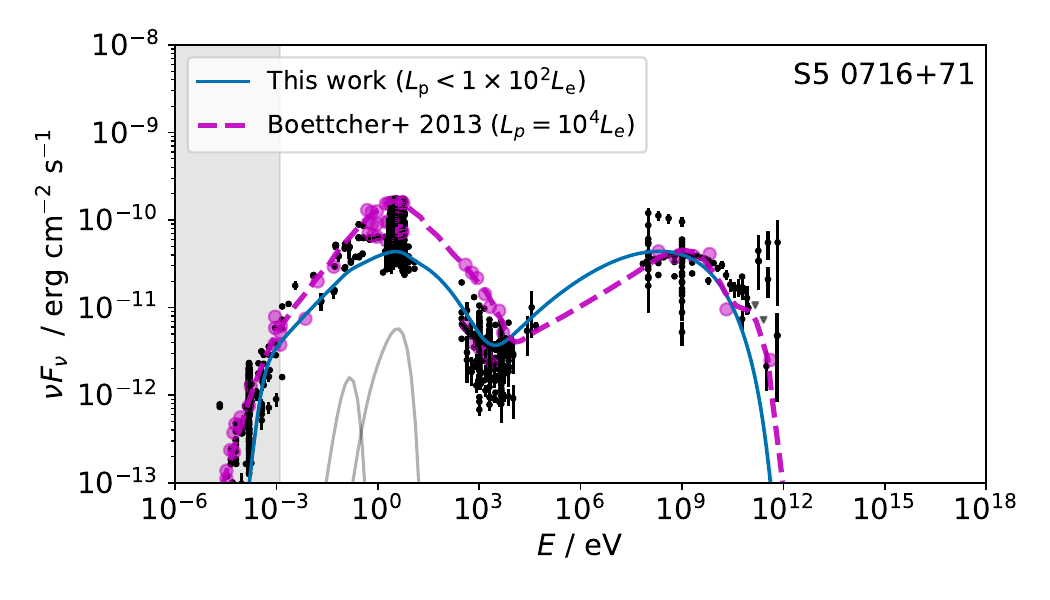}

\includegraphics[width=0.47\textwidth,trim={0 12mm 0 5mm}, clip]{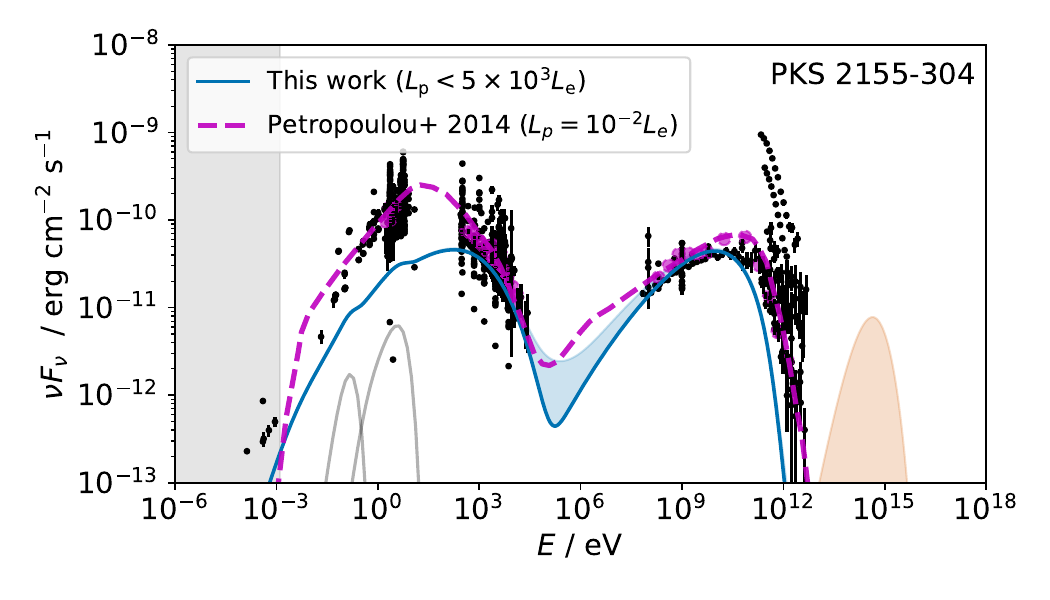}
\includegraphics[width=0.47\textwidth,trim={0 12mm 0 5mm}, clip]{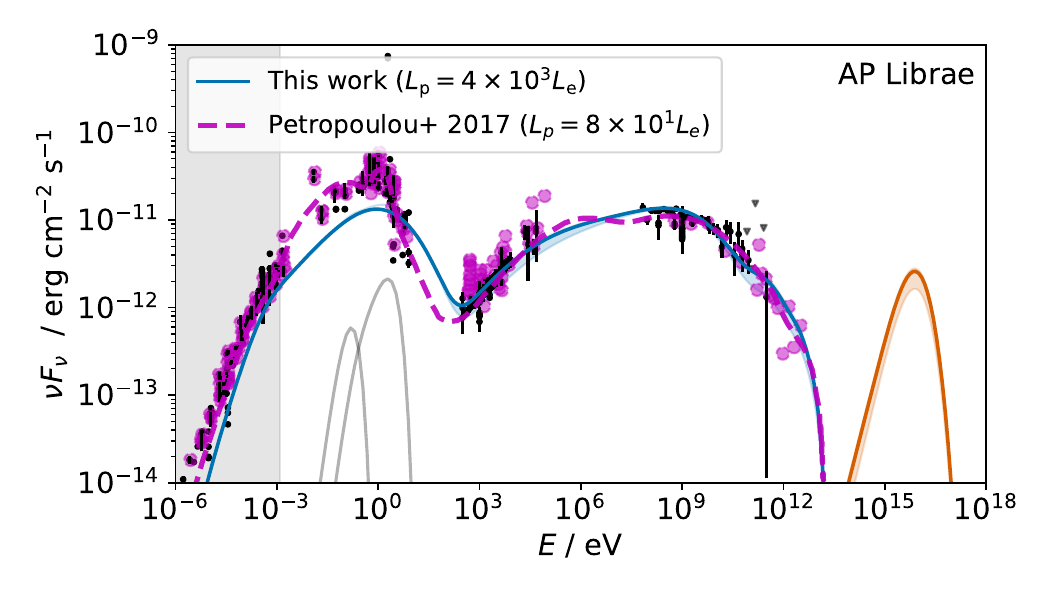}

\includegraphics[width=0.47\textwidth,trim={0 12mm 0 5mm}, clip]{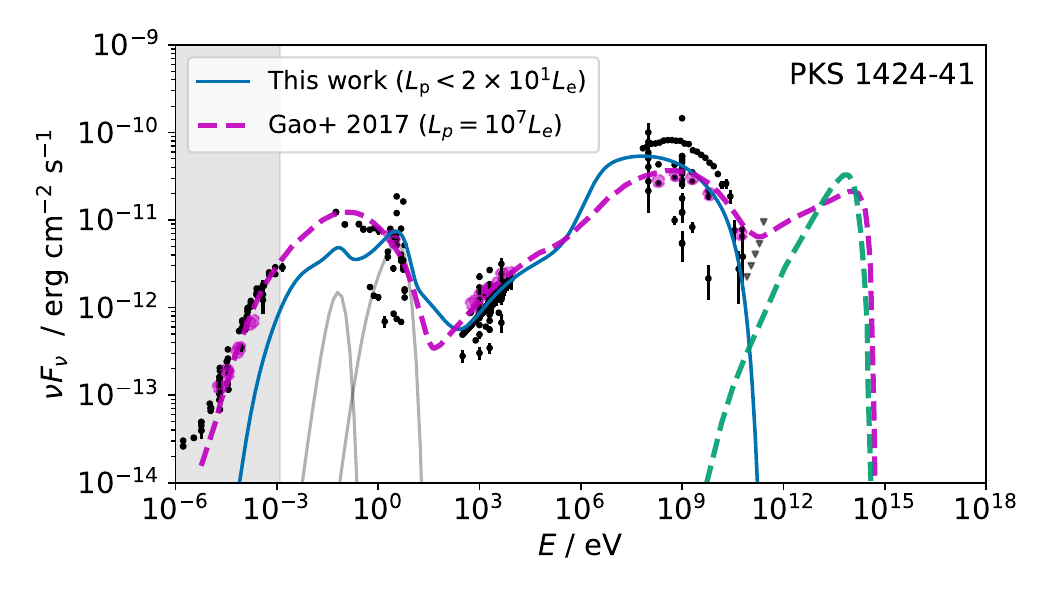}
\includegraphics[width=0.47\textwidth,trim={0 12mm 0 5mm}, clip]{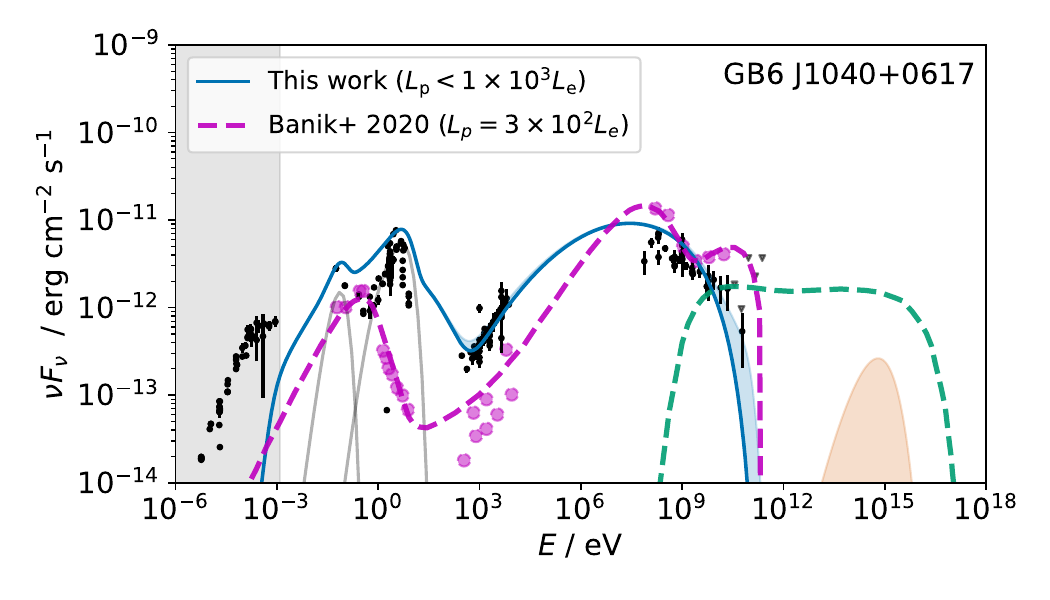}


\includegraphics[width=0.47\textwidth,trim={0 12mm 0 5mm}, clip]{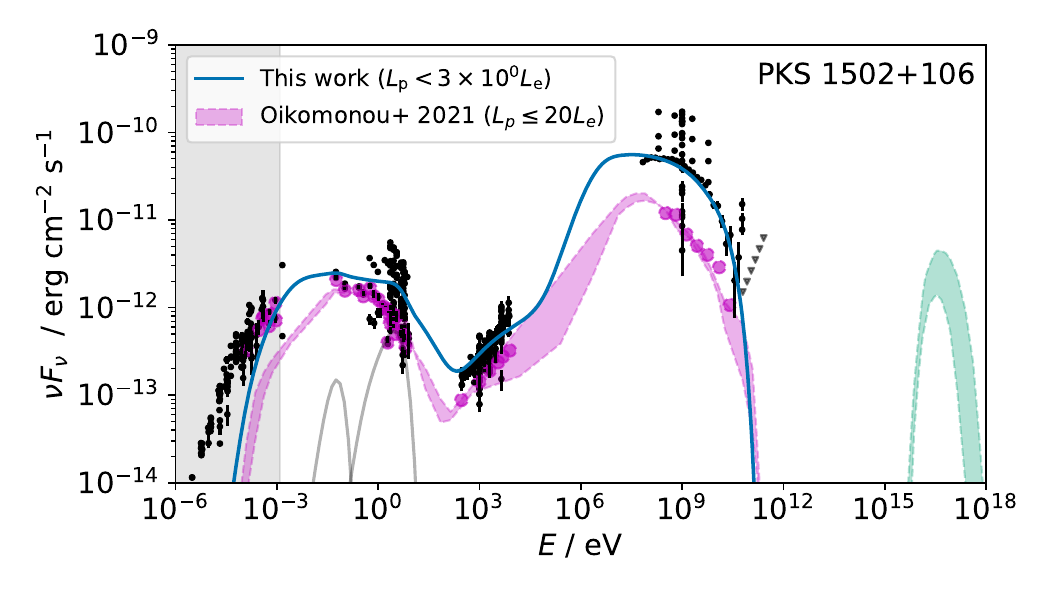}
\includegraphics[width=0.47\textwidth,trim={0 12mm 0 5mm}, clip]{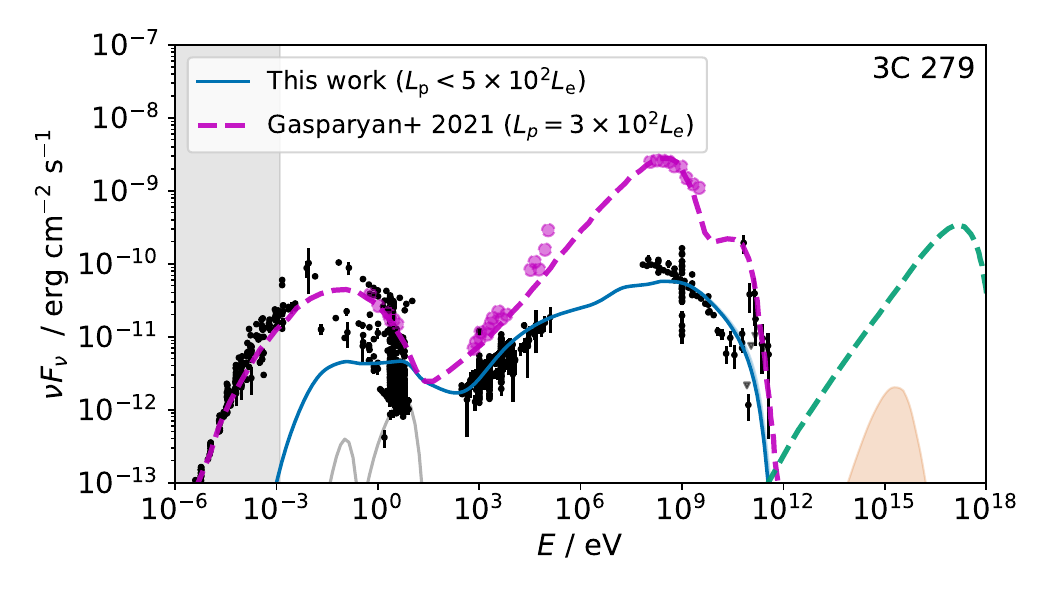}


\includegraphics[width=0.47\textwidth,trim={0 5mm 0 5mm}, clip]{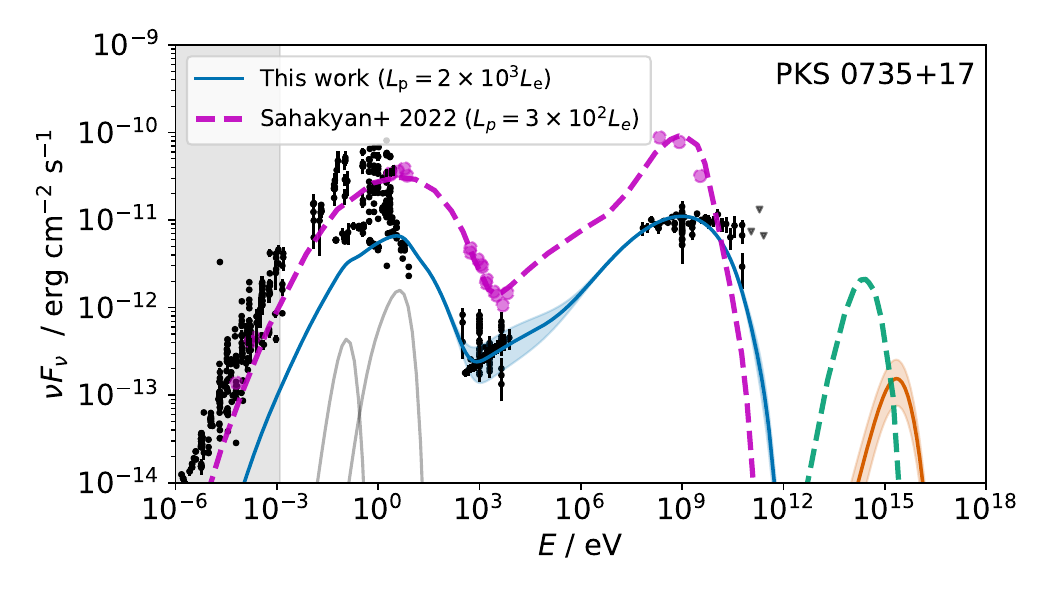}
\includegraphics[width=0.47\textwidth,trim={0 5mm 0 5mm}, clip]{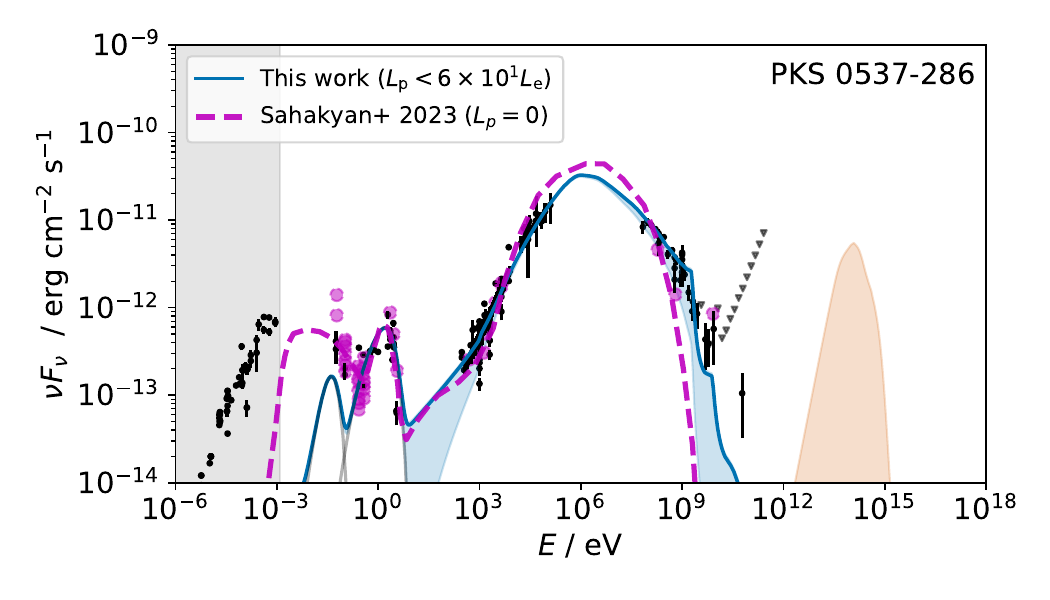}

\caption{Comparison between the results obtained in this work and ten selected results from recent multiwavelength studies, in chronological order of publication~\citep{Boettcher:2013wxa,Petropoulou:2014pfa,Gao:2016uld,Petropoulou:2016tro,Banik:2019twt,Oikonomou:2021akf,Gasparyan:2021oad,Sahakyan:2022nbz,Sahakyan:2023cas}. In blue and orange we show respectively the best-fit photon and neutrino fluxes from the current model; the dashed magenta and green curves show the photon and neutrino spectra (when available) predicted by previous works. The magenta points represent the data fitted in the respective study, which in some cases varies significantly from the current work because of time-domain selection. The best-fit baryonic loading values are given in each caption.}
\label{fig:literature}
\end{figure*}

Multiple sources in this sample have also been independently modeled with numerical, time-dependent frameworks in previous studies. In this section we compare our results with some of those in the literature, a comparison that is summarized in \Fig\ref{fig:literature}.

On the upper row we see two results by~\citet{Boettcher:2013wxa}, who applied  leptonic and leptohadronic models to a group of six FSRQs, four LBLs and two IBLs. Of these 12 sources, nine are also included in the sample studied in this work. On the left we see the fits to the IBL object W Comae. The dashed magenta curve shows the best-fit SED by~\citet{Boettcher:2013wxa} and the solid blue curve the one from this work. The datasets that are fitted are shown as magenta and black data points, respectively. As we can see, the authors fit a time-selected dataset that captures a state of relatively enhanced activity compared to the solution found in this work, which corresponds to a quiescent state. The high gamma-ray fluxes above 10~GeV are better described by the model by~\citet{Boettcher:2013wxa}. It is possible that the low state in optical captured by our model should correspond to a lower state in high-energy gamma rays, which is the case in our best-fit result. However, further analysis of the quiescent state data would be necessary in order to confirm this difference in the gamma-ray peak between the two models. 

In the right plot we see the two results of the modeling of S5~0716+71. In this case, the synchrotron peaks predicted by the two models correspond again to different activity states, but the gamma-ray emission is at a comparable level. In both these cases the current model predicts a lower hadronic contribution, with only an upper limit on the proton luminosity. In comparison, the results by~\citet{Boettcher:2013wxa} suggest a large proton loading of $\sim10^4$, although the peak of the gamma-ray emission is dominated by inverse Compton scattering like in our case.

In the left panel on the second row of \Fig\ref{fig:literature} we compare our result for the HBL object PKS~2155-304 with that by \citet{Petropoulou:2014pfa}. This is a good example of a strongly variable source from optical, X-rays, and gamma rays. As we can see, the gamma-ray data points reach as high as $10^{-9}\,\mathrm{erg}\,\mathrm{cm}^{-2}\,\mathrm{s}^{-1}$. In our model, as was the case for W Comae, we see that our result reproduces a state of both low synchrotron flux and low gamma-ray flux. On the other hand,~\citet{Petropoulou:2014pfa} have  modeled quasi-simultaneous data form the source (magenta points) and have shown that those fluxes cannot be reproduced with a single-zone leptonic model. Instead, it requires either two distinct emission zones or a proton population co-accelerated with the electrons, which is the case shown here as a dashed magenta curve. In our approach we do not distinguish between data from different epochs (see discussion in \Sec\ref{sec:limitations}), so our method cannot produce this kind of statement regarding any specific epoch. Instead, the only statement that can be derived from our method for this source is that the SED shown in blue is the best fit to the overall dataset from the source (in black) that can be achieved with a one-zone model, which is achieved without any necessary proton component.

On the same row, we show on the right the result of our leptohadronic modeling of the IBL object AP Librae as well as the result by~\citet{Petropoulou:2016tro}. With their model, \citet{Petropoulou:2016tro} have shown that TeV emission from AP Librae can be well described with a component from hadronic interactions. As we can see by comparing the dashed magenta and blue curves in the plot, our result supports this finding. The best-fit proton populations have a higher luminosity in our case, but a similar maximum energy within a factor of 5. Although the predicted  neutrino spectrum is not shown explicitly by \citet{Petropoulou:2016tro}, the predicted flux should lie within the same order of magnitude of that predicted by this model, given the level of cascade emission seen in the right plot of~\Fig~1 of that work.

The major difference in terms of multiwavelength predictions lies in the nonthermal optical emission. This is because, as shown also in \Fig\ref{fig:fits} (e.g., for PKS~0426-380 and B2~2234+28A), the best-fit SED corresponds to a low state in optical emission. However, the predicted synchrotron peak frequency is compatible with the observed one and in this case also with that predicted by \citet{Petropoulou:2016tro}, at $\nu_\mathrm{syn}\approx3\times10^{14}\,\mathrm{Hz}$. This means that in spite of the different flux level, the model still captures the IBL nature of the source.

In the third row of~\Fig\ref{fig:literature}, we see on the left the photon spectrum of PKS 1424-41 predicted by \citet{Gao:2016uld}, together with the respective neutrino flux as a dashed green curve. This was in the context of the detection of a high-energy IceCube event in 2016 \citep{2016NatPh..12..807K} from the direction of the source. In terms of the multiwavelength fluxes, the first difference between the two models lies in the flux level of the synchrotron emission, since the search algorithm used in this work has selected a low state compared to that by \citet{Gao:2016uld}. This is because 1) we treat radio fluxes as upper limits, while in this case they can in fact be described by a single-zone model as shown by \citet{Gao:2016uld}; 2) the large variability in the ultraviolet range of the dataset used in this work, which in this case disfavors solutions with high synchrotron fluxes. In contrast, in the previous study the data was time-selected, as represented by the magenta points, which means that a particular activity state is targeted. The second major difference is the large neutrino flux predicted by \citet{Gao:2016uld}, while here the best-fit result is compatible with a purely leptonic solution. This is because 1) \citet{Gao:2016uld} were explaining an IceCube association, for which high neutrino fluxes are necessary; and 2) the same study derived a baryonic loading value of $10^7$, which lies considerably above the range tested in this work (cf.~\Tab\ref{tab:parameters}).

In the case of GB6~J1040+0617, spatially coincident with an IceCube alert in 2014~\citep{Fermi-LAT:2019hte}, we see that the neutrino flux predicted by~\citet{Banik:2019twt} has a drastically different spectral shape compared to this work (third row of \Fig\ref{fig:literature}, right panel). This difference is due to the fact that that model includes proton interactions with high-density clouds surrounding the relativistic jet. This leads to a soft neutrino spectrum typical of proton-proton models and a gamma-ray emission that is harder at high energies compared to that found in this work. 

In the left plot on the fourth row of \Fig\ref{fig:literature} we compare our result for the FSRQ PKS~1502+106 with that found by~\citep{Oikonomou:2021akf} \citep[see also][where the same model used in this work is applied to time-selected datasets from this source]{Rodrigues:2020fbu}. We can see that in spite of a baryonic loading only a factor of a few higher, \citet{Oikonomou:2021akf} predict a much stronger neutrino emission, which is possible due to protons with energies above 10~PeV, which interact efficiently with infrared emission from a dust torus. In comparison, as listed in \Tab\ref{tab:all_parameters}, we set a limit on the neutrino flux of at most $3\times10^{-16}\,\mathrm{erg}\,\mathrm{cm}^{-2}\mathrm{s}^{-1}$, a factor of $10^4$ lower. Two aspects are worth noting in this case: firstly, a solution like that obtained by \citet{Oikonomou:2021akf} should in principle be possible in the current  framework and within the parameter set explored in this work, with a jet emission zone outside the BLR but within the radiation field of the dust torus. The reason why the genetic algorithm converged on this parameter set may be because it provides a larger fraction of parameter space with acceptable chi-squared values, therefore gathering more solutions over time and being therefore statistically preferred. The second aspect is that in the solution by \citet{Oikonomou:2021akf} the emission that originates in  hadronic interactions is actually subdominant in the X-ray range compared to the leptonic component and dominates the observations only for gamma rays above tens of GeV, near the cut off of the SED (cf.~\Fig~8 of that reference). In that sense, this proton component may easily be neglected by a large-scale parameter search like the one used in the current work, when no requirements on neutrino emission are enforced.  

On the right we show the result for 3C 279, an extremely variable quasar present in our sample that was also modeled recently by~\citet{Gasparyan:2021oad}. Comparing the dashed green curve with the orange shaded area, we see that the neutrino fluxes predicted by the two models are vastly different. The first factor for this difference is the fact that~\citet{Gasparyan:2021oad} model a state of extremely high gamma ray fluxes corresponding to a flaring event in 2015. This allows for strong gamma-ray emission from efficient hadronic interactions, compared to the dataset modeled in this work. The second interesting fact is that this extremely high neutrino flux is achieved with a baryon loading of $L^\prime_\mathrm{p}/L^\prime_\mathrm{e}=300$, while our best-fit result has a higher value of $L^\prime_\mathrm{p}/L^\prime_\mathrm{e}=500$ and yet places only an upper limit on neutrino emission. This is due to the different nature of the model by~\citet{Gasparyan:2021oad}, particularly the high values of maximum proton energy (about 200~PeV) and magnetic field strength (70~gauss), both of which are above our search range. This illustrates the point discussed in \Sec\ref{sec:limitations} on the existence of alternative scenarios that lie outside the parameter space considered in this work that can lead to different predictions in the neutrino sector.  

In the case of PKS~0735+178 (lower left plot), \citet{Sahakyan:2022nbz} describe multiwavelength observations surrounding the detection of a high-energy IceCube event from the direction of the source in December 2021, during which the source was undergoing a multiwavelength flare of unprecedented magnitude. In contrast, the dataset modeled in this work is not time-selected and is an order of magnitude lower in flux.  Additionally to this difference, the authors also use a masquerading BL Lac model, based on observations that suggest that in spite of its classification as  BL Lac, the source actually possesses a BLR whose emission is outshone by the synchrotron peak \citep[see also][]{Padovani:2019xcv}. In contrast, in this work we do not consider the effect of a BLR for BL Lacs, which leads to a slightly lower Compton dominance and a different spectral shape of the predicted X-ray and gamma-ray spectra. Furthermore, in the current model the only target photons for photo-pion production are those from synchrotron emission; compared to an external BLR field, these synchrotron photons have lower densities, leading to a slightly lower ratio $L_\nu/L_\gamma$, and lower frequencies, leading to a higher energy of the interacting protons and the emitted neutrinos.

Finally, in the recent study by \citet{Sahakyan:2023cas}, the authors describe time-selected multiwavelength datasets from the source PKS~0537 by with a purely leptonic model. One of their fits to multiwavelength dataset is shown in the lower right panel of \Fig\ref{fig:literature}, compared to that obtained in this work. As we can see, the X-ray and gamma-ray spectral shapes are comparable, but the additional component from hadronic interactions can potentially help describe the slightly harder gamma-ray fluxes at the highest energies. At the same time, as we can see in the blue shaded band, the SED by \citet{Sahakyan:2023cas} is consistent with the result that we find in the purely leptonic limit, which is contained within the $1\sigma$ uncertainty band, shown in blue.

\section{Best-fit parameters}
\label{app:parameters}

\begin{figure*}[htpb!]
\includegraphics[width=\textwidth]{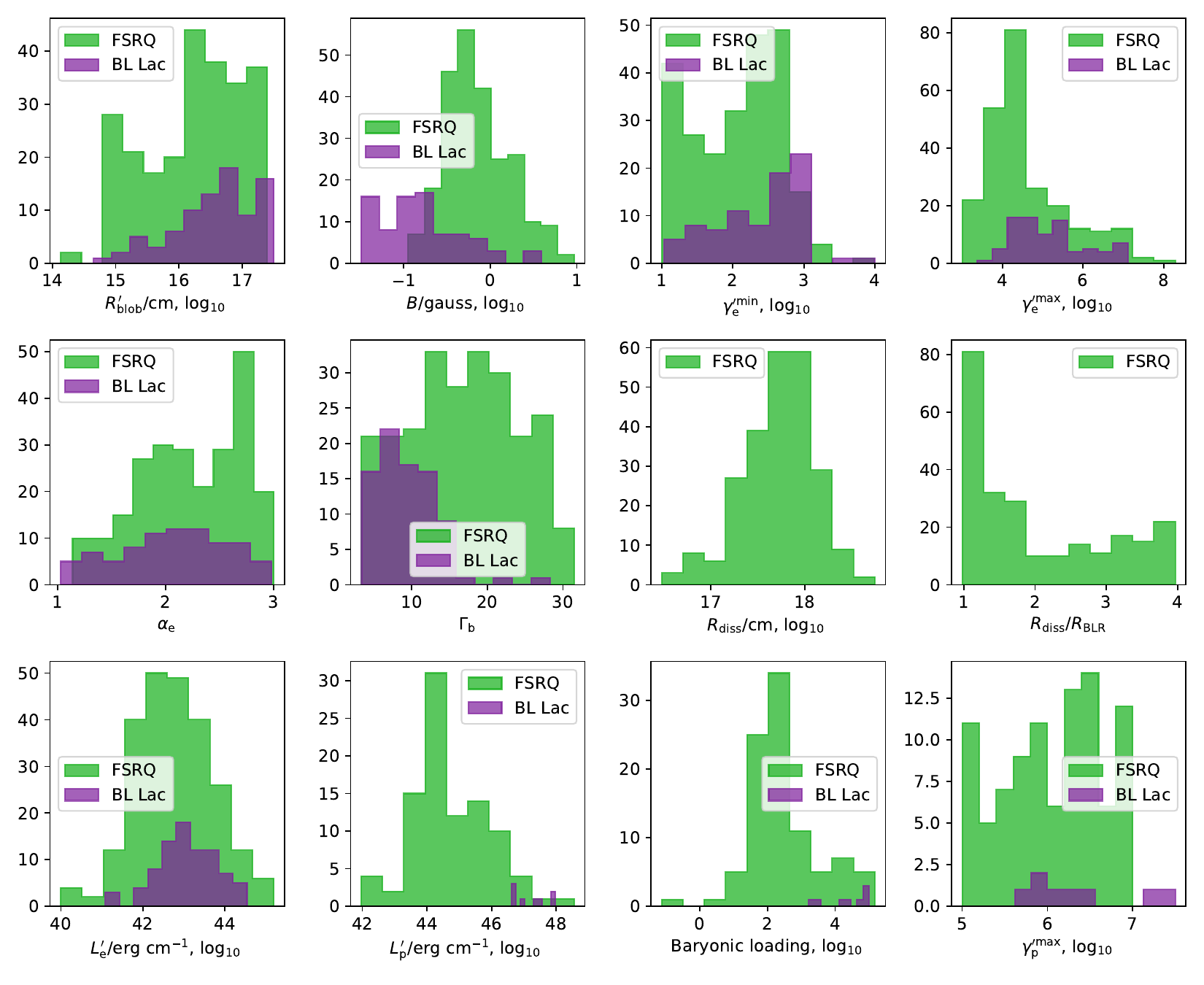}
\caption{Binned distribution of the best-fit parameter values for the entire sample. In green we show the results for the FSRQs in the sample and in purple the BL Lacs. The best-fit parameter values are also listed in \Tab\ref{tab:all_parameters}, as well as in the online repository \protect\hyperlink{https://github.com/xrod/lephad-blazars}{https://github.com/xrod/lephad-blazars}.}
\label{fig:lephad_histrograms}
\end{figure*}

In \Fig\ref{fig:lephad_histrograms} we show the distribution of the best-fit parameters for FSRQs (green) and BL Lacs (purple). The parameters are those listed in \Tab\ref{tab:parameters}.

In the case of $R_\mathrm{diss}$, only FSRQs are included in the histogram since the BL Lac model is not sensitive to this parameter. The last four histograms relate to hadronic variables; for these cases, we only include the sources whose best fit has a nonzero baryonic loading. 


In \Tab\ref{tab:all_parameters} we provide the best-fit parameter values of the leptohadronic model for the entire sample. Additionally, for each source we also provide the predicted energy-integrated muon neutrino flux, $F_{\nu_\mu}$, the predicted number of events per year in IceCube, $N_{\nu_\mu}$, and the peak energy of the neutrino spectrum in the observer's frame, $E_\nu^\mathrm{peak}$. 

As explained in the main text, the errors in the neutrino flux $F_{\nu_\mu}$ were obtained by varying the proton injection luminosity until the change in the emission violates the observations by 1$\sigma$ compared to the best-fit result. The corresponding uncertainties in $N_{\nu_\mu}$ and $L^\prime_\mathrm{p}$ can be extrapolated from those in $F_{\nu_\mu}$, since these variables are directly proportional to each other. 

The sources in \Tab\ref{tab:all_parameters} are given in descending order of the predicted neutrino flux, starting with those for which the flux is incompatible with zero within the uncertainty. Then follow those for which the best-fit neutrino flux is nonzero but compatible with zero within the uncertainties, also provided in descending order of flux. Finally, we list those sources for which the best-fit solution is purely leptonic. These are sorted by the upper limit on the neutrino flux.

\begin{table*}
  \centering
  \caption{\label{tab:all_parameters}List of best-fit parameter values for each source, ordered by the predicted muon neutrino flux $F_{\nu_\mu}$. The parameter values are given in the form $\log_{10}$(best-fit parameter value / CGS units), with the following exceptions: the predicted event rate in IceCube, as shown in \Fig\ref{fig:neutrino_events}, is given as $\log10(N_{\mu_\nu}/\mathrm{year})$; the predicted neutrino peak energy is given as $\log_{10}(E_{\nu_\mu}/\mathrm{GeV})$; and for $\alpha_\mathrm{e,p}$ and $\Gamma_\mathrm{b}$ the actual parameter value is given, rather than the logarithm. The values of $L_\mathrm{disk}$ and $R_\mathrm{BLR}$ are fixed to those deduced by \citet{Paliya:2017xaq}. For sources modeled as BL Lacs a best-fit  dissipation radius is not given, because the fit is not sensitive to it. The full table can be found in machine-readable format in the online repository \protect\hyperlink{https://github.com/xrod/lephad-blazars}{https://github.com/xrod/lephad-blazars}, together with the model results for each individual source.}
\resizebox{\textwidth}{!}{
  \begin{tabular}{llllllllllllllllll}
  
  \toprule
Source & Class & $z$ & $F_{\nu_\mu}$ & $N_{\nu_\mu}$ & $E_\nu^\mathrm{peak}$ & $L_\mathrm{disk}$ & $R_\mathrm{BLR}$ & $R_\mathrm{diss}$ & $R_\mathrm{b}^\prime$ & $B^\prime$ & $\Gamma$ & $\gamma_\mathrm{p}^\mathrm{max}$ & $L_\mathrm{p}^\prime$ & $L_\mathrm{e}^\prime$ & $\gamma_\mathrm{e}^\mathrm{min}$ & $\gamma_\mathrm{e}^\mathrm{max}$ & $\alpha_\mathrm{e}$ \\
\midrule
PKS 0426-380&LBL&  1.1&-11.2$^{+0.1}_{-1.5}$&-0.6$^{+0.1}_{-1.5}$&  6.0&46.0&17.5&-&16.2&-1.3&13.6&6.2&48.0$^{+0.1}_{-1.0}$&43.7&2.1&4.6&1.5\\
PKS B0802-010&FSRQ&  1.4&-11.5$^{+0.1}_{-0.3}$&-0.5$^{+0.1}_{-0.3}$&  5.3&46.4&17.7&17.7&15.2&-0.3&19.3&5.1&46.1$^{+0.1}_{-0.2}$&43.0&1.0&3.0&3.0\\
GB6 J0654+5042&FSRQ&  1.2&-11.5$^{+0.0}_{-1.3}$&-1.2$^{+0.0}_{-1.3}$&  6.1&45.0&17.0&17.0&15.8&0.4&24.9&6.2&44.1$^{+0.0}_{-1.3}$&41.4&2.8&3.1&1.9\\
S4 1849+67&FSRQ&  0.7&-11.6$^{+0.0}_{-0.2}$&-1.6$^{+0.0}_{-0.2}$&  6.4&45.5&17.2&17.4&16.6&-0.1&18.5&6.4&44.5$^{+0.0}_{-0.2}$&42.2&1.3&3.9&2.0\\
AP Librae&IBL&  0.1&-11.7$^{+0.0}_{-0.2}$&-2.1$^{+0.0}_{-0.2}$&  6.9&43.2&16.1&-&17.3&-1.3&4.0&7.1&46.6$^{+0.0}_{-0.2}$&43.0&2.5&4.8&2.1\\
PKS 1954-388&FSRQ&  0.6&-11.9$^{+0.4}_{-0.3}$&-0.9$^{+0.4}_{-0.3}$&  5.3&46.0&17.5&18.1&15.0&0.0&24.5&5.1&46.7$^{+0.2}_{-0.2}$&41.7&1.8&3.8&1.9\\
TXS 0938-133&FSRQ&  0.6&-12.0$^{+0.1}_{-0.3}$&-1.0$^{+0.1}_{-0.3}$&  5.4&45.6&17.3&17.3&15.0&-0.2&17.9&5.0&45.0$^{+0.1}_{-0.2}$&41.2&1.6&3.3&2.6\\
B2 2234+28A&FSRQ&  0.8&-12.0$^{+0.3}_{-0.8}$&-1.4$^{+0.3}_{-0.8}$&  5.8&45.4&17.2&17.3&16.8&-0.3&19.9&5.6&44.3$^{+0.2}_{-0.8}$&41.9&1.4&3.6&1.4\\
PKS 2201+171&FSRQ&  1.1&-12.0$^{+0.0}_{-0.7}$&-3.1$^{+0.0}_{-0.7}$&  7.2&45.8&17.4&18.0&17.3&-0.2&19.1&7.0&43.9$^{+0.0}_{-0.7}$&42.4&3.2&3.7&2.2\\
PKS 1104-445&FSRQ&  1.6&-12.1$^{+0.1}_{-0.3}$&-2.9$^{+0.1}_{-0.3}$&  7.0&47.0&18.0&18.6&17.3&-0.4&12.9&7.0&45.6$^{+0.1}_{-0.3}$&43.3&2.8&4.0&2.2\\
S4 1030+41&FSRQ&  1.1&-12.2$^{+0.1}_{-0.5}$&-3.3$^{+0.1}_{-0.5}$&  7.2&46.0&17.5&18.1&17.3&-0.3&17.4&7.0&44.0$^{+0.1}_{-0.5}$&42.4&2.6&4.1&2.8\\
TXS 2331+073&FSRQ&  0.4&-12.3$^{+0.0}_{-0.5}$&-1.3$^{+0.0}_{-0.5}$&  5.2&45.6&17.3&17.3&15.8&0.3&6.8&5.3&46.6$^{+0.0}_{-0.4}$&42.6&2.9&3.0&2.6\\
S5 1053+81&FSRQ&  0.7&-12.3$^{+0.1}_{-0.3}$&-2.8$^{+0.1}_{-0.3}$&  7.1&45.5&17.3&17.5&16.6&0.0&21.8&6.7&44.1$^{+0.1}_{-0.3}$&41.5&2.0&4.3&2.7\\
PMN J0850-1213&FSRQ&  0.6&-12.3$^{+0.0}_{-0.4}$&-3.5$^{+0.0}_{-0.4}$&  7.3&45.0&17.0&17.5&17.2&-0.2&19.3&6.8&43.4$^{+0.0}_{-0.4}$&41.6&2.8&3.8&2.2\\
PKS 1145-071&FSRQ&  1.3&-12.4$^{+0.2}_{-1.8}$&-2.3$^{+0.2}_{-1.8}$&  6.4&46.1&17.6&17.8&16.6&-0.5&14.1&6.6&45.5$^{+0.2}_{-1.8}$&42.5&3.0&4.2&1.9\\
4C +31.63&FSRQ&  0.3&-12.4$^{+0.0}_{-0.4}$&-1.4$^{+0.0}_{-0.4}$&  5.1&46.2&17.6&17.6&16.5&1.0&8.1&5.0&45.1$^{+0.0}_{-0.4}$&42.5&1.0&3.5&2.3\\
S4 1842+68&FSRQ&  0.5&-12.4$^{+0.4}_{-0.4}$&-1.5$^{+0.4}_{-0.4}$&  5.4&45.1&17.0&17.6&15.2&0.4&14.5&5.4&46.5$^{+0.2}_{-0.2}$&41.4&2.2&3.8&1.8\\
TXS 0214+083&IBL&  0.1&-12.5$^{+0.2}_{-0.3}$&-1.9$^{+0.2}_{-0.3}$&  5.9&43.1&16.1&-&17.1&-1.3&7.4&6.1&47.5$^{+0.1}_{-0.2}$&42.6&2.8&4.0&1.8\\
PKS 2209+236&FSRQ&  1.1&-12.5$^{+0.2}_{-0.7}$&-1.8$^{+0.3}_{-0.7}$&  5.7&45.7&17.3&17.5&15.5&-0.4&15.0&5.7&46.6$^{+0.2}_{-0.5}$&42.4&1.1&4.5&1.9\\
4C +51.37&FSRQ&  1.4&-12.5$^{+0.0}_{-0.3}$&-2.7$^{+0.0}_{-0.3}$&  6.9&46.3&17.6&17.9&17.1&-0.0&28.6&6.5&43.8$^{+0.0}_{-0.3}$&41.8&2.5&3.5&1.8\\
PKS 2155-152&FSRQ&  0.7&-12.6$^{+0.1}_{-0.3}$&-3.6$^{+0.1}_{-0.3}$&  7.1&45.5&17.3&17.7&16.3&0.0&27.4&6.6&43.9$^{+0.1}_{-0.2}$&41.5&1.5&5.4&2.6\\
PKS 2355-534&FSRQ&  1.0&-12.7$^{+0.2}_{-0.5}$&-2.3$^{+0.2}_{-0.5}$&  6.1&46.2&17.6&17.8&16.8&0.1&23.4&5.9&44.2$^{+0.2}_{-0.5}$&41.6&2.1&5.5&2.5\\
1H 1914-194&HBL&  0.1&-12.7$^{+0.4}_{-1.2}$&-2.0$^{+0.4}_{-1.1}$&  5.8&43.9&16.4&-&16.9&-1.3&8.8&5.9&47.4$^{+0.2}_{-0.6}$&42.4&1.6&4.4&1.4\\
TXS 1308+554&FSRQ&  0.9&-12.7$^{+0.4}_{-0.5}$&-1.8$^{+0.4}_{-0.5}$&  5.5&45.5&17.3&17.5&15.0&-0.0&18.3&5.5&46.5$^{+0.2}_{-0.2}$&41.5&2.5&3.6&1.5\\
3C 371&LBL&  0.1&-12.8$^{+0.2}_{-0.2}$&-2.2$^{+0.2}_{-0.2}$&  5.9&42.9&16.0&-&17.0&-1.4&9.2&6.0&47.0$^{+0.1}_{-0.1}$&42.3&2.5&4.1&1.7\\
B2 2113+29&FSRQ&  1.5&-12.9$^{+0.1}_{-0.3}$&-2.8$^{+0.1}_{-0.3}$&  6.4&46.0&17.5&17.7&16.5&-0.3&23.1&6.4&44.4$^{+0.1}_{-0.3}$&42.1&1.5&3.8&1.7\\
OJ 451&FSRQ&  0.2&-12.9$^{+0.2}_{-1.0}$&-2.8$^{+0.2}_{-1.0}$&  6.4&43.7&16.3&16.5&16.6&0.3&25.6&6.0&42.4$^{+0.2}_{-1.0}$&40.0&1.9&3.4&1.5\\
PKS 0735+17&IBL&  0.4&-12.9$^{+0.2}_{-0.3}$&-2.7$^{+0.3}_{-0.4}$&  6.3&45.1&17.0&-&17.4&-1.1&5.8&6.6&46.8$^{+0.2}_{-0.3}$&43.5&2.4&4.6&1.2\\
PKS 1451-375&FSRQ&  0.3&-12.9$^{+0.1}_{-0.3}$&-2.7$^{+0.1}_{-0.3}$&  6.2&44.1&16.5&16.7&16.9&0.0&14.8&6.0&44.6$^{+0.1}_{-0.3}$&41.3&1.6&3.6&1.7\\
PKS 0118-272&HBL&  0.6&-13.0$^{+0.1}_{-0.4}$&-2.1$^{+0.1}_{-0.4}$&  5.5&45.1&17.1&-&17.5&-1.3&10.5&5.6&48.0$^{+0.1}_{-0.2}$&43.0&2.8&4.8&1.8\\
B2 1348+30B&FSRQ&  0.7&-13.0$^{+0.2}_{-0.7}$&-2.4$^{+0.2}_{-0.7}$&  6.0&45.8&17.4&17.9&15.4&-0.2&21.6&5.9&45.6$^{+0.1}_{-0.4}$&41.7&2.6&3.5&2.1\\
PKS 0347-211&FSRQ&  2.9&-13.0$^{+0.2}_{-0.6}$&-3.5$^{+0.2}_{-0.6}$&  6.8&46.9&17.9&18.4&17.3&-0.6&26.2&6.6&44.1$^{+0.2}_{-0.6}$&42.3&3.4&3.7&1.8\\
PMN J1903-6749&FSRQ&  0.2&-13.1$^{+0.2}_{-0.9}$&-2.7$^{+0.2}_{-0.9}$&  6.1&43.9&16.4&16.6&17.1&-0.7&7.2&6.1&45.3$^{+0.2}_{-0.9}$&42.2&2.4&3.6&1.1\\
OI 280&FSRQ&  0.9&-13.2$^{+0.1}_{-0.2}$&-3.7$^{+0.1}_{-0.2}$&  7.0&46.4&17.7&18.2&16.9&-0.1&16.2&6.6&45.0$^{+0.1}_{-0.2}$&42.4&2.3&4.8&2.7\\
PKS 0208-512&FSRQ&  1.0&-13.3$^{+0.0}_{-0.2}$&-3.4$^{+0.0}_{-0.2}$&  6.7&46.3&17.6&17.9&17.3&-0.3&19.2&6.4&44.3$^{+0.0}_{-0.2}$&42.5&2.3&3.9&2.1\\
4C +01.02&FSRQ&  2.1&-13.3$^{+0.0}_{-0.8}$&-3.8$^{+0.0}_{-0.8}$&  6.8&46.9&18.0&18.5&17.3&-0.6&21.4&6.5&44.5$^{+0.0}_{-0.8}$&43.4&2.3&4.2&2.3\\
OK 630&FSRQ&  1.4&-13.6$^{+0.1}_{-0.5}$&-4.2$^{+0.1}_{-0.5}$&  6.9&46.0&17.5&18.0&16.4&-0.2&27.9&6.5&44.0$^{+0.1}_{-0.5}$&42.5&1.6&4.3&2.7\\
PKS 0502+049&FSRQ&  0.9&-13.7$^{+0.0}_{-0.9}$&-4.3$^{+0.0}_{-1.0}$&  6.9&45.9&17.4&18.0&16.8&-0.3&29.8&6.2&44.0$^{+0.0}_{-0.9}$&41.7&1.9&3.9&2.1\\
OI 275&FSRQ&  0.4&-13.7$^{+0.0}_{-0.8}$&-3.7$^{+0.0}_{-0.8}$&  6.4&44.0&16.5&16.8&17.1&-0.2&16.0&6.2&44.1$^{+0.0}_{-0.8}$&41.6&2.1&4.3&2.7\\
PKS 1329-049&FSRQ&  2.1&-13.7$^{+0.0}_{-0.7}$&-3.7$^{+0.0}_{-0.7}$&  6.7&46.8&17.9&18.3&16.8&-0.4&25.0&6.3&44.9$^{+0.0}_{-0.6}$&43.1&1.1&3.5&1.7\\
B2 1846+32A&FSRQ&  0.8&-13.9$^{+0.1}_{-0.1}$&-4.4$^{+0.1}_{-0.1}$&  6.9&45.6&17.3&17.6&16.5&-0.3&25.9&6.3&43.8$^{+0.1}_{-0.1}$&41.8&2.3&4.8&2.8\\
PKS 1335-127&FSRQ&  0.5&-13.9$^{+0.1}_{-0.2}$&-3.2$^{+0.1}_{-0.2}$&  5.8&45.5&17.2&17.7&16.4&-0.2&20.8&5.7&45.0$^{+0.1}_{-0.2}$&42.4&1.5&4.4&2.7\\
PKS 0736+01&FSRQ&  0.2&-13.9$^{+0.0}_{-0.1}$&-4.2$^{+0.0}_{-0.1}$&  7.0&45.0&17.0&17.3&16.7&-0.1&17.6&6.3&43.8$^{+0.0}_{-0.1}$&41.3&1.3&3.9&2.0\\
3C 345&FSRQ&  0.6&-14.0$^{+0.1}_{-0.1}$&-4.6$^{+0.1}_{-0.1}$&  6.9&45.8&17.4&17.8&17.0&0.0&25.7&6.2&43.7$^{+0.1}_{-0.1}$&41.6&2.1&4.8&2.6\\
OG 050&FSRQ&  1.2&-14.0$^{+0.1}_{-0.1}$&-4.5$^{+0.1}_{-0.1}$&  6.9&46.6&17.8&18.4&17.3&-0.4&21.9&6.4&44.1$^{+0.1}_{-0.1}$&42.5&1.6&4.5&2.1\\
PKS 0252-549&FSRQ&  0.5&-14.2$^{+0.2}_{-0.4}$&-4.8$^{+0.2}_{-0.4}$&  7.0&45.8&17.4&18.0&17.3&0.1&14.2&6.5&44.2$^{+0.2}_{-0.3}$&41.7&1.5&4.7&2.0\\
PKS 0035-252&FSRQ&  1.2&-14.4$^{+0.2}_{-0.3}$&-3.6$^{+0.2}_{-0.3}$&  5.5&44.9&17.0&17.2&17.3&0.0&20.1&5.3&44.7$^{+0.2}_{-0.2}$&41.9&1.0&4.1&1.8\\
4C +05.64&FSRQ&  1.4&-14.4$^{+0.2}_{-0.5}$&-4.3$^{+0.2}_{-0.6}$&  6.7&46.3&17.6&18.2&16.7&-0.2&21.7&6.3&44.7$^{+0.2}_{-0.4}$&42.4&1.3&3.8&2.1\\
PKS 1034-293&FSRQ&  0.3&-14.5$^{+0.2}_{-0.2}$&-3.8$^{+0.2}_{-0.2}$&  5.8&43.8&16.4&16.9&16.7&-0.2&11.8&5.9&45.2$^{+0.1}_{-0.1}$&41.8&2.2&4.6&2.7\\
PKS 2255-282&FSRQ&  0.9&-14.5$^{+0.2}_{-1.9}$&-3.9$^{+0.2}_{-1.9}$&  6.0&46.5&17.7&18.2&17.2&-0.4&11.3&6.2&45.5$^{+0.1}_{-1.5}$&43.2&1.8&4.0&1.9\\
OP 313&FSRQ&  1.0&-14.7$^{+0.0}_{-0.7}$&-4.4$^{+0.1}_{-0.8}$&  6.7&45.7&17.3&17.8&16.4&-0.0&31.5&6.0&44.2$^{+0.0}_{-0.6}$&41.9&1.9&4.0&2.4\\
PKS 0135-247&FSRQ&  0.8&-14.9$^{+0.0}_{-0.2}$&-5.1$^{+0.0}_{-0.2}$&  6.8&46.4&17.7&18.3&17.3&-0.4&23.9&6.1&43.8$^{+0.0}_{-0.2}$&42.1&1.6&4.1&2.7\\
PKS 0454-46&FSRQ&  0.9&-15.0$^{+0.4}_{-0.1}$&-4.7$^{+0.5}_{-0.2}$&  6.7&46.3&17.6&18.2&17.3&-0.3&12.5&6.4&44.9$^{+0.2}_{-0.1}$&42.7&2.0&4.1&1.9\\
Ton 599&FSRQ&  0.7&-15.0$^{+0.1}_{-0.2}$&-5.5$^{+0.1}_{-0.2}$&  6.9&45.6&17.3&17.9&17.0&-0.3&27.6&6.1&43.4$^{+0.1}_{-0.2}$&41.7&2.2&4.5&2.3\\
\bottomrule
\end{tabular}
}
\end{table*}

\begin{table*}
  \centering
\resizebox{\textwidth}{!}{
  \begin{tabular}{llllllllllllllllll}
  
  \toprule
Source & Class & $z$ & $F_{\nu_\mu}$ & $N_{\nu_\mu}$ & $E_\nu^\mathrm{peak}$ & $L_\mathrm{disk}$ & $R_\mathrm{BLR}$ & $R_\mathrm{diss}$ & $R_\mathrm{b}^\prime$ & $B^\prime$ & $\Gamma$ & $\gamma_\mathrm{p}^\mathrm{max}$ & $L_\mathrm{p}^\prime$ & $L_\mathrm{e}^\prime$ & $\gamma_\mathrm{e}^\mathrm{min}$ & $\gamma_\mathrm{e}^\mathrm{max}$ & $\alpha_\mathrm{e}$ \\
\midrule
PKS 0400-319&FSRQ&  1.3&-15.0$^{+0.3}_{-0.9}$&-5.5$^{+0.3}_{-1.0}$&  6.8&45.0&17.0&17.4&17.0&-0.1&26.4&6.2&43.3$^{+0.2}_{-0.9}$&41.5&2.2&5.2&2.7\\
PKS 0514-459&FSRQ&  0.2&-15.0$^{+0.2}_{-0.3}$&-4.7$^{+0.2}_{-0.3}$&  6.1&44.6&16.8&17.1&17.1&-0.2&13.1&5.9&44.0$^{+0.2}_{-0.3}$&40.9&2.3&4.7&2.6\\
PKS 0048-071&FSRQ&  2.0&-15.1$^{+0.2}_{-0.4}$&-4.7$^{+0.2}_{-0.5}$&  6.5&46.0&17.5&17.9&17.3&-0.7&14.3&6.4&45.0$^{+0.2}_{-0.4}$&43.5&2.2&4.8&2.5\\
PKS 0002-478&FSRQ&  0.9&-15.3$^{+0.2}_{-0.9}$&-4.3$^{+0.2}_{-0.9}$&  5.0&45.1&17.1&17.3&16.5&0.0&19.6&5.0&45.6$^{+0.2}_{-0.7}$&41.7&1.7&4.0&1.8\\
S4 1726+45&FSRQ&  0.7&-15.4$^{+0.2}_{-0.3}$&-5.0$^{+0.2}_{-0.3}$&  6.3&45.9&17.4&18.0&17.0&-0.5&12.8&6.3&44.7$^{+0.1}_{-0.2}$&42.4&2.4&4.0&1.8\\
GB6 J1604+5714&FSRQ&  0.7&-15.4$^{+0.2}_{-0.4}$&-5.6$^{+0.2}_{-0.5}$&  6.8&45.9&17.4&18.0&17.2&-0.3&17.9&6.2&43.9$^{+0.2}_{-0.4}$&41.7&1.1&3.8&1.5\\
GB6 J0937+5008&FSRQ&  0.3&-15.7$^{+0.0}_{-0.4}$&-6.3$^{+0.0}_{-0.4}$&  6.9&44.0&16.5&17.0&17.3&0.0&26.2&6.0&42.3$^{+0.0}_{-0.4}$&40.1&2.0&4.0&2.4\\
PKS 0336-01&FSRQ&  0.8&-16.4$^{+0.1}_{-0.1}$&-6.0$^{+0.1}_{-0.1}$&  6.7&46.2&17.6&18.1&17.3&-0.3&26.8&5.9&43.3$^{+0.1}_{-0.1}$&41.7&1.3&4.1&2.1\\
PKS 0805-07&FSRQ&  1.8&-11.3$^{+0.0}_{-11.3}$&-0.2$^{+0.0}_{-0.2}$&  4.6&46.6&17.8&18.1&16.4&-0.6&13.2&5.0&48.6$^{+0.0}_{-48.6}$&43.9&3.1&4.1&1.6\\
PKS 2326-502&FSRQ&  0.5&-11.3$^{+0.0}_{-11.3}$&-1.0$^{+0.0}_{-1.0}$&  6.2&45.3&17.2&17.2&16.3&-0.2&19.7&5.9&44.2$^{+0.0}_{-44.2}$&41.9&1.7&3.3&1.9\\
PKS 2227-08&FSRQ&  1.6&-11.4$^{+0.0}_{-11.4}$&-1.1$^{+0.0}_{-1.1}$&  6.1&46.5&17.7&17.8&16.0&0.4&30.3&6.2&44.1$^{+0.0}_{-44.1}$&43.0&1.1&4.0&3.0\\
PKS 0605-08&FSRQ&  0.9&-11.4$^{+0.0}_{-11.4}$&-0.8$^{+0.0}_{-0.8}$&  5.8&46.7&17.8&17.9&17.2&0.2&22.5&5.6&43.7$^{+0.0}_{-43.7}$&41.9&1.4&6.4&2.8\\
PKS 0537-286&FSRQ&  3.1&-11.5$^{+0.0}_{-11.5}$&-0.4$^{+0.0}_{-0.4}$&  5.1&46.8&17.9&17.9&16.7&-0.5&22.6&5.2&45.3$^{+0.0}_{-45.3}$&43.5&1.2&3.0&2.6\\
PKS 2345-16&FSRQ&  0.6&-11.6$^{+0.2}_{-11.6}$&-0.7$^{+0.2}_{-0.7}$&  5.4&45.8&17.4&17.4&15.0&0.5&12.6&5.4&46.4$^{+0.2}_{-46.4}$&42.2&2.4&4.9&2.7\\
B2 1015+35B&FSRQ&  1.2&-11.7$^{+0.4}_{-11.7}$&-1.1$^{+0.4}_{-1.1}$&  6.1&46.5&17.7&17.7&15.0&-0.4&3.6&7.0&46.4$^{+0.3}_{-46.4}$&44.7&3.0&3.9&1.7\\
TXS 1700+685&FSRQ&  0.3&-11.7$^{+0.0}_{-11.7}$&-1.4$^{+0.0}_{-1.4}$&  6.2&44.5&16.7&16.8&16.1&-0.1&17.4&5.9&44.3$^{+0.0}_{-44.3}$&41.7&1.7&3.5&2.7\\
OJ 535&FSRQ&  1.4&-11.8$^{+0.2}_{-11.8}$&-1.6$^{+0.2}_{-1.6}$&  6.4&46.3&17.7&17.7&15.7&0.2&7.3&7.0&45.5$^{+0.2}_{-45.5}$&43.6&2.6&3.6&2.0\\
OX 036&FSRQ&  1.9&-12.0$^{+0.1}_{-12.0}$&-1.2$^{+0.1}_{-1.2}$&  5.5&46.9&17.9&18.0&16.0&0.6&14.1&5.8&45.8$^{+0.1}_{-45.8}$&43.2&2.0&4.4&2.8\\
PKS 2355-106&FSRQ&  1.6&-12.0$^{+0.1}_{-12.0}$&-0.9$^{+0.1}_{-0.9}$&  4.8&46.7&17.8&18.0&15.6&-0.5&9.4&5.2&47.4$^{+0.1}_{-47.4}$&44.1&1.6&3.2&2.5\\
S5 1357+76&FSRQ&  1.6&-12.0$^{+0.0}_{-12.0}$&-1.4$^{+0.0}_{-1.4}$&  5.7&45.6&17.3&17.4&16.5&0.3&25.1&5.6&44.0$^{+0.0}_{-44.0}$&41.3&1.9&3.1&1.5\\
PKS 0524-485&FSRQ&  1.3&-12.0$^{+0.0}_{-12.0}$&-2.5$^{+0.0}_{-2.5}$&  7.1&46.1&17.5&17.7&17.3&-0.5&17.3&7.0&44.1$^{+0.0}_{-44.1}$&42.4&2.5&4.0&2.1\\
S4 0805+41&FSRQ&  1.4&-12.1$^{+0.1}_{-12.1}$&-1.2$^{+0.1}_{-1.2}$&  5.4&46.1&17.6&17.8&15.2&0.2&18.5&5.5&46.9$^{+0.1}_{-46.9}$&41.7&1.2&4.1&2.1\\
B2 0202+31&FSRQ&  1.5&-12.1$^{+0.3}_{-12.1}$&-1.4$^{+0.3}_{-1.4}$&  5.6&46.8&17.9&17.9&16.6&0.3&20.5&5.6&43.8$^{+0.2}_{-43.8}$&42.8&1.2&6.0&2.9\\
PKS 2143-156&FSRQ&  0.7&-12.2$^{+0.3}_{-12.2}$&-2.2$^{+0.3}_{-2.2}$&  6.6&46.1&17.6&17.6&16.3&0.2&9.1&7.0&43.9$^{+0.3}_{-43.9}$&43.7&1.1&4.2&2.8\\
PKS 0047-579&FSRQ&  1.8&-12.2$^{+0.2}_{-12.2}$&-3.1$^{+0.2}_{-3.1}$&  7.1&46.9&17.9&18.4&17.3&-0.1&26.3&6.9&43.5$^{+0.2}_{-43.5}$&42.0&1.0&3.7&2.0\\
PKS 0451-28&FSRQ&  2.6&-12.2$^{+0.0}_{-12.2}$&-1.4$^{+0.0}_{-1.4}$&  5.5&47.1&18.0&18.2&17.3&-0.3&19.4&5.5&45.6$^{+0.0}_{-45.6}$&43.4&2.1&6.4&2.8\\
OL 318&FSRQ&  1.4&-12.3$^{+0.0}_{-12.3}$&-1.5$^{+0.0}_{-1.5}$&  5.6&46.0&17.5&17.5&16.1&0.5&19.9&5.5&44.5$^{+0.0}_{-44.5}$&42.4&1.9&3.1&2.7\\
PKS 1903-80&FSRQ&  1.8&-12.4$^{+0.5}_{-12.4}$&-2.5$^{+0.5}_{-2.5}$&  6.9&46.0&17.5&17.7&16.1&-0.4&23.1&6.6&45.1$^{+0.5}_{-45.1}$&42.5&2.7&4.2&2.8\\
PMN J1303-4621&FSRQ&  1.7&-12.4$^{+0.2}_{-12.4}$&-2.1$^{+0.2}_{-2.1}$&  6.2&45.2&17.1&17.2&15.6&0.4&29.7&6.2&44.0$^{+0.2}_{-44.0}$&41.1&2.7&3.5&2.7\\
B2 1732+38A&FSRQ&  1.0&-12.4$^{+0.2}_{-12.4}$&-3.5$^{+0.2}_{-3.5}$&  7.2&45.0&17.0&17.6&17.0&-0.2&26.5&6.8&43.1$^{+0.2}_{-43.1}$&41.8&2.4&8.3&2.9\\
B2 1324+22&FSRQ&  1.4&-12.5$^{+0.5}_{-12.5}$&-1.5$^{+0.5}_{-1.5}$&  5.2&46.3&17.6&17.8&15.1&-0.1&28.2&5.1&46.2$^{+0.5}_{-46.2}$&42.3&1.2&3.6&1.9\\
PKS 0102-245&FSRQ&  1.8&-12.5$^{+0.2}_{-12.5}$&-2.2$^{+0.2}_{-2.2}$&  6.2&46.3&17.6&17.8&17.0&-0.3&15.3&6.5&44.3$^{+0.2}_{-44.3}$&43.1&1.8&4.4&2.6\\
MG2 J174803+3403&FSRQ&  2.8&-12.5$^{+0.4}_{-12.5}$&-3.3$^{+0.4}_{-3.3}$&  7.1&46.6&17.8&18.1&15.9&-0.1&28.5&7.0&44.4$^{+0.4}_{-44.4}$&42.0&2.5&4.2&2.3\\
TXS 0907+230&FSRQ&  2.7&-12.5$^{+0.1}_{-12.5}$&-2.4$^{+0.2}_{-2.4}$&  6.7&46.2&17.6&18.2&15.5&-0.5&18.7&6.8&46.3$^{+0.1}_{-46.3}$&42.9&3.0&4.1&1.9\\
PKS 1725+044&FSRQ&  0.3&-12.5$^{+0.2}_{-12.5}$&-1.7$^{+0.2}_{-1.7}$&  5.6&45.2&17.1&17.1&16.1&0.4&6.3&5.6&45.7$^{+0.2}_{-45.7}$&42.6&2.4&4.0&2.5\\
PMN J0113-3551&FSRQ&  1.2&-12.7$^{+0.2}_{-12.7}$&-3.2$^{+0.2}_{-3.2}$&  7.0&45.8&17.4&17.6&16.7&-0.1&19.9&6.7&44.4$^{+0.2}_{-44.4}$&41.6&2.6&4.1&2.5\\
PKS 2335-027&FSRQ&  1.1&-12.8$^{+0.1}_{-12.8}$&-2.6$^{+0.1}_{-2.6}$&  6.4&46.0&17.5&17.7&16.3&-0.3&16.4&6.4&44.7$^{+0.1}_{-44.7}$&42.6&2.4&3.3&1.5\\
87GB 122531.6+494958&FSRQ&  1.4&-12.8$^{+0.2}_{-12.8}$&-3.0$^{+0.2}_{-3.0}$&  6.8&45.0&17.0&17.0&15.6&0.7&28.6&7.0&42.0$^{+0.2}_{-42.0}$&40.4&2.6&3.0&2.8\\
PKS B1413+135&BCU&  0.2&-12.9$^{+0.0}_{-12.9}$&-3.4$^{+0.0}_{-3.4}$&  7.3&43.6&16.3&-&16.9&-1.0&10.2&7.5&46.6$^{+0.0}_{-46.6}$&42.3&2.9&3.4&2.2\\
5C 12.291&FSRQ&  1.1&-12.9$^{+0.2}_{-12.9}$&-2.1$^{+0.2}_{-2.1}$&  5.4&44.5&16.8&16.9&15.9&-0.2&19.1&5.3&46.0$^{+0.2}_{-46.0}$&42.0&1.7&3.2&1.2\\
PKS B1434+235&FSRQ&  1.5&-13.0$^{+0.2}_{-13.0}$&-2.9$^{+0.3}_{-2.9}$&  6.7&46.1&17.6&18.1&15.3&0.0&23.5&6.6&45.4$^{+0.2}_{-45.4}$&42.0&1.3&3.7&1.7\\
PMN J0257-1211&FSRQ&  1.4&-13.0$^{+0.3}_{-13.0}$&-1.9$^{+0.3}_{-1.9}$&  5.1&46.1&17.6&17.6&15.0&0.3&18.5&5.0&46.1$^{+0.2}_{-46.1}$&42.1&2.8&5.3&2.7\\
PKS 0346-27&FSRQ&  1.0&-13.2$^{+0.1}_{-13.2}$&-4.1$^{+0.1}_{-4.1}$&  7.0&45.7&17.3&17.9&17.2&-0.1&21.5&6.6&43.4$^{+0.1}_{-43.4}$&41.6&1.8&4.7&2.2\\
PKS 0334-131&FSRQ&  1.3&-13.4$^{+0.1}_{-13.4}$&-3.3$^{+0.1}_{-3.3}$&  6.5&45.8&17.4&17.6&17.3&0.0&30.1&6.2&42.9$^{+0.1}_{-42.9}$&41.1&1.1&3.3&1.7\\
B3 0307+380&FSRQ&  0.9&-13.7$^{+0.6}_{-13.7}$&-3.3$^{+0.6}_{-3.3}$&  6.2&44.8&16.9&17.3&16.7&-0.4&10.4&6.5&45.7$^{+0.4}_{-45.7}$&43.1&2.4&4.5&2.4\\
MRC 1659-621&FSRQ&  1.8&-13.7$^{+0.3}_{-13.7}$&-4.1$^{+0.3}_{-4.1}$&  6.8&46.8&17.9&18.2&17.2&-0.2&22.7&6.4&44.3$^{+0.2}_{-44.3}$&42.6&2.5&4.2&2.2\\
TXS 0919-052&FSRQ&  1.0&-14.2$^{+0.3}_{-14.2}$&-3.7$^{+0.3}_{-3.7}$&  5.9&45.0&17.0&17.2&17.1&0.1&29.1&5.7&42.5$^{+0.2}_{-42.5}$&40.8&2.8&5.2&2.8\\
B3 0020+446&FSRQ&  1.1&-14.4$^{+0.3}_{-14.4}$&-3.3$^{+0.3}_{-3.3}$&  4.9&45.0&17.0&17.5&16.5&-0.5&14.3&5.0&46.9$^{+0.2}_{-46.9}$&42.4&2.7&4.1&2.3\\
OD 166&FSRQ&  2.7&-14.4$^{+0.4}_{-14.4}$&-3.6$^{+0.4}_{-3.6}$&  5.7&46.0&17.5&17.7&16.9&-0.3&20.7&5.6&45.0$^{+0.4}_{-45.0}$&42.9&2.3&4.9&2.7\\
PMN J0134-3843&FSRQ&  2.1&-15.0$^{+0.6}_{-15.0}$&-4.1$^{+0.6}_{-4.1}$&  5.5&47.3&18.1&18.7&17.0&-0.1&15.2&5.7&45.6$^{+0.3}_{-45.6}$&42.9&1.6&5.1&2.3\\
B2 1040+24A&FSRQ&  0.6&-15.8$^{+0.5}_{-15.8}$&-5.1$^{+0.5}_{-5.1}$&  5.9&44.6&16.8&17.4&16.8&-0.2&14.6&5.9&44.4$^{+0.3}_{-44.4}$&42.0&2.4&5.6&2.6\\
MG1 J221916+1806&FSRQ&  1.1&-16.0$^{+4.0}_{-16.0}$&-4.9$^{+4.0}_{-4.9}$&  4.2&45.9&17.4&17.6&15.0&-0.5&3.4&5.0&43.6$^{+3.9}_{-43.6}$&44.7&1.1&5.5&1.3\\
Mkn 501&HBL&  0.0&<-10.8&<0.1&  5.8&43.2&16.1&-&16.4&-1.3&5.8&6.0&<46.7&42.2&1.8&6.9&2.1\\
1ES 1959+650&HBL&  0.1&<-11.1&<0.5&  5.9&43.3&16.1&-&16.8&-1.4&8.2&6.0&<46.7&42.0&2.8&6.5&2.0\\
PKS 2155-304&HBL&  0.1&<-11.2&<0.4&  5.6&44.5&16.7&-&17.3&-0.9&4.0&6.0&<47.2&43.5&2.8&5.7&1.4\\
3C 273&FSRQ&  0.2&<-11.4&<0.6&  5.7&46.3&17.7&17.7&16.2&0.6&4.6&6.0&<45.7&44.1&2.8&5.1&2.6\\
PKS J1829-5813&FSRQ&  1.5&<-11.4&<0.6&  5.6&46.6&17.8&17.8&15.2&-0.1&8.8&6.0&<46.5&44.0&2.7&4.2&1.9\\
PKS J2129-1538&FSRQ&  3.3&<-11.4&<0.6&  5.5&47.8&18.4&18.4&17.4&-0.9&11.1&6.0&<44.7&44.2&1.2&3.9&2.7\\
PKS 0925-203&FSRQ&  0.3&<-11.5&<1.1&  6.1&45.9&17.4&17.4&16.1&0.5&22.5&6.0&<42.0&40.4&1.8&7.0&1.5\\
TXS 0616-116&BCU&  1.0&<-11.5&<0.5&  5.2&45.1&17.1&-&14.7&-0.8&3.4&6.0&<47.2&44.5&2.3&5.3&1.9\\
PMN J0017-0512&FSRQ&  0.2&<-11.5&<0.7&  5.5&44.8&16.9&16.9&14.3&0.7&4.3&6.0&<45.6&43.0&1.3&4.5&2.2\\
PKS 2144+092&FSRQ&  1.1&<-11.6&<0.6&  5.3&46.2&17.6&18.0&15.2&-0.7&4.3&6.0&<46.8&45.0&1.2&6.1&1.9\\
4C +38.41&FSRQ&  1.8&<-11.6&<0.9&  5.9&47.0&18.0&18.0&15.1&-0.3&15.2&6.0&<45.7&43.6&2.3&3.7&1.7\\
\bottomrule
\end{tabular}
}
\end{table*}

\begin{table*}
  \centering
\resizebox{\textwidth}{!}{
  \begin{tabular}{llllllllllllllllll}
  
  \toprule
Source & Class & $z$ & $F_{\nu_\mu}$ & $N_{\nu_\mu}$ & $E_\nu^\mathrm{peak}$ & $L_\mathrm{disk}$ & $R_\mathrm{BLR}$ & $R_\mathrm{diss}$ & $R_\mathrm{b}^\prime$ & $B^\prime$ & $\Gamma$ & $\gamma_\mathrm{p}^\mathrm{max}$ & $L_\mathrm{p}^\prime$ & $L_\mathrm{e}^\prime$ & $\gamma_\mathrm{e}^\mathrm{min}$ & $\gamma_\mathrm{e}^\mathrm{max}$ & $\alpha_\mathrm{e}$ \\
\midrule
GB6 J0225+1846&FSRQ&  2.7&<-11.6&<0.8&  5.6&46.9&18.0&17.9&17.0&-0.0&12.3&6.0&<44.7&44.3&1.1&6.8&2.7\\
PG 1553+113&HBL&  0.4&<-11.6&<0.9&  5.7&44.6&16.8&-&17.3&-1.4&6.0&6.0&<47.2&43.7&2.8&5.8&1.2\\
4C +40.24&FSRQ&  1.2&<-11.7&<0.7&  5.1&46.5&17.7&17.7&15.0&-0.2&3.3&6.0&<46.6&45.0&4.0&6.7&1.4\\
OX 169&FSRQ&  0.2&<-11.7&<1.2&  6.1&45.6&17.3&17.3&15.1&0.5&7.8&6.0&<45.0&42.4&2.9&5.4&2.7\\
3C 279&FSRQ&  0.5&<-11.8&<1.4&  6.3&45.3&17.2&17.3&15.0&-0.2&27.8&6.0&<45.0&42.3&2.4&5.6&2.8\\
GB6 B0642+4454&FSRQ&  3.4&<-11.8&<0.9&  5.5&47.1&18.1&18.0&16.1&0.3&10.2&6.0&<46.1&44.1&1.8&3.0&1.2\\
PKS 1730-13&FSRQ&  0.9&<-11.8&<1.0&  5.8&46.3&17.6&17.9&15.5&-0.8&15.2&6.0&<46.5&43.3&1.4&3.6&1.3\\
4C +49.22&FSRQ&  1.1&<-11.8&<0.8&  5.3&46.6&17.8&17.8&15.1&-0.6&5.0&6.0&<47.0&44.2&2.3&5.8&2.0\\
PKS 1933-400&FSRQ&  1.0&<-11.8&<1.0&  5.7&46.3&17.6&17.6&15.0&0.3&10.4&6.0&<45.8&43.2&2.8&5.4&2.7\\
PKS 2149-306&FSRQ&  2.4&<-11.8&<1.1&  5.8&47.1&18.0&18.1&17.0&-0.3&17.3&6.0&<44.0&44.4&1.2&3.8&2.8\\
S5 0633+73&FSRQ&  1.9&<-11.8&<0.9&  5.3&46.9&18.0&18.0&15.5&-0.2&6.2&6.0&<47.0&44.2&1.7&3.9&1.3\\
PKS 0528+134&FSRQ&  2.1&<-11.8&<0.9&  5.4&47.3&18.1&18.1&15.0&-0.3&8.1&6.0&<46.1&44.6&2.9&4.2&1.6\\
I Zw 187&HBL&  0.1&<-11.8&<1.3&  6.0&42.8&16.0&-&15.7&-0.2&9.1&6.0&<45.4&41.2&2.9&6.8&2.0\\
PKS 0637-75&FSRQ&  0.7&<-11.9&<1.1&  5.6&46.9&17.9&18.0&16.7&0.2&5.3&6.0&<46.6&44.0&2.1&3.4&1.3\\
PKS 1346-112&FSRQ&  0.3&<-11.9&<1.1&  5.6&45.5&17.2&17.3&15.3&0.2&5.6&6.0&<46.1&43.1&2.5&6.1&2.5\\
4C +14.23&FSRQ&  1.0&<-12.0&<1.3&  5.8&46.1&17.5&17.6&15.6&-0.7&9.7&6.0&<46.5&43.5&2.6&5.7&2.1\\
PKS 0646-306&FSRQ&  1.1&<-12.0&<1.1&  5.5&46.2&17.6&17.6&15.1&0.5&6.0&6.0&<46.2&43.9&2.9&7.0&2.2\\
PKS 0754+100&LBL&  0.3&<-12.0&<1.3&  5.8&44.1&16.6&-&16.7&-0.7&7.9&6.0&<47.2&43.1&2.7&6.4&2.8\\
OQ 253&FSRQ&  1.4&<-12.0&<1.1&  5.3&46.3&17.6&17.7&15.0&-0.9&4.9&6.0&<47.0&44.4&2.1&6.0&2.0\\
PKS 0202-17&FSRQ&  1.7&<-12.0&<1.2&  5.6&47.0&18.0&18.0&15.0&0.6&11.8&6.0&<45.7&43.5&2.6&4.1&1.9\\
OS 562&FSRQ&  0.8&<-12.1&<1.6&  6.0&46.5&17.8&17.8&15.0&0.7&14.5&6.0&<44.8&42.5&1.5&5.3&2.8\\
PKS 2204-54&FSRQ&  1.2&<-12.1&<1.4&  5.8&46.5&17.7&17.8&15.2&0.3&13.0&6.0&<45.7&42.9&2.7&7.4&2.7\\
PKS 2145+06&FSRQ&  1.0&<-12.1&<1.2&  5.3&47.0&18.0&18.0&16.3&0.6&4.0&6.0&<46.6&44.8&2.9&4.4&2.2\\
ON 246&HBL&  0.1&<-12.1&<1.6&  6.0&43.9&16.4&-&15.4&0.5&9.5&6.0&<45.4&41.8&2.7&5.5&2.4\\
PKS B1339-206&FSRQ&  1.6&<-12.1&<1.3&  5.5&46.5&17.7&17.8&15.1&0.3&10.0&6.0&<46.0&43.5&2.7&3.7&1.4\\
PKS 2326-477&FSRQ&  1.3&<-12.2&<1.2&  5.3&46.8&17.9&17.9&16.0&0.3&4.4&6.0&<46.9&44.5&2.7&5.4&2.0\\
Mkn 180&HBL&  0.0&<-12.2&<1.5&  5.9&42.5&15.8&-&16.8&-1.3&6.2&6.0&<46.3&41.9&3.0&6.8&2.3\\
4C +40.25&FSRQ&  1.2&<-12.2&<1.3&  5.6&46.2&17.6&17.6&15.2&0.4&8.5&6.0&<45.9&43.6&2.8&6.2&2.6\\
PKS 0437-454&LBL&  2.0&<-12.2&<1.5&  5.8&46.0&17.5&-&15.4&0.1&21.2&6.0&<46.3&42.6&2.7&5.5&2.5\\
PKS 2052-47&FSRQ&  1.5&<-12.2&<1.5&  5.8&46.6&17.8&18.3&15.2&0.2&20.7&6.0&<45.6&42.9&1.9&3.9&1.3\\
4C +39.23&FSRQ&  1.2&<-12.2&<1.5&  5.9&46.1&17.6&17.6&15.3&0.2&11.7&6.0&<45.7&42.9&1.1&3.7&1.5\\
PKS 1610-77&FSRQ&  1.7&<-12.2&<1.5&  5.9&46.7&17.8&17.9&15.0&0.4&18.6&6.0&<45.2&42.9&2.2&3.8&1.7\\
PKS 1016-311&FSRQ&  0.8&<-12.2&<1.4&  5.5&46.0&17.5&17.5&15.9&0.4&5.2&6.0&<45.9&43.6&2.9&6.1&2.6\\
S5 1044+71&FSRQ&  1.1&<-12.2&<1.6&  5.9&46.7&17.8&17.9&15.2&-0.1&15.7&6.0&<45.7&43.0&2.4&4.3&1.8\\
S4 0954+65&IBL&  0.4&<-12.2&<1.4&  5.7&44.8&16.9&-&15.8&-0.2&7.3&6.0&<46.1&43.1&2.5&6.8&2.5\\
S4 0003+38&FSRQ&  0.2&<-12.2&<1.3&  5.5&43.8&16.4&16.5&15.3&-0.3&4.8&6.0&<46.1&43.1&2.4&4.0&1.8\\
OJ 508&FSRQ&  1.4&<-12.2&<1.3&  5.3&46.4&17.7&17.8&15.0&0.8&4.9&6.0&<45.9&44.4&3.0&5.9&2.2\\
OK 290&FSRQ&  0.7&<-12.2&<1.6&  5.8&45.9&17.5&17.5&16.3&0.0&7.9&6.0&<45.6&43.2&2.5&4.5&2.7\\
4C +04.42&FSRQ&  1.0&<-12.2&<2.0&  6.3&46.2&17.6&17.7&16.4&0.3&27.6&6.0&<43.3&42.6&1.1&6.4&2.9\\
PKS 1057-79&LBL&  0.6&<-12.3&<1.5&  5.7&45.6&17.3&-&17.5&-1.3&6.1&6.0&<48.0&43.9&2.9&6.0&2.4\\
TXS 0322+222&FSRQ&  2.1&<-12.3&<1.4&  5.6&46.5&17.7&17.9&15.8&-0.3&15.0&6.0&<46.4&43.6&3.0&4.5&2.4\\
PKS 1519-273&IBL&  1.3&<-12.3&<1.7&  6.0&44.5&16.8&-&15.8&-0.9&28.3&6.0&<46.5&42.7&2.9&6.1&3.0\\
S5 0212+73&FSRQ&  2.4&<-12.3&<1.7&  6.0&47.3&18.2&18.3&17.3&-0.2&24.3&6.0&<43.5&43.4&1.5&3.8&2.8\\
PKS 0436-129&FSRQ&  1.3&<-12.3&<1.8&  6.2&46.0&17.5&17.6&15.0&0.2&20.5&6.0&<45.6&42.1&2.2&3.8&1.9\\
PKS B1319-093&FSRQ&  1.9&<-12.3&<1.7&  5.9&46.5&17.7&17.8&16.2&0.1&23.1&6.0&<43.6&43.3&1.2&6.8&2.9\\
4C +25.05&FSRQ&  2.4&<-12.3&<1.4&  5.5&46.9&18.0&18.0&16.1&-0.1&7.7&6.0&<46.1&44.6&1.4&4.4&2.6\\
PKS 0420+022&FSRQ&  2.3&<-12.3&<1.5&  5.8&46.6&17.8&17.8&15.7&-0.5&10.6&6.0&<46.1&43.9&1.8&7.1&2.7\\
4C +01.28&LBL&  0.9&<-12.3&<1.5&  5.6&45.5&17.3&-&17.3&-0.9&7.1&6.0&<47.6&44.2&3.0&5.6&2.3\\
PKS 0332-403&LBL&  1.4&<-12.3&<1.5&  5.7&45.5&17.2&-&17.0&-1.1&14.2&6.0&<47.5&43.7&2.9&5.2&2.4\\
OL 220&FSRQ&  0.6&<-12.3&<1.8&  6.0&45.8&17.4&17.5&15.9&0.2&10.7&6.0&<45.4&42.5&2.2&6.2&2.8\\
PKS 2325+093&FSRQ&  1.8&<-12.3&<1.7&  5.9&46.5&17.7&17.8&16.7&-0.3&15.6&6.0&<43.9&43.8&1.0&5.3&2.5\\
PKS 0310+013&FSRQ&  0.7&<-12.3&<2.1&  6.3&45.4&17.2&17.3&15.0&0.0&27.5&6.0&<44.4&41.6&1.9&4.1&2.6\\
PKS 1532+01&FSRQ&  1.4&<-12.3&<1.3&  5.2&46.1&17.6&17.6&17.1&-1.0&3.3&6.0&<48.0&45.2&3.3&3.7&1.2\\
PKS 1717+177&IBL&  0.1&<-12.3&<1.7&  6.0&43.7&16.3&-&15.6&-0.9&11.7&6.0&<46.3&42.6&1.2&4.4&2.2\\
PMN J0127-0821&LBL&  0.4&<-12.3&<1.7&  5.9&44.0&16.5&-&15.5&-0.4&12.5&6.0&<46.5&42.1&2.8&4.7&2.8\\
B3 1417+385&FSRQ&  1.8&<-12.3&<1.6&  5.8&46.4&17.7&18.3&15.3&-0.2&19.6&6.0&<46.4&42.5&2.5&7.5&2.7\\
PMN J0948+0022&NLS1&  0.6&<-12.4&<1.5&  5.6&45.7&17.3&-&15.2&0.5&7.6&6.0&<45.4&43.5&1.7&3.9&1.1\\
Mkn 421&HBL&  0.0&<-12.4&<1.6&  5.7&43.6&16.3&-&16.6&-0.9&5.0&6.0&<44.8&42.3&2.0&6.3&1.6\\
1H 1013+498&HBL&  0.2&<-12.4&<1.8&  5.9&44.6&16.8&-&16.9&-0.7&7.3&6.0&<46.1&42.8&2.0&5.6&1.6\\
PKS 1034-374&FSRQ&  1.8&<-12.4&<1.6&  5.9&46.5&17.7&17.8&15.1&0.6&12.6&6.0&<45.6&42.7&2.5&4.8&2.6\\
PKS 1546+027&FSRQ&  0.4&<-12.4&<1.6&  5.6&45.7&17.3&17.5&16.5&-0.3&4.6&6.0&<46.9&43.8&1.7&4.5&1.7\\
PKS 1004-217&FSRQ&  0.3&<-12.4&<2.0&  6.1&45.4&17.2&17.3&15.2&0.1&12.6&6.0&<45.1&42.0&2.2&4.2&2.2\\
\bottomrule
\end{tabular}
}
\end{table*}

\begin{table*}
  \centering
\resizebox{\textwidth}{!}{
  \begin{tabular}{llllllllllllllllll}
  
  \toprule
Source & Class & $z$ & $F_{\nu_\mu}$ & $N_{\nu_\mu}$ & $E_\nu^\mathrm{peak}$ & $L_\mathrm{disk}$ & $R_\mathrm{BLR}$ & $R_\mathrm{diss}$ & $R_\mathrm{b}^\prime$ & $B^\prime$ & $\Gamma$ & $\gamma_\mathrm{p}^\mathrm{max}$ & $L_\mathrm{p}^\prime$ & $L_\mathrm{e}^\prime$ & $\gamma_\mathrm{e}^\mathrm{min}$ & $\gamma_\mathrm{e}^\mathrm{max}$ & $\alpha_\mathrm{e}$ \\
\midrule
S3 0827+24&FSRQ&  0.9&<-12.4&<2.0&  6.2&46.5&17.7&17.8&16.3&0.2&21.0&6.0&<43.5&42.9&1.1&6.7&2.8\\
PMN J2345-1555&FSRQ&  0.6&<-12.4&<2.1&  6.1&45.4&17.2&17.2&15.2&-0.3&23.6&6.0&<43.2&41.8&2.3&5.8&2.4\\
S5 1803+784&LBL&  0.7&<-12.4&<1.8&  5.9&45.6&17.3&-&16.8&-1.0&14.1&6.0&<47.0&43.2&2.6&4.9&2.4\\
PMN J1959-4246&FSRQ&  2.2&<-12.4&<1.9&  6.0&46.1&17.6&17.6&15.6&0.1&26.3&6.0&<44.5&42.5&2.6&4.0&2.7\\
PKS 0142-278&FSRQ&  1.1&<-12.4&<1.9&  5.9&46.3&17.6&17.7&16.1&-0.0&12.3&6.0&<45.3&43.0&2.2&4.0&2.4\\
PKS 0539-543&FSRQ&  1.2&<-12.4&<1.7&  5.7&46.5&17.7&17.8&15.9&0.2&8.7&6.0&<45.4&43.4&2.2&3.4&1.4\\
OK 492&FSRQ&  1.9&<-12.5&<1.7&  5.8&46.6&17.8&18.0&14.9&0.3&19.4&6.0&<45.9&42.4&2.3&4.2&1.9\\
MG2 J153938+2744&FSRQ&  2.2&<-12.5&<1.9&  6.0&45.6&17.3&17.4&15.1&0.2&27.1&6.0&<44.7&42.1&1.2&3.8&1.6\\
PKS 2312-505&LBL&  0.8&<-12.5&<1.8&  5.8&44.6&16.8&-&16.2&-0.8&15.3&6.0&<47.1&42.8&2.3&4.1&2.1\\
PKS 2233-148&IBL&  0.3&<-12.5&<1.5&  5.4&44.9&16.9&-&16.5&-1.1&3.4&6.0&<47.0&44.3&2.7&6.4&2.2\\
1REX J061757+7816.1&FSRQ&  1.4&<-12.5&<1.8&  5.8&45.5&17.2&17.3&14.1&0.8&13.7&6.0&<45.4&42.1&2.5&6.9&2.3\\
GB6 J0941+2721&FSRQ&  1.3&<-12.5&<1.7&  5.7&45.7&17.4&17.4&15.0&0.4&11.2&6.0&<45.8&42.8&2.8&6.3&2.5\\
PKS 0047-051&FSRQ&  0.9&<-12.5&<2.0&  5.9&45.4&17.2&17.2&16.3&0.1&12.1&6.0&<44.5&42.3&2.3&5.4&2.5\\
S4 1800+44&FSRQ&  0.7&<-12.5&<1.8&  5.6&45.8&17.4&17.4&16.2&0.2&5.3&6.0&<45.5&43.9&1.8&4.1&1.8\\
TXS 0800+618&FSRQ&  3.0&<-12.5&<1.8&  5.8&46.6&17.8&17.8&16.6&-0.6&13.7&6.0&<45.3&44.7&1.5&4.4&2.8\\
GB 1310+487&BCU&  0.6&<-12.6&<1.8&  5.7&44.0&16.5&-&15.4&-0.9&12.9&6.0&<46.1&43.1&3.0&5.1&2.8\\
4C +54.15&LBL&  0.2&<-12.6&<2.0&  6.0&43.6&16.3&-&15.4&-0.2&10.1&6.0&<45.9&42.0&2.7&6.1&2.8\\
S4 1315+34&FSRQ&  1.1&<-12.6&<1.8&  5.7&46.1&17.5&18.0&15.7&0.1&8.8&6.0&<46.4&43.0&2.3&4.9&2.0\\
GB6 J1040+0617&LBL&  1.3&<-12.6&<1.9&  5.8&46.8&17.9&-&15.8&0.1&14.8&6.0&<46.1&43.0&1.3&4.1&1.5\\
4C +56.27&LBL&  0.7&<-12.6&<1.7&  5.6&44.0&16.5&-&17.3&-0.9&5.9&6.0&<47.5&44.1&2.9&5.3&2.3\\
PKS 1958-179&FSRQ&  0.7&<-12.6&<1.9&  5.8&45.0&17.0&17.4&16.1&-0.2&9.0&6.0&<46.4&43.1&1.1&4.5&1.8\\
PMN J0157-4614&FSRQ&  2.3&<-12.6&<2.1&  6.2&45.7&17.4&17.5&15.1&-0.3&29.8&6.0&<45.6&42.2&2.8&4.1&3.0\\
S4 0814+42&LBL&  0.5&<-12.7&<1.9&  5.8&44.7&16.9&-&16.7&-1.3&8.1&6.0&<47.3&43.5&2.3&4.9&2.0\\
PKS 0215+015&FSRQ&  1.7&<-12.7&<1.9&  5.7&46.3&17.6&18.1&16.3&-0.5&14.6&6.0&<46.6&43.4&1.9&4.4&1.9\\
PKS 0834-20&FSRQ&  2.8&<-12.7&<2.0&  5.8&46.8&17.9&17.9&17.3&0.2&23.9&6.0&<42.6&42.9&1.4&4.5&2.8\\
PG 1246+586&HBL&  0.8&<-12.7&<2.1&  5.9&45.7&17.3&-&17.4&-1.2&10.2&6.0&<47.3&43.4&2.1&5.3&1.8\\
PMN J0726-4728&FSRQ&  1.7&<-12.7&<2.1&  5.8&46.0&17.5&17.6&16.3&-0.6&12.6&6.0&<45.9&43.5&2.6&3.9&1.8\\
S4 0110+49&FSRQ&  0.4&<-12.7&<2.3&  6.2&44.3&16.6&16.8&15.0&0.1&21.4&6.0&<45.1&41.6&2.4&5.6&2.9\\
TXS 0404+075&LBL&  1.1&<-12.7&<1.9&  5.4&45.7&17.3&-&16.2&-0.3&8.9&6.0&<46.7&43.6&2.8&4.1&1.4\\
PMN J2206-0031&LBL&  1.1&<-12.8&<2.0&  5.7&45.1&17.1&-&15.8&-0.4&12.8&6.0&<46.5&42.9&2.9&5.4&2.7\\
PMN J0625-5438&FSRQ&  2.0&<-12.8&<2.1&  5.9&46.5&17.7&17.8&16.3&-0.3&14.1&6.0&<44.5&44.0&1.3&4.5&2.9\\
87GB 080551.6+535010&FSRQ&  2.1&<-12.8&<2.0&  5.7&45.3&17.1&17.3&15.6&0.1&12.8&6.0&<46.3&43.1&1.9&4.2&1.5\\
GB6 J0929+5013&LBL&  0.4&<-12.8&<2.2&  6.0&44.0&16.5&-&16.6&-0.9&10.7&6.0&<46.9&42.7&2.7&5.4&2.6\\
OT 081&LBL&  0.3&<-12.8&<2.1&  5.8&44.5&16.8&-&17.1&-1.0&8.3&6.0&<46.8&43.4&2.8&6.9&2.8\\
4C +14.60&LBL&  0.6&<-12.8&<1.9&  5.5&44.3&16.6&-&16.2&0.6&3.9&6.0&<46.4&44.0&1.4&4.1&1.1\\
1H 1720+117&HBL&  0.0&<-12.8&<2.2&  6.0&42.2&15.5&-&15.6&-0.5&7.9&6.0&<44.6&41.1&3.0&6.5&2.5\\
TXS 0730+504&FSRQ&  0.7&<-12.8&<1.9&  5.5&45.5&17.2&17.7&16.3&-0.3&5.1&6.0&<46.8&43.7&1.9&4.5&1.7\\
4C +06.21&FSRQ&  0.4&<-12.8&<2.0&  5.6&45.3&17.2&17.7&16.6&-0.4&5.0&6.0&<46.8&43.3&2.2&5.2&2.2\\
PKS 0823+033&LBL&  0.5&<-12.8&<2.2&  5.8&44.4&16.7&-&16.8&-0.9&11.4&6.0&<46.8&43.1&2.6&5.6&2.9\\
OS 319&FSRQ&  1.4&<-12.8&<2.5&  6.4&46.5&17.7&17.9&16.5&0.0&28.6&6.0&<44.2&42.2&1.0&5.0&2.4\\
PKS 0219-164&FSRQ&  0.7&<-12.9&<2.0&  5.6&45.7&17.3&17.5&16.3&-0.4&6.3&6.0&<46.7&43.5&1.0&4.6&1.9\\
4C +10.45&FSRQ&  1.2&<-12.9&<2.1&  5.7&46.1&17.5&18.0&15.8&-0.2&13.0&6.0&<45.9&43.2&1.9&4.0&1.4\\
W Comae&IBL&  0.1&<-12.9&<2.2&  5.8&43.6&16.3&-&16.7&-0.7&5.5&6.0&<46.4&42.6&2.9&5.3&2.4\\
PKS 0139-09&LBL&  0.7&<-12.9&<2.3&  5.9&45.1&17.1&-&17.5&-1.5&10.9&6.0&<47.4&43.5&2.9&6.9&2.7\\
TXS 1951-115&LBL&  0.7&<-12.9&<2.0&  5.4&45.6&17.3&-&16.3&-0.5&6.1&6.0&<46.8&43.8&2.0&4.1&1.0\\
PMN J1808-5011&FSRQ&  1.6&<-12.9&<2.3&  6.0&45.9&17.4&17.9&15.1&-0.0&24.9&6.0&<45.6&42.0&2.5&6.9&2.8\\
PKS 0516-621&IBL&  2.0&<-12.9&<2.2&  5.7&45.3&17.2&-&17.5&-1.3&12.0&6.0&<47.7&44.0&1.8&5.1&2.0\\
S4 1749+70&IBL&  0.8&<-12.9&<2.4&  6.0&45.6&17.3&-&17.1&-1.3&13.4&6.0&<46.8&43.4&2.8&6.1&2.6\\
MS 1402.3+0416&LBL&  3.2&<-13.0&<2.1&  5.6&46.6&17.8&-&16.2&-0.5&16.7&6.0&<46.7&43.3&2.9&7.1&2.5\\
S4 1716+68&FSRQ&  0.8&<-13.0&<2.6&  6.1&44.7&16.9&17.0&16.5&-0.2&18.1&6.0&<44.7&42.4&1.1&4.5&2.5\\
S5 1039+81&FSRQ&  1.2&<-13.0&<2.1&  5.4&46.3&17.6&18.1&16.7&-0.3&8.3&6.0&<46.6&43.8&2.9&4.5&2.4\\
PMN J2145-3357&FSRQ&  1.4&<-13.0&<2.7&  6.3&45.2&17.1&17.2&15.6&-0.1&29.1&6.0&<44.3&41.7&2.6&5.2&2.9\\
PKS 1005-333&FSRQ&  1.8&<-13.0&<2.5&  5.9&46.0&17.5&17.5&16.2&0.2&25.3&6.0&<42.7&42.4&1.1&4.7&2.4\\
PKS 0235-618&FSRQ&  0.5&<-13.0&<2.4&  5.9&45.3&17.2&17.8&16.0&-0.2&12.1&6.0&<45.6&42.7&2.5&6.2&3.0\\
PKS 0906+01&FSRQ&  1.0&<-13.0&<2.2&  5.5&46.5&17.8&18.2&16.8&-0.5&7.8&6.0&<46.6&43.9&1.4&4.4&1.6\\
OM 235&FSRQ&  1.6&<-13.1&<2.4&  5.8&46.0&17.5&18.0&15.6&-0.2&16.0&6.0&<46.1&42.6&2.7&5.5&2.5\\
PMN J1226-1328&IBL&  0.5&<-13.1&<2.5&  6.0&44.4&16.7&-&16.0&-0.8&10.9&6.0&<46.4&42.6&3.0&5.4&2.6\\
PKS 2240-260&LBL&  0.8&<-13.1&<2.3&  5.8&45.3&17.1&-&17.3&-1.2&8.7&6.0&<47.5&43.5&2.9&5.0&2.2\\
GB6 J0712+5033&IBL&  0.5&<-13.1&<2.4&  6.0&44.6&16.8&-&16.5&-1.0&12.8&6.0&<46.6&42.8&2.9&5.5&2.8\\
S2 0109+22&HBL&  0.3&<-13.1&<2.5&  5.9&44.7&16.9&-&16.3&-0.1&6.8&6.0&<45.6&43.1&1.4&4.6&1.6\\
B2 2214+24B&LBL&  0.5&<-13.1&<2.5&  6.0&44.4&16.7&-&16.7&-0.9&13.0&6.0&<46.6&42.8&2.5&5.3&2.6\\
PKS 0829+046&IBL&  0.2&<-13.1&<2.4&  5.8&43.7&16.3&-&16.8&-0.9&7.1&6.0&<46.7&43.0&1.4&4.4&1.9\\
\bottomrule
\end{tabular}
}
\end{table*}

\begin{table*}
  \centering
\resizebox{\textwidth}{!}{
  \begin{tabular}{llllllllllllllllll}
  
  \toprule
Source & Class & $z$ & $F_{\nu_\mu}$ & $N_{\nu_\mu}$ & $E_\nu^\mathrm{peak}$ & $L_\mathrm{disk}$ & $R_\mathrm{BLR}$ & $R_\mathrm{diss}$ & $R_\mathrm{b}^\prime$ & $B^\prime$ & $\Gamma$ & $\gamma_\mathrm{p}^\mathrm{max}$ & $L_\mathrm{p}^\prime$ & $L_\mathrm{e}^\prime$ & $\gamma_\mathrm{e}^\mathrm{min}$ & $\gamma_\mathrm{e}^\mathrm{max}$ & $\alpha_\mathrm{e}$ \\
\midrule
PKS 0047+023&LBL&  1.5&<-13.1&<2.4&  5.8&46.0&17.5&-&16.7&-0.9&12.8&6.0&<46.9&43.4&3.0&5.7&2.5\\
PKS 1514+197&LBL&  1.1&<-13.1&<2.4&  5.8&44.6&16.8&-&16.1&-0.1&11.4&6.0&<46.3&42.8&2.5&4.6&2.1\\
PKS B1908-201&FSRQ&  1.1&<-13.1&<2.5&  5.9&46.1&17.6&17.9&15.7&-0.1&20.9&6.0&<45.3&42.7&2.2&4.1&1.9\\
PKS 0113-118&FSRQ&  0.7&<-13.2&<2.3&  5.6&45.8&17.4&17.9&17.0&-0.5&6.9&6.0&<46.6&43.7&2.0&4.7&2.1\\
PKS 0048-09&HBL&  0.6&<-13.2&<2.5&  5.7&45.3&17.1&-&16.7&-0.2&7.6&6.0&<45.8&43.4&2.9&5.8&2.0\\
OJ 287&IBL&  0.3&<-13.2&<2.5&  5.8&44.7&16.9&-&17.4&-1.1&9.1&6.0&<46.6&43.5&2.8&5.1&2.5\\
PKS 1256-220&FSRQ&  1.3&<-13.2&<2.6&  5.9&46.0&17.5&18.0&15.3&-0.2&23.3&6.0&<45.5&42.3&2.3&4.2&2.0\\
PKS 0627-199&LBL&  1.7&<-13.2&<2.5&  5.8&46.1&17.6&-&16.7&-0.6&15.8&6.0&<46.6&43.3&2.2&4.6&1.9\\
PKS 0507+17&FSRQ&  0.4&<-13.2&<2.6&  6.0&44.9&16.9&17.3&15.8&-0.3&12.3&6.0&<45.4&42.8&1.2&4.6&2.2\\
S4 1738+49&FSRQ&  1.5&<-13.2&<2.8&  6.1&45.8&17.4&17.6&16.0&-0.0&24.7&6.0&<45.0&42.2&2.4&5.1&2.6\\
PKS B1130+008&LBL&  1.6&<-13.2&<2.4&  5.7&46.4&17.7&-&15.0&0.2&11.2&6.0&<45.5&43.4&2.9&4.3&1.2\\
PKS 1313-333&FSRQ&  1.2&<-13.3&<2.9&  6.1&45.9&17.4&17.7&16.3&-0.2&19.3&6.0&<45.1&42.7&2.7&7.0&2.8\\
B2 1128+38&FSRQ&  1.7&<-13.3&<2.6&  5.7&46.4&17.7&17.9&17.1&-0.5&9.9&6.0&<46.0&43.6&2.1&4.2&2.0\\
S3 0013-00&FSRQ&  1.6&<-13.3&<2.7&  5.9&45.9&17.5&17.6&16.1&-0.5&15.0&6.0&<45.5&42.9&2.7&4.5&2.9\\
B3 1708+433&FSRQ&  1.0&<-13.4&<3.0&  6.1&45.0&17.0&17.2&16.2&-0.4&19.2&6.0&<44.3&42.6&1.8&4.0&2.0\\
PKS 2329-16&FSRQ&  1.1&<-13.4&<2.5&  5.5&45.5&17.2&17.8&16.8&-0.7&7.1&6.0&<46.9&43.7&2.1&4.6&2.1\\
TXS 1055+567&IBL&  0.1&<-13.4&<2.8&  5.9&43.8&16.4&-&16.7&-0.5&6.3&6.0&<45.5&42.3&3.7&4.8&1.3\\
GB6 J1001+2911&LBL&  0.6&<-13.4&<2.6&  5.6&44.1&16.5&-&17.3&-0.7&4.3&6.0&<46.9&44.1&2.2&5.0&1.8\\
S5 0716+71&IBL&  0.1&<-13.4&<2.9&  6.1&44.5&16.7&-&16.3&-0.5&10.4&6.0&<44.9&42.8&1.1&4.8&2.1\\
SBS 0846+513&NLS1&  0.6&<-13.5&<2.6&  5.6&44.8&16.9&-&16.4&-0.8&7.0&6.0&<46.4&43.5&1.8&4.5&1.7\\
S3 2150+17&IBL&  0.9&<-13.5&<2.7&  5.7&44.4&16.7&-&16.8&-0.8&10.8&6.0&<46.8&43.1&2.9&4.9&2.5\\
PKS 1244-255&FSRQ&  0.6&<-13.5&<3.3&  6.3&45.9&17.4&17.7&16.4&-0.2&21.9&6.0&<44.1&42.3&2.1&4.6&2.6\\
B2 1215+30&HBL&  0.1&<-13.5&<3.0&  6.0&44.1&16.6&-&16.6&-1.0&8.6&6.0&<45.4&42.8&1.0&5.3&2.1\\
PKS 0235+164&IBL&  0.9&<-13.5&<2.8&  5.7&45.0&17.0&-&17.3&-1.4&12.9&6.0&<46.4&43.9&2.8&4.8&2.2\\
GB6 J0814+6431&IBL&  0.2&<-13.5&<2.8&  5.8&44.0&16.5&-&16.5&-0.3&5.2&6.0&<46.2&43.0&2.0&4.5&1.6\\
GB2 1217+348&IBL&  0.2&<-13.6&<2.8&  5.9&44.0&16.5&-&16.6&-0.7&5.8&6.0&<46.5&42.6&2.1&4.7&2.1\\
OQ 530&LBL&  0.1&<-13.6&<2.8&  5.7&43.5&16.3&-&16.6&-0.1&4.7&6.0&<46.0&43.0&1.8&4.1&1.4\\
PKS 0116-219&FSRQ&  1.2&<-13.6&<3.1&  6.0&45.9&17.4&17.7&16.6&-0.5&15.7&6.0&<45.0&43.0&1.4&4.2&1.9\\
PKS 1502+036&NLS1&  0.4&<-13.6&<2.7&  5.5&44.8&16.9&-&16.5&-0.9&5.8&6.0&<46.6&43.5&1.6&4.2&1.4\\
TXS 1318+225&FSRQ&  0.9&<-13.6&<3.4&  6.3&45.6&17.3&17.5&16.3&-0.3&27.2&6.0&<43.9&41.9&2.0&5.4&2.8\\
B2 2155+31&FSRQ&  1.5&<-13.6&<2.8&  5.8&45.7&17.4&17.6&17.0&-0.7&8.2&6.0&<46.5&44.1&1.0&4.1&1.2\\
PKS 0338-214&IBL&  0.2&<-13.6&<3.0&  5.9&43.6&16.3&-&16.9&-0.7&8.8&6.0&<46.5&42.6&1.5&4.4&2.0\\
PKS 1454-354&FSRQ&  1.4&<-13.7&<3.2&  6.1&46.6&17.8&18.0&16.3&-0.4&21.0&6.0&<44.9&42.9&1.2&4.4&1.9\\
PKS 0700-465&FSRQ&  0.8&<-13.7&<3.0&  5.8&45.1&17.1&17.3&16.4&-0.4&11.1&6.0&<45.7&43.1&2.6&4.8&2.5\\
PKS 2131-021&LBL&  1.3&<-13.7&<2.8&  5.3&45.5&17.3&-&17.0&-0.3&5.4&6.0&<46.5&44.4&2.2&4.3&1.3\\
B3 0650+453&FSRQ&  0.9&<-13.7&<3.0&  5.8&45.0&17.0&17.2&16.4&-0.6&10.7&6.0&<45.8&43.2&2.4&5.0&2.2\\
PKS 2029+121&LBL&  1.2&<-13.7&<3.0&  5.8&44.8&16.9&-&16.4&-0.7&15.7&6.0&<46.1&42.9&2.2&4.4&2.0\\
PKS 2005-489&HBL&  0.1&<-13.8&<3.4&  6.2&43.3&16.1&-&16.7&-0.7&12.0&6.0&<44.1&41.4&4.0&5.2&1.9\\
PKS 0402-362&FSRQ&  1.4&<-13.8&<3.1&  5.7&46.6&17.8&18.1&16.7&-0.4&14.6&6.0&<45.2&43.6&2.5&4.5&2.4\\
4C +47.44&FSRQ&  0.7&<-13.8&<3.3&  6.0&45.6&17.3&17.5&16.6&-0.5&11.7&6.0&<44.9&43.0&2.5&7.0&2.8\\
PKS 0414-189&FSRQ&  1.4&<-13.9&<3.5&  6.1&46.0&17.5&17.7&16.3&-0.3&22.1&6.0&<44.4&42.6&1.2&4.4&2.4\\
S5 2023+760&LBL&  0.6&<-14.0&<3.2&  5.8&44.9&16.9&-&16.4&-0.5&10.7&6.0&<45.4&43.1&1.9&4.4&2.0\\
S4 1144+40&FSRQ&  1.1&<-14.0&<3.3&  5.8&46.0&17.5&17.8&16.7&-0.6&13.6&6.0&<45.7&43.2&2.7&4.1&2.1\\
PKS 0306+102&FSRQ&  0.9&<-14.0&<3.4&  5.9&45.2&17.1&17.5&16.4&-0.3&14.8&6.0&<45.2&42.9&2.6&6.4&2.9\\
4C +28.07&FSRQ&  1.2&<-14.1&<3.5&  6.0&46.4&17.7&17.9&16.8&-0.7&13.3&6.0&<45.2&43.7&1.5&4.2&2.0\\
GB6 J1439+4958&LBL&  0.2&<-14.1&<3.4&  5.8&44.0&16.5&-&16.0&-0.3&5.1&6.0&<45.3&42.4&2.6&5.1&2.3\\
MG2 J133305+2725&FSRQ&  2.1&<-14.1&<3.2&  5.5&46.6&17.8&18.0&16.8&-0.7&9.4&6.0&<46.2&43.8&2.0&4.5&2.1\\
PKS 0308-611&FSRQ&  1.5&<-14.1&<3.8&  6.1&46.2&17.6&17.8&16.4&-0.3&24.3&6.0&<44.3&42.5&2.3&3.8&2.1\\
B2 1436+37B&FSRQ&  2.4&<-14.2&<3.4&  5.7&46.4&17.7&18.3&15.9&-0.7&19.7&6.0&<45.4&43.0&2.7&6.0&2.7\\
MG2 J043337+2905&IBL&  1.0&<-14.2&<3.6&  6.0&45.6&17.3&-&17.1&-1.4&12.1&6.0&<46.5&43.4&1.4&5.0&1.9\\
OJ 014&LBL&  1.1&<-14.2&<3.3&  5.4&45.5&17.2&-&17.4&-1.5&6.8&6.0&<47.0&44.2&1.4&4.7&1.7\\
PKS 1144-379&LBL&  1.1&<-14.2&<3.3&  5.6&45.6&17.3&-&17.2&-0.8&7.4&6.0&<46.0&44.1&1.2&4.6&1.8\\
S4 1030+61&FSRQ&  1.4&<-14.2&<3.4&  5.7&45.7&17.3&17.8&16.8&-0.5&10.4&6.0&<45.6&43.8&1.1&4.5&1.7\\
PKS 0149+21&FSRQ&  1.3&<-14.2&<3.7&  6.1&45.4&17.2&17.6&15.6&-0.0&27.2&6.0&<44.6&42.0&2.1&3.8&2.6\\
S3 0458-02&FSRQ&  2.3&<-14.4&<3.6&  5.8&46.8&17.9&18.5&16.4&-0.4&20.9&6.0&<45.0&43.1&2.7&6.4&2.9\\
PKS 1424-41&FSRQ&  1.6&<-14.4&<3.8&  5.9&47.0&18.0&18.4&16.4&-0.5&22.1&6.0&<44.6&43.2&2.4&4.4&2.1\\
PKS 0420-01&FSRQ&  0.9&<-14.4&<3.8&  6.0&46.2&17.6&18.0&16.8&-0.2&17.6&6.0&<45.0&42.8&2.5&4.4&2.7\\
OC 457&FSRQ&  0.9&<-14.5&<4.0&  6.0&45.8&17.4&17.9&16.1&-0.2&22.7&6.0&<44.4&42.4&2.3&4.1&2.3\\
S4 0917+44&FSRQ&  2.2&<-14.5&<3.7&  5.5&46.9&17.9&18.3&17.2&-0.8&11.9&6.0&<45.6&44.3&2.5&4.2&1.7\\
TXS 1920-211&FSRQ&  0.9&<-14.5&<3.8&  5.7&46.0&17.5&18.1&16.9&-0.4&10.3&6.0&<45.3&43.5&2.0&4.3&1.7\\
PKS 0302-623&FSRQ&  1.4&<-14.6&<3.9&  5.9&46.4&17.7&18.2&16.2&-0.2&20.3&6.0&<44.7&42.9&1.1&4.4&2.4\\
PKS 1124-186&FSRQ&  1.1&<-14.6&<4.2&  6.1&46.1&17.6&17.8&16.7&-0.3&18.7&6.0&<43.8&42.7&2.4&5.7&2.5\\
\bottomrule
\end{tabular}
}
\end{table*}

\begin{table*}
  \centering
\resizebox{\textwidth}{!}{
  \begin{tabular}{llllllllllllllllll}
  
  \toprule
Source & Class & $z$ & $F_{\nu_\mu}$ & $N_{\nu_\mu}$ & $E_\nu^\mathrm{peak}$ & $L_\mathrm{disk}$ & $R_\mathrm{BLR}$ & $R_\mathrm{diss}$ & $R_\mathrm{b}^\prime$ & $B^\prime$ & $\Gamma$ & $\gamma_\mathrm{p}^\mathrm{max}$ & $L_\mathrm{p}^\prime$ & $L_\mathrm{e}^\prime$ & $\gamma_\mathrm{e}^\mathrm{min}$ & $\gamma_\mathrm{e}^\mathrm{max}$ & $\alpha_\mathrm{e}$ \\
\midrule
PKS 0250-225&FSRQ&  1.4&<-14.6&<3.9&  5.8&46.2&17.6&18.1&16.4&-0.8&17.5&6.0&<45.2&43.1&2.7&4.9&2.6\\
PKS 2320-035&FSRQ&  1.4&<-14.6&<3.9&  5.9&46.2&17.6&18.0&16.6&-0.4&17.3&6.0&<44.9&43.2&1.2&4.9&2.2\\
PKS 0130-17&FSRQ&  1.0&<-14.6&<3.8&  5.6&46.0&17.5&17.9&17.3&-0.5&8.9&6.0&<45.7&43.7&2.5&4.4&2.0\\
PKS B1149-084&FSRQ&  2.4&<-14.8&<3.9&  5.6&46.3&17.6&17.9&17.0&-0.5&10.7&6.0&<45.9&43.7&2.6&4.4&1.5\\
PMN J1344-1723&FSRQ&  2.5&<-15.0&<4.3&  5.9&46.0&17.5&17.8&16.1&-0.4&22.3&6.0&<44.9&42.8&2.0&4.2&1.5\\
B2 1504+37&FSRQ&  0.7&<-15.2&<5.0&  6.2&44.3&16.6&16.9&17.2&-0.6&20.4&6.0&<42.3&42.3&1.5&3.8&2.5\\
PKS B1921-293&FSRQ&  0.3&<-15.2&<4.7&  6.0&45.0&17.0&17.4&17.0&0.0&16.7&6.0&<43.8&42.3&1.1&6.9&2.5\\
PKS 1741-03&FSRQ&  1.1&<-15.2&<4.6&  5.9&46.5&17.7&18.3&16.7&-0.0&19.4&6.0&<44.2&42.6&1.0&6.3&2.5\\
3C 454.3&FSRQ&  0.9&<-15.3&<4.6&  5.8&46.3&17.6&18.0&16.9&-0.6&14.7&6.0&<43.2&44.0&2.6&4.0&1.8\\
B2 1520+31&FSRQ&  1.5&<-15.4&<4.8&  6.0&45.9&17.4&18.0&16.2&-0.8&26.9&6.0&<44.2&42.8&2.7&4.7&2.7\\
PKS 0440-00&FSRQ&  0.8&<-15.4&<4.9&  6.1&45.9&17.4&17.9&16.4&-0.5&19.4&6.0&<43.9&42.7&2.3&5.4&2.8\\
NRAO 512&FSRQ&  1.7&<-15.5&<5.0&  6.0&46.3&17.6&18.2&16.2&-0.5&27.3&6.0&<44.0&42.5&2.6&5.8&2.9\\
PKS 0454-234&FSRQ&  1.0&<-15.5&<4.9&  6.1&45.6&17.3&17.7&16.6&-0.5&20.9&6.0&<44.1&42.8&2.5&4.4&2.1\\
S4 1851+48&FSRQ&  1.2&<-15.6&<5.0&  6.0&45.0&17.0&17.5&16.4&-0.5&16.9&6.0&<44.6&42.6&2.3&4.6&2.1\\
RX J0011.5+0058&FSRQ&  1.5&<-15.8&<5.1&  5.8&45.5&17.2&17.8&16.6&-0.5&17.3&6.0&<44.6&42.7&2.5&4.2&2.8\\
PKS 0446+11&FSRQ&  2.1&<-15.8&<4.9&  5.5&46.5&17.8&18.3&17.3&-0.7&13.7&6.0&<44.8&43.7&1.1&4.3&1.8\\
PKS 1622-253&FSRQ&  0.8&<-15.9&<5.3&  5.8&45.7&17.4&17.9&16.8&-0.5&18.8&6.0&<43.6&42.7&1.2&4.4&2.2\\
PKS 1502+106&FSRQ&  1.8&<-16.0&<5.3&  5.9&46.2&17.6&18.0&16.8&-0.6&22.1&6.0&<43.8&43.3&2.4&4.3&2.0\\
S5 1053+70&FSRQ&  2.5&<-16.2&<5.5&  6.4&46.5&17.7&18.3&17.2&-0.3&24.8&6.0&<43.9&42.7&1.3&4.9&2.4\\
S5 2007+77&LBL&  0.3&<-16.3&<5.7&  6.0&44.5&16.8&-&16.4&-0.8&13.7&6.0&<43.3&42.6&1.9&4.2&1.9\\
PMN J2036-2830&FSRQ&  2.3&<-16.3&<5.7&  5.9&45.8&17.4&17.9&17.1&-0.4&20.4&6.0&<44.3&42.0&3.0&6.8&2.5\\
\bottomrule
\end{tabular}
}
\end{table*}

\end{appendix}

\end{document}